\documentclass[12pt]{iopart}
\makeatletter
\renewcommand\@appendixstar{\@@par
 \ifnumbysec 
 \@addtoreset{table}{section}
 \@addtoreset{figure}{section}\fi
 \setcounter{section}{0}
 \setcounter{subsection}{0}
 \setcounter{subsubsection}{0}
 \setcounter{equation}{0}
 \setcounter{figure}{0}
 \setcounter{table}{0}
 \def\thesection{\Alph{section}} % this line has been \def\thesection{Appendix \Alph{section}} before
 \def\theequation{\ifnumbysec
      \Alph{section}.\arabic{equation}\else
      \Alph{section}\arabic{equation}\fi}
 \def\thetable{\ifnumbysec
      \Alph{section}\arabic{table}\else
      A\arabic{table}\fi}
 \def\thefigure{\ifnumbysec
      \Alph{section}\arabic{figure}\else
      A\arabic{figure}\fi}}
\makeatother

\usepackage{iopams} 
\expandafter\let\csname equation*\endcsname\relax

\expandafter\let\csname endequation*\endcsname\relax

\newcommand{\apropto}{\mathrel{\vcenter{
  \offinterlineskip\halign{\hfil$##$\cr
    \propto\cr\noalign{\kern2pt}\sim\cr\noalign{\kern-2pt}}}}}

\usepackage{qcircuit}
\usepackage{amsmath} 
\usepackage{graphicx}% Include figure files
\usepackage{dcolumn}% Align table columns on decimal point
\usepackage{bm}% bold math
\usepackage[markings, customcolors]{hf-tikz}
\usepackage{circuitikz}
\usetikzlibrary{shapes}
\usepackage{caption}
\usepackage{subcaption}
\usepackage{booktabs}
\usepackage{xcolor}
\usepackage{fancybox}
\usepackage{hyperref}
\usepackage{listings}
\usepackage{braket}
\usepackage{hyperref}

\definecolor{DarkGreen}{RGB}{0,75,0}

\newcommand{\bigvartheta}{\scalebox{1.5}{$\vartheta$}}

\begin{document}

\title{Towards Scalable Bosonic Quantum Error Correction}

\author{B. M. Terhal$^{1,2}$, J. Conrad$^{3,4}$, C. Vuillot$^{5}$}
\address{$^1$QuTech, Delft University of Technology, P.O. Box 5046, 2600 GA
    Delft,
  The Netherlands}
\address{$^2$ JARA Institute for Quantum Information, Forschungszentrum Juelich, D-52425 Juelich, Germany}
\address{$^3$ Dahlem Center for Complex Quantum Systems, Physics Department, Freie Universit\"at Berlin, Arnimallee 14, 14195 Berlin, Germany }  
\address{$^4$ Helmholtz-Zentrum Berlin f\"ur Materialien und Energie, Hahn-Meitner-Platz 1, 14109 Berlin, Germany }  
\address{$^5$ Inria Paris, 2 rue Simone Iff, 75012 Paris, France}

%\address{IOP Publishing, Temple Circus, Temple Way, Bristol BS1 6HG, UK}
%\ead{submissions@iop.org}
%\vspace{10pt}
%\begin{indented}
%\[]August 2017
%\end{indented}

% photon parity \Pi_{\rm photon}, Pauli =P, pump is {\cal E}

\begin{abstract}
We review some of the recent efforts in devising and engineering bosonic qubits for superconducting devices, with emphasis on the Gottesman-Kitaev-Preskill (GKP) qubit. We present some new results on decoding repeated GKP error correction using finitely-squeezed GKP ancilla qubits, exhibiting differences with previously studied stochastic error models. We discuss circuit-QED ways to realize CZ gates between GKP qubits and we discuss different scenarios for using GKP and regular qubits as building blocks in a scalable superconducting surface code architecture.
\end{abstract}

\maketitle

\tableofcontents

\clearpage

\markboth{}{}

\section{Introduction}

There has been a recent surge in interest in bosonic error correction, both from the experimental as well as from the theoretical side. By bosonic quantum error correction we mean the representation of a qubit as a two-dimensional subspace of an oscillator, a means of performing some error correction on this qubit, as well as a suite of techniques to perform universal computation on the qubit.

We review some of these recent developments and older proposals, with an eye towards integration of the ideas into a scalable (code) architecture. To be concrete, we concentrate on superconducting devices as physical realizations, due to the excellent control and engineerability of strong non-linearities, as described by the formalism of circuit quantum electrodynamics (circuit-QED). For more background, we refer the reader to a recent review of circuit-QED \cite{BGGW:cQED} and also the realization of quantum error correction in circuit-QED \cite{BGO:cQED}.

Due to the commonality of the quantum optics language, some of our discussion applies more generally to other physical systems realizing oscillators, such as optical modes or mechanical oscillators. Our paper does not aim to be comprehensive in reviewing all possible bosonic codes, but rather seeks to identify some promising approaches and future work to be undertaken, in particular emphasizing scalable bosonic GKP error correction.

A first condition to even consider encoding a qubit into an oscillator is that a high-Q oscillator is available \footnote{The $Q$ of the oscillator captures the number of oscillations until it is fully damped and is given by $Q=\frac{\omega}{\kappa}$ with $\omega=2 \pi f$ the angular frequency of the oscillator and $\kappa$ the oscillator decay rate.}. Examples of such high-Q oscillators are microwave cavity modes, of 3D or co-planar resonators in a frequency range $f=3-10$ GHz where single-photon life-times can be up to $\tau=1/\kappa=1-10$ ms \cite{reagor+:3D, reagor+:10ms}. Thus, without additional couplings and drives, the native noise model of such microwave cavities is simply photon loss, governed by the cavity decay rate $\kappa$. 

In order to prepare and manipulate an encoded qubit as prescribed by some code, one induces additional errors which the chosen code should, ideally, also be able to correct. It is thus important to pick a code in which computational manipulations and error corrections are relatively simple and the chosen code can also handle the errors which occur in these processes. As is well known, no finite code can correct all errors, and hence the goal of bosonic quantum error correction is simply to provide a logical qubit which can be used as a building block in a further coding scheme. The repetition or surface codes are the simplest examples of such further encoding steps, using then multiple oscillators.

Even though we can embrace the surface code as the simplest scalable 2D coding scheme \cite{terhal:review}, variants on who is playing the role of data and ancilla qubit and what additional error correction on these qubits takes place, are important in actually getting the very demanding engineering it, to pan out. If we have learned anything over the past 20 years of Hamiltonian engineering it is that partially-coherent dynamics can be implemented in many quantum systems, while very few to none may allow for the high-precision control and scalability needed for quantum error correction.

%The platform of multi-layer microwave-integrated quantum circuits \cite{brecht+:MMIQC, lei+:MMIQC} in which 3D cavities each storing high-Q mode play a central role is the platform of choice to pursue scalable bosonic error correction.

%It has been previously suggested that the use of continuous information may be boon in error correction \cite{}, leading to higher thresholds as more information is available. We believe that this point of view is partially misleading: in fact we will show that when using continuous-variable encodings extracting only discrete information leads to the impossibility of quantum information storage. Better said, using bosonic encoding {\em requires} continuous error information to be extracted.

% also true for other bosonic codes, comment is specific to GKP here..

%Beyond how a bosonic code can handle photon loss, it may be thus be equally or even more important to pick a code for which preparation and computational manipulation are simple and robust,.
% comments on bosonic cat & binomial not being so useful in this respect
% also encoding has a disadvantage of making gates intrinsically more complicated not every rotation is allowed..

% give frequencies of oscillator, give Q

Overall, the challenges of efficiently using a bosonic qubit encoding are in (1) keeping the harmonicity of the oscillators as high as possible while temporally coupling to this mode, with high on/off ratio, to create and manipulate non-classical code states, (2) finding a photon number regime in which approximations of engineered Hamiltonians are accurate while the error correcting properties and benefits of the encoding are valid. When using bosonic qubits as basic qubits in a code architecture, it may be advantageous to choose data qubits differently than ancilla qubits and we will give some examples of such choices. The simplest encoding of a single qubit into a bosonic mode can be done using Fock states: the vacuum state represents the logical $\ket{0}$, denoted as $\ket{\overline{0}}$, and a single-photon state represents the logical $\ket{1}$, denoted as $\ket{\overline{1}}$. 
%If we view the transmon qubit simply as slightly anharmonic oscillator, we may also see a transmon qubit as realizing such Fock encoding. 

For superconducting devices one can view the difference between bosonic encoding versus the regular transmon qubit encoding \cite{koch+:transmon} as an interchange between the roles played by the anharmonic and the harmonic oscillator. Using transmon qubits to store information, resonators are used as couplers and for read-out.  Using bosonic qubits to store information, anharmonic oscillators can be used for state preparation and couplers generating effective nonlinearities to realize gates. 
In this review we will refer to systems in which the lowest two energy eigenstates (in the absence of couplers) are used as {\em regular qubits}: this definition covers a Fock encoding as well as a transmon or a fluxonium qubit.

\subsection{Preliminaries $\&$ Notation}

Here we collect a few definitions and mathematical identities that are used throughout the paper. Additionally, useful textbooks for quantum optics and its mathematical description are \cite{book:HR}, \cite{book:GK} and \cite{book:carm}. We use $\hat{q}=\frac{1}{\sqrt{2}} (a+a^{\dagger})$ and $\hat{p}=\frac{i}{\sqrt{2}}(a^{\dagger}-a)$, where $a$ ($a^{\dagger}$) are annihilation (creation) operators,
so that $[\hat{q},\hat{p}]=i I$ and we sometimes refer to $\hat{p}$ and $\hat{q}$ as quadratures. A displacement in phase space is denoted as $D(\alpha)=\exp(\alpha a^{\dagger}-\alpha^*a)$ and acts as $D^{\dagger}(\alpha) a D(\alpha)=a+\alpha$, while a coherent state is defined as $D(\alpha) \ket{0}=\ket{\alpha}=\exp(-|\alpha|^2/2) \sum_{n=0}^{\infty} \frac{\alpha^n}{n!}\ket{n}$. We have $e^{i \theta a^{\dagger} a} a e^{-i \theta a^{\dagger}a}=a e^{-i \theta}$ so that $e^{i \theta a^{\dagger} a} D(\alpha) e^{-i \theta a^{\dagger}a}=D(\alpha e^{i \theta})$. The following identities hold
\begin{align}
\exp(-i v \hat{q}) \ket{p}=\ket{p-v},\; \exp(-i u \hat{p}) \ket{q}=\ket{q+u},\
\ket{p}=\frac{1}{\sqrt{2\pi}} \int_{\mathbb{R}} dq \,e^{ i p q} \ket{q},
\label{eq:usef}
\end{align}
so that
\begin{align}
\exp(i v \hat{q}) \hat{p} \exp(-i v \hat{q})=\hat{p}-v, \;\;\exp(i u \hat{p}) \hat{q} \exp(-i u \hat{p})=\hat{q}+u.
\label{eq:id}
\end{align}

A single-mode squeezing transformation is given by $\exp(-i H_{\rm sq} t)={\rm Sq}(\xi)=\exp(\frac{1}{2}(\xi^* a^2-\xi {a^{\dagger}}^2))$ with Hamiltonian $H_{\rm sq}={\cal E}{a^{\dagger}}^2+{\cal E}^* a^2$ with $\xi=2 i {\cal E} t$. The squeezer enacts the mode transformation $a_{\rm out}={\rm Sq}^{\dagger}(\xi) a {\rm Sq}(\xi)=a \cosh(r)-a^{\dagger} e^{i \theta} \sinh(r)$ with $r=|\xi|$ and $\theta=\arg(\xi)$. 

In Lindblad equations we use the notation ${\cal D}(A)(\rho)=A\rho A^{\dagger}-\frac{1}{2}\{ A^{\dagger} A, \rho \}$ for some operator $A$. 

It is standard to denote gates acting on a logical qubit subspace with overlines, i.e. $\overline{\rm CNOT}$ etc. In order to avoid notation clutter, only in Section \ref{sec:log-gates} we denote logical gates on the GKP codewords without it, i.e. CNOT and $Z$ instead of $\overline{\rm CNOT}$ and $\overline{Z}$.

%enacting $\hat{p} \rightarrow e^r \hat{p}$, and $\hat{q} \rightarrow e^{-r} \hat{q}$. \xi=r e^{i \theta}, squeezing parameter r.
% rotated quadratures \hat{q}_{\theta}=\cos(\theta/2)\hat{q}+\sin(\theta/2) \hat{p}, \hat{p}_{\theta}=-\sin(\theta/2) \hat{q}+\cos(\theta/2)\hat{p}
% 
%A general beam-splitter between modes $a$ and $b$ is the evolution ${\rm BS}(\theta,\phi)=\exp(-i g t (e^{i \phi} a^{\dagger} b+e^{-i \phi} b^{\dagger} a))$ with $-ig t=\theta$.
% include?

\section{Bosonic Qubits and Their Components}

\subsection{Early Birds $\&$ Cats and Their Generalizations}

%\section{Bosonic Qubits Modern Qubits Beyond The Transmon: (Heavy)-Fluxonium, 0-$\pi$, GKP, Kerr-Cat}

% history, first proposals

The first bosonic codes were formulated in \cite{CLY:bos} and designed to protect against photon loss. Of particular interest is a two-mode code with codewords
\begin{equation}
\ket{\overline{0}}=\frac{1}{\sqrt{2}}(\ket{40}+\ket{04})\; ,\;\ket{\overline{1}}=\ket{22},
\label{eq:2mode-cat}
\end{equation}
with $\ket{k}$ denoting a Fock state with $k$ photons. If either $\ket{\overline{0}}$ or $\ket{\overline{1}}$ (or both) were hit by the loss of a single photon on any one of the two modes, we can readily see that the resulting states would still be orthogonal. This orthogonality is a prerequisite for being able to correct the photon loss error, but it is not a sufficient condition. To examine the error correction capability of a (bosonic) code, one asks whether a set of dominant errors satisfies the quantum error correction (QEC) conditions \cite{book:NC} of the code: if this holds (approximately) then there is an (approximate) recovery operation undoing these dominant errors. For a set of errors ${\sf E}=\{E_1, \ldots, E_k\}$ acting on the encoding of a single qubit, the quantum error conditions are as follows. $\forall i,j$, we require
\begin{align}
\bra{\overline{0}} E_i^{\dagger} E_j \ket{\overline{0}}=\bra{\overline{1}} E_i^{\dagger} E_j \ket{\overline{1}} \mbox{ logical states are indistinguishable}, 
\label{eq:qec1} \\
\bra{\overline{0}} E_i^{\dagger} E_j \ket{\overline{1}}=\bra{\overline{1}} E_i^{\dagger} E_j \ket{\overline{0}}=0 \mbox{ orthogonality remains.}
\label{eq:qec2}
\end{align}
To examplify the use of these conditions, let us first look at the single-mode version of the code in Eq.~(\ref{eq:2mode-cat}):
\begin{equation}
\ket{\overline{0}}=\frac{1}{\sqrt{2}}(\ket{0}+\ket{4})\;,\; \ket{\overline{1}}=\ket{2}.
\label{eq:kitten}
\end{equation}
This code was introduced in \cite{michael+:binomial} as the smallest member of a family of so-called binomial codes, hence its name kitten or `baby-binomial' code. This code and its logical gates has been implemented using a superconducting microwave cavity mode as an oscillator in Ref. \cite{hu+:baby}, but the life-time of the encoded qubit was comparable to that of a Fock state encoding.
One can easily check that for the error set ${\sf E}=\{I,\sqrt{\gamma} a\}$, the QEC conditions in Eqs.~(\ref{eq:qec1}),(\ref{eq:qec2}) for this code are met.  However, these errors are only an approximation of the real noise. A photon loss channel with photon decay rate $\kappa$ lasting for time $t$ with $\gamma \equiv \kappa t \ll 1$, can be modeled by a superoperator ${\cal N}_{\gamma}$ with Kraus operators $E_0=I-\frac{\gamma}{2} a^{\dagger} a+O(\gamma^2)\approx e^{-\gamma a^{\dagger}a /2}$ and $E_1=\sqrt{\gamma} a$, or
\begin{equation}
{\cal N}_{\gamma}(\rho)=E_0 \rho E_0^{\dagger}+E_1 \rho E_1^{\dagger}, \; E_0^{\dagger}E_0+E_1^{\dagger} E_1 =I + O(\gamma^2).
\label{eq:channel-pl}
\end{equation}
For the Kraus operators $E_0$ and $E_1$ the QEC conditions in Eqs.~(\ref{eq:qec1}),(\ref{eq:qec2}) are not quite met. In particular, we have 
\begin{equation}
\bra{\overline{0}} E_0^{\dagger} E_0 \ket{\overline{0}}-\bra{\overline{1}} E_0^{\dagger} E_0 \ket{\overline{1}}=O(\gamma^2) (\bra{\overline{0}} (a^{\dagger} a)^2 \ket{\overline{0}}-\bra{\overline{1}} (a^{\dagger} a)^2 \ket{\overline{1}}) \neq 0,
\end{equation}
as $\ket{\overline{1}}$ is an eigenstate of $a^{\dagger} a$, while $\ket{\overline{0}}$ is not. This means that upon the detection of no photon loss (corresponding to $E_0$) the code states undergo an irreversible distortion. The two-mode version of this code, Eq.~(\ref{eq:2mode-cat}), improves on this distortion issue as the quantum error correction conditions for the two mode code are met for the error set ${\sf E}=\{ \sqrt{\gamma} a, \sqrt{\gamma}b, \exp(-\frac{\gamma}{2}(\hat{n}_a+\hat{n}_b))\approx I-\frac{\gamma}{2}(\hat{n}_a+\hat{n}_b)\}$. These error operators can be viewed as the three Kraus operators of a process in which there is either photon loss on mode a, photon loss on mode b, or no photon loss on either modes. For the states in Eq.~(\ref{eq:2mode-cat}) we have no distortion upon not detecting a photon from either modes as $\ket{\overline{0}}$ and $\ket{\overline{1}}$ are both eigenstates of 
$\exp(-\frac{\gamma}{2}(\hat{n}_a+\hat{n}_b))$ with eigenvalue $\exp(-2\gamma)$. As far as we know, this two-mode code is still awaiting experimental realization.

% symmetrize via phase flip..? Debbie idee, drop here, BMT debbie idea

By allowing ourselves code states with higher average photon number, we can correct for more loss errors, as well as gain and dephasing errors. More precisely, Ref.~\cite{michael+:binomial} has introduced families of binomial and so-called cat codes which correct against the set of errors ${\sf E}=\{I,a, \ldots, a^L, a^{\dagger}, \ldots, {a^{\dagger}}^G, a^{\dagger} a, \ldots,  (a^{\dagger} a)^D\}$ for arbitrary $L,G$ and $D$. For example, the idea behind the binomial codes can be understood as follows. Using the Holstein-Primakoff transformation $a^{\dagger} a \rightarrow J_z+J$, the binomial code words $\ket{\overline{0}}$ and $\ket{\overline{1}}$ can be seen as spin-eigenstates of $J_x=\pm J$ with $2J=N+1$, where one defines $N=\max(L,G,2D)$. Dephasing errors $(a^{\dagger} a)^m$, $m=1, \ldots, D$ thus lead to a change in $J_x$ by at most $D \leq \lfloor J\rfloor$, hence keeping codewords orthogonal. At the same time, protection against photon loss and gain is achieved by using a subspace of sufficiently separated Fock states stabilized by the operator $\Pi_S=e^{i 2\pi a^{\dagger} a/(S+1)}$ with $S=L+G$. For $S=1$ this gives the photon parity operator $\Pi_{S=1} \equiv \Pi_{\rm photon}=e^{i \pi a^{\dagger} a}$: the even-photon codewords in Eq.~(\ref{eq:kitten}) are clearly $+1$ eigenstates of this photon parity operator.

% cat hierarchy and its generalizations

Another family of single-mode codes are the cat codes. A very simple encoding is $\ket{\overline{0}}\approx \ket{\alpha}$ and $\ket{\overline{1}} \approx \ket{-\alpha}$  with coherent state $\ket{\alpha}$, first proposed in \cite{CMM:cat, mirrahimi+:cat}. Since $\ket{\alpha}$ and $\ket{-\alpha}$ are not orthogonal, it is more appropriate to define the code states as $\ket{C_{\alpha}^{\pm}}=\frac{1}{\sqrt{N_{\pm}}}(\ket{\alpha}\pm \ket{-\alpha})$ with $N_{\pm}=2(1\pm \exp(-2 |\alpha|^2))$. These states are orthogonal for all $\alpha$ and we can define $\ket{\pm} \equiv \ket{C_{\alpha}^{\pm}}$. On this encoding, photon loss induces immediate phase-flip errors since $a \ket{C_{\alpha}^{\pm}} \propto \ket{C_{\alpha}^{\mp}}$. Thus the phase-flip error rate (probability per unit time) is proportional to $\kappa |\alpha|^2$ with $\kappa$ the photon loss rate of the encoding mode. 

On the other hand, for large enough $\alpha$, bit-flips, $\alpha \leftrightarrow -\alpha$, can be expected to occur at a much lower error rate as they correspond to a large change of the state in phase space. Particularly interesting is the engineering of Hamiltonians or dissipative processes which have these code states $\ket{\pm \alpha}$ as degenerate fixed-points, so that there is a  `macrosopic' energy barrier to transition between them, leading to a bit flip rate exponentially small in $|\alpha|^2$. This design can lead to a qubit for which the noise is biased as phase-flip errors are more prominent than bit-flip errors. We will discuss this noise-biased qubit in more detail in Section \ref{sec:bias-cat}.

The next-level cat encoding was introduced in \cite{mirrahimi+:cat, leghtas+:cat}, and is sometimes referred to as the 4-legged cat code since its codewords have four blobs in phase space:
\begin{align}
\ket{\overline{0}}=\frac{1}{\sqrt{N_0}}\left(\ket{\alpha}+\ket{-\alpha}+\ket{i\alpha}+\ket{-i\alpha}\right), & \;\;\ket{\overline{1}} =\frac{1}{\sqrt{N_1}}\left(\ket{\alpha}+\ket{-\alpha}-\ket{i\alpha}-\ket{-i\alpha}\right), & \notag \\
N_{b}=8 e^{-\alpha^2}(\cosh \alpha^2 + (-1)^b \cos \alpha^2), \alpha \in \mathbb{R}. 
\label{eq:cat}
\end{align}
Using the standard identity $\bra{\alpha} \beta \rangle=e^{-(|\alpha|^2+|\beta|^2)/2}e^{\alpha^*\beta}$ one can verify the orthogonality of these two states. As for the kitten code, we can verify that both states are $+1$ eigenstates of the photon parity operator $\Pi_{\rm photon}=e^{i \pi a^{\dagger} a}$ using that $\Pi_{\rm photon} \ket{\alpha}=\ket{-\alpha}$. The photon parity thus functions as a check operator, taking eigenvalue $+1$ on the code space and measuring it is a natural way to detect photon loss and perform error correction. 

The states $\ket{\overline{0}}$ and $\ket{\overline{1}}$, both having even photon parity, are however distinguished by their photon parity modulo 4, expressed as the $\pm 1$ eigenvalues of the operator $\Pi_{\rm photon}^{1/2}=\exp(i \pi a^{\dagger} a/2)$. To measure the photon parity operator via an ancilla qubit, a cavity mode-qubit dispersive interaction $-\chi Z a^{\dagger} a/2 $ can be used \cite{book:HR}.  In circuit-QED the interaction $Z a^{\dagger} a$ comes about naturally as the effective interaction between, say, a cavity mode and a linearly-coupled, off-resonant, transmon qubit mode \cite{koch+:transmon}. The measurement of $\Pi_{\rm photon}$ then proceeds by preparing the ancilla qubit in $\ket{+}$, letting the interaction take place for time $t=\pi/\chi$ and subsequently measuring the qubit in the $\ket{\pm}$ basis. Using a transmon qubit and cavity mode, Ref. \cite{ofek+:cat} has shown that tracking the photon parity by repeated measurements of $\Pi_{\rm photon}$ makes for a logical qubit which has a longer life-time than a Fock qubit without error correction in the same cavity mode. This result has essentially been the first demonstration of quantum error correction lengthening the life-time as compared to that of native qubits (transmon and/or Fock encoding) in the hardware.

Before we discuss further generalizations, let us examine the quantum error correction conditions, Eq.~(\ref{eq:qec1}),(\ref{eq:qec2}) for this cat code with respect to the set of errors ${\sf E}=\{I, \sqrt{\gamma} a\}$. One can quickly observe that all conditions are obeyed except $\bra{\overline{0}} a^{\dagger} a \ket{\overline{0}} \stackrel{?}{=} \bra{\overline{1}} a^{\dagger} a \ket{\overline{1}}$. Besides the uninteresting case of taking $\alpha$ very large (so that all $\ket{\pm \alpha}, \ket{\pm i \alpha}$ are orthogonal), this last condition is {\em exactly} met at sweet spots given by the equation $\tan \alpha^2=-\tanh \alpha^2$. The smallest sweet-spot at $|\alpha|^2=2.34$ lies close to the number of photons $\bar{n}=2$ of the cat code used in the experiment \cite{ofek+:cat}.

There are several error channels which impact the performance of the cat code using repeated photon parity measurements. First of all, the code cannot fully correct against the photon loss channel as it cannot correct the distortion Kraus operator $E_0=I-\gamma a^{\dagger}a/2$. Secondly, two photon-loss events $\propto a^2$ implement a logical bit-flip $\ket{\overline{0}} \leftrightarrow \ket{\overline{1}}$. Thirdly, photon loss in combination with the inevitable Kerr nonlinearity $\sim (a^{\dagger} a)^2$ on the cavity mode causes incorrectable dephasing: the Kerr interaction makes the cavity rotation speed depend on the number of photons in the cavity, but this number becomes indeterminate in the presence of photon loss. Last but not least, transmon qubit decay during the qubit controlled-$a^{\dagger}a$ interaction, is a serious source of {\em feedback error}. For example, when the qubit decays half-way through the interaction, $\ket{1} \rightarrow \ket{0}$, it applies only half the rotation on the cavity mode. The result is that the eigenvalues of $\Pi_{\rm photon}^{1/2}=\overline{Z}$ are measured via the qubit measurement, collapsing the logical state. 

This last feedback error problem is an important issue for any bosonic qubit, and it has been a central theme in the theory of fault-tolerant computing in general \cite{preskill:faulttol}. A disadvantage of the theoretical schemes for fault-tolerant quantum error correction is that they typically require additional hardware resources, such as logical ancilla qubits or (verified) multi-qubit GHZ states. Instead, we may seek hardware-efficient mitigation of the feedback error problem. As an example, Ref.~\cite{rosenblum+:ft} has addressed the feedback error due to transmon relaxation by drive-engineering the dispersive coupling Hamiltonian to equal $-\chi (\ket{2}\bra{2}+\ket{1}\bra{1}) a^{\dagger} a$ and starting the ancilla transmon qubit in the state $\frac{1}{\sqrt{2}}(\ket{0}+\ket{2})$. Transmon qubit decay from $2 \rightarrow 1$ then commutes with the transmon-cavity interaction and does not cause errors on the cavity mode. The decay does, --as in the normal case--, affect the reliability of the transmon qubit measurement outcome. All-in all, this has led to an overall factor 5 in improvement of the life-time of the encoded cat qubit \cite{rosenblum+:ft}.

Another way of minimizing feedback errors on a bosonic code is to use a biased-noise ancilla qubit (Section \ref{sec:bias-cat}) as an ancilla qubit. As proposed in Ref.~\cite{puri+:Kerr-cat}, the goal is then to let the strong-noise error channel affect the ancilla qubit measurement, while the low-noise (bit-flip) channel on the ancilla feeds back low-noise to the bosonic code. 

% some generalizations, of course there are many more..

Single-mode cat codes with higher-photon numbers can be formulated and form a class of codes \cite{michael+:binomial, BL:cats}. Ref.~\cite{albert+:photon-loss} studied the performance of binomial, cat and GKP codes (see Section \ref{sec:GKP}), against photon loss, assuming optimal noisefree recovery as permitted by the quantum error correcting conditions. Ref.~\cite{GCB:rot} has formulated a general framework of rotation-symmetric codes of which the binomial and cat codes are subclasses: the unifying theme is rotation symmetry of the code states in phase space captured by invariance under the operator $\Pi_S$. Another interesting class of bosonic codes uses a 3-wave mixing $\chi^{(2)}$-interaction, Eq.~(\ref{eq:H-3wave}), as the central element for defining the code and correcting photon loss \cite{NCS:chi2-code}. Various classes of multi-mode codes against photon loss exist, see for example \cite{BL:noon, OR:multiloss, albert+:photon-loss} and references therein.

% Naturally, other generalizations of binomial, cat or GKP codes to two or more modes exist, encoding qubit(s) or qudits in several modes. 

%Two-mode bosonic qubits, \cite{wang+:2mode-cat} 
%$\ket{\overline{\pm}}=\ket{\alpha} \ket{\alpha} \pm \ket{-\alpha} \ket{-\alpha}$
% more of fundamental interest, invariant under parity operator for both modes

%In fact, it is partially for this reason that the surface code has emerged as one of the most viable code platforms as the ancilla feedback errors for surface code parity check measurements do not break fault-tolerance and hence can be tolerated. Whether such trade-offs between having fewer feedback errors versus (potentially) more ancilla qubit measurement errors are useful partially depends on what we do next with the bosonic qubit.  Whether this is a good idea partially depends on the trade-off of having less incoming or feed-back errors what is more harmful when using a surface code: more incorrect measurement outcomes or more incoming errors.In any case \cite{andrist+:pq-boundary}.

% concluding thoughts on use of bosonic qubits

A challenge in using a bosonic qubit is that some computational manipulations can be more involved than for a regular qubit. For example, on a regular qubit such as a transmon qubit, rotations by an angle $\theta$ around axes $X$ or $Y$ are easily accomplished by temporarily supplying microwave radiation. On a bosonic qubit, these simple single-qubit gates can be non-trivial. An advocated solution in Ref.~\cite{LP:universal} is to always use a dual-rail (dr) encoding of a bosonic qubit with $\ket{\overline{0}}_{\rm dr}=\ket{\overline{0}\overline{1}}$ and $\ket{\overline{1}}_{\rm dr}=\ket{\overline{1}\overline{0}}$, where $\ket{\overline{0}}$ and $\ket{\overline{1}}$ are the states of (an arbitrary) bosonic qubit itself. Having mapped the Bloch-sphere of a qubit onto a two-mode state space, the exponential mode-SWAP operator $\exp(i \theta \;{\rm SWAP}_{a,b})$ becomes a universal gate to do single-qubit and two-qubit gates, and has been realized in \cite{gao+:eswap}. Here the linear ${\rm SWAP}_{a,b}$ transformation interchanges two modes a and b, i.e. its action on quadrature operators for the modes is given by $q_a \leftrightarrow q_b$ and $p_a \leftrightarrow p_b$.
If we envision using a bosonic qubit as a building block qubit in a stabilizer code, it is however not necessary to perform any gate, but rather we can focus on performing CNOT or CZ, Hadamard ($H$) and $T$ gates possibly using ancilla qubits, see e.g. Ref.~\cite{terhal:review}.

% BMT CV I do not define T gate as review barely deals with it. CNOT, CZ and Hadamard should be standard I think

\subsection{Noise-Biased Cat Qubit}
\label{sec:bias-cat}

A method to set up a dissipative process which stabilizes the coherent states $\ket{\pm \alpha}$ was devised in \cite{mirrahimi+:cat}. The idea is to engineer the Lindblad equation (in a frame rotating at the mode frequency):
\begin{equation}
\dot{\rho}=-i [H_{\rm sq},\rho]+\kappa_{2ph} {\cal D}(a^2)(\rho) \equiv {\cal L}(\rho), 
\label{eq:lb-cat}
\end{equation}
with $H_{\rm sq}={\cal E} {a^{\dagger}}^2+{\cal E}^* a^2$ where ${\cal E}=i |{\cal E}|$ with $|{\cal E}|$ proportional to the strength of a pumped microwave mode acting as a classical field. To understand the fixed points of this evolution, --$\rho$ for which ${\cal L}(\rho)=0$--, we can write the Lindblad equation as 
\begin{equation}
\dot{\rho}=-i (H_{\rm eff} \rho-\rho H_{\rm eff}^{\dagger})+\kappa_{2ph} a^2(\rho){a^{\dagger}}^2, 
\label{eq:lb-complex}
\end{equation}
with $H_{\rm eff}=H_{\rm sq} -\frac{i\kappa_{2ph}}{2} {a^{\dagger}}^2 a^2$. We can then use, with $K=\frac{\kappa_{2ph}}{2}$:
\begin{align}
-i H_{\rm eff} = -K {a^{\dagger}}^2 a^2+ |{\cal E}|({a^{\dagger}}^2 -a^2)  = & -K \tilde{M}_{\alpha}^{\dagger} M_{\alpha}-\frac{|{\cal E}|^2}{K}, \notag \\
{\rm with}\;M_{\alpha} = a^2-\alpha^2 I,\;\; \tilde{M}_{\alpha}=a^2+\alpha^2 I,\; & \alpha=\sqrt{\frac{|{\cal E}|}{K}}. 
\label{eq:Kwrite}
\end{align}
This immediately implies that the states $\ket{\pm \alpha=\pm \sqrt{|{\cal E}|/K}}$ are fixed points of the Lindblad evolution, as $M_{\alpha} \ket{\pm \alpha}=0$, and the last term in Eq.~(\ref{eq:lb-complex}) is canceled by the constant $\frac{-|{\cal E}|^2}{K}$ which remains from the first term. Hence, any linear combination of the states $\ket{\alpha=\pm \sqrt{|{\cal E}|/K}}$ is a fixed point of the dynamics.

When the pump inducing the squeezing Hamiltonian $H_{\rm sq}$ is off, ${\cal E}=0$, we can observe that the Fock states $\ket{0}=\lim_{\alpha \rightarrow 0} \ket{C_{\alpha}^+}$ 
and $\ket{1}=\lim_{\alpha \rightarrow 0}\ket{C_{\alpha}^-}$ are fixed points, distinguished by their photon parity. Thus when ${\cal E}$ is gradually increased, we can smoothly change from a Fock encoding into the cat $\ket{C_{\alpha}^{\pm}}$ encoding.
Photon loss at rate $\kappa$, which can be modeled by introducing an additional term $\kappa {\cal D}(a)(\rho)$ in Eq.~(\ref{eq:lb-cat}), causes phase-flip errors, i.e. flipping between the states $\ket{C^{\pm}_{\alpha}}$, but does not interfere with the stabilization itself as $\ket{\pm \alpha}$ are eigenstates of $a$ so that ${\cal D}(a)(\ket{\pm \alpha}\bra{\pm \alpha})=0$. One can add a drive term $H_{\rm drive}=\epsilon(t) a^{\dagger}+\epsilon^*(t) a$ to the Lindblad equation and observe that the annihilation operator $a$ will generate rotations around the $Z$-axis (periodically interchanging $\ket{C_{\alpha}^{\pm}}$). At the same time, $a^{\dagger}$ in principle leads to a departure from the qubit subspace spanned by $\ket{\pm \alpha}$ corresponding to leakage. However, due the $\sim |\alpha|^2$ gap of the Lindbladian, such departure from the eigenvalue-0 manifold is exponentially suppressed and the effect of the driving term can be analyzed by projecting it onto the stabilized subspace. In this subspace it then induces Rabi oscillations around an axis which is exponentially-closely aligned with the $Z$-axis, with Rabi frequency $\Omega \propto |\epsilon| |\alpha|$, experimentally demonstrated in \cite{touzard+:diss} \footnote{In Refs.~\cite{mirrahimi+:cat}, \cite{grimm+:Kerr-cat} and some other papers a different convention is used, namely $\ket{C_{\alpha}^+}$ (resp. $\ket{C_{\alpha}^-}$) is the $Z$ eigenstate $\ket{0}$ (resp. $\ket{1}$), thus interchanging what is called $Z$ and $X$ here.}. A measurement in the $X$-basis can be accomplished by measuring the photon parity through a coupled transmon qubit.
% The Pauli $X$ rotations is harder, check epsilon Rabi oscillations
The (pumped) squeezing interaction and the required two-photon dissipative process have first been experimentally realized in \cite{leghtas+:conf}.
This was achieved by coupling a 3D storage cavity (at frequency $\omega_a$) via a bridging transmon to a lossy cavity (at different frequency $\omega_b$) and applying a two-tone drive on the lossy cavity so as to set up a process to convert two storage photons to one lossy cavity photon which is subsequently lost (the 
$\kappa_{2ph} {\cal D}(a^2)$ process). The lossy cavity is driven at pump frequency $\omega_p=2 \omega_a-\omega_b$ as well as close to its own frequency $\omega_b$, generating, through the transmon nonlinearity, an effective degenerate parametric oscillator with resonant terms of the form ${a^{\dagger}}^2 b+ b^{\dagger} a^2$. 
% what about the squeezing part.

% alpha limited by large kappa_2ph? E_J and needing lots of pumping??
% slow gate
% gap/driving 

%A similar dissipative mechanism to stabilize the cat code states in Eq.~(\ref{eq:cat}) was also introduced in \cite{mirrahimi+:cat}, requiring however a challenging 4-photon dissipator $\kappa_{4ph} {\cal D}(a^4)$. Stabilization of a two-mode version of the (4-legged) cat code might then be easier and was considered in \cite{albert+:pair-cat}.

A more recent experimental realization in Ref.~\cite{lescanne+:cat} has been able to cleanly generate the desired interactions (via an effective 3-wave mixing, see also Sec.~\ref{sec:GKP-CZ}) and observe the exponential decrease of the bit-flip error rate in $|\alpha|^2$ as well as the linear increase of the phase-flip error rate with $|\alpha|^2$.
% BMT check that they do it the dissipative way, why can they do these CNOT then..no discussion on CNOT

An alternative, non-dissipative, route towards a noise-biased qubit was first proposed in \cite{PBB:Kerr-cat}. Instead of invoking dissipation, the idea is to engineer a Hamiltonian which has $\ket{\pm \alpha}$ as degenerate eigenstates, using a Kerr nonlinearity and squeezing. The two-photon dissipation is then considered an optional add-on which helps in mitigating leakage, i.e. a departure from the subspace spanned by $\ket{\pm \alpha}$. The target Hamiltonian (in the rotating frame of the cavity mode) is
\begin{align}
H=-K {a^{\dagger}}^2 a^2+ {\cal E}{a^{\dagger}}^2 +{\cal E}^* a^2 =-K M_{\alpha}^{\dagger} M_{\alpha}+\frac{|{\cal E}|^2}{K}, \notag \\
M_{\alpha}=a^2-\alpha^2, \; \alpha=\sqrt{\frac{|{\cal E}|}{K}}e^{i\varphi}, {\cal E}=|{\cal E}|e^{2 i \varphi}.
\label{eq:cassH}
\end{align}
The spectrum of $H$ has eigenvalues running from $\frac{|{\cal E}|^2}{K}$ downwards as the first term in $H$ is negative-semi-definite. Omitting the factor $\frac{|{\cal E}|^2}{K}$, the highest eigenstates are the states $\ket{\pm \alpha}$ with degenerate zero eigenvalues. We can observe the similarity and difference with Eq.~(\ref{eq:Kwrite}): here we consider a Hermitian matrix and the phase of the pump amplitude ${\cal E}$ is variable and determines the phase of the coherent states which are the zero energy eigenstates. Thus, by adiabatically changing the phase of ${\cal E}$ we move to different zero energy eigenstates, allowing us to transform $\alpha \rightarrow -\alpha$ and hence realize a $X$ gate on $\ket{C_{\alpha}^{\pm}}$.  For the stability of the encoded space it is important to understand the spectrum of $H$ and the gap below these degenerate zero eigenstates, see the analysis in \cite{PBB:Kerr-cat, puri+:Kerr-cat}. 
To understand this, assume that the phase $\varphi=0$ for simplicity. We can displace the Hamiltonian by $D(\pm \alpha)$ with $\alpha=\sqrt{|{\cal E}|/K}$. For large $|{\cal E}|/K$, one can approximate $D^{\dagger}(\pm \alpha) H D(\pm \alpha)\approx -4 K |\alpha|^2 a^{\dagger} a$, a harmonic oscillator Hamiltonian. This shows that for large $\alpha$, the spectrum approximately has the gap $4 K |\alpha|^2$ and the first excited states below $\ket{\pm \alpha}$ are roughly equal to $D(\pm \alpha)\ket{1}$. The so-called `Cassinian' Hamiltonian in Eq.~(\ref{eq:cassH}) was first studied in \cite{WB:cass}: the surfaces of constant classical energy are described by Cassinian ovals in $\langle p \rangle$ and $\langle q\rangle$ with the focii of the ovals at $\langle q\rangle =\pm \sqrt{|{\cal E}|/K}$. As a quantum system the spectrum is that of an inverted double-well (`double-oscillator') with the well maxima at zero energy for the states $\ket{\pm \alpha}$. We can consider the effect of driving and several dissipative processes for the Hamiltonian in Eq.~(\ref{eq:cassH}). For example, when one includes photon loss $\kappa {\cal D}(\rho)$ in the Lindblad equation and the pump amplitude is sufficiently large, i.e. $16|{\cal E}|^2 > \kappa^2$ \cite{Meaney2014, WS:para, PBB:Kerr-cat}, the fixed point of the Lindblad equation is the state $p\ket{\tilde{\alpha}}\bra{\tilde{\alpha}}+(1-p)\ket{-\tilde{\alpha}}\bra{-\tilde{\alpha}}$ with modified $\tilde{\alpha}$, $|\tilde{\alpha}|^2 < \frac{|{\cal E}|}{K}$. In this regime the system neatly represent the dissipative storage of a classical bit.
% BMT JC CV https://arxiv.org/pdf/1605.09408.pdf says on page 6 that p=1/2 is the only fixed point which seems incorrect

% P complex or real
% meaney: parametric drive = twice coherent drive

The effect of other sources of noise such as dephasing ($\kappa_{\rm deph}{\cal D}(a^{\dagger} a)(\rho)$, see also Eq.~(\ref{eq:stoch-deph})), photon gain ($\kappa \bar{n}_{\rm therm}{\cal D}(a^{\dagger})(\rho)$) due to the coupling with a finite temperature heat bath, as well coupling with baths with other spectral densities are discussed in detail in \cite{mirrahimi+:cat,PBB:Kerr-cat, puri+:Kerr-cat, puri+:bias}.

% BMT mode a is storage mode

Ref.~\cite{grimm+:Kerr-cat} has implemented the Kerr-cat Hamiltonian in Eq.~(\ref{eq:cassH}) and the corresponding qubit in the resonant mode of a so-called SNAIL element (see Section \ref{sec:GKP-CZ}), coupled to a read-out cavity mode. The fourth-order nonlinearity of the SNAIL element gives the wanted $-K {a^{\dagger 2}}a^2$ term, while one can drive the mode at twice its frequency so as to use the third-order SNAIL term $\propto {a^{\dagger}}^3+a^3$ to turn on squeezing. The experiment generated cat states with $|\alpha|^2 \approx 2.5$ with a dephasing life-time of $3\, \mu$s, and an enhanced decay life-time of $105\, \mu$s, and a $\pi/2$ rotation around the $Z$-axis obtained by driving took 24 ns.
The ability to convert the noise-biased qubit to a Fock encoding by turning off the squeezing drive allows to measure Pauli $X$ via a standard dispersive measurement \cite{grimm+:Kerr-cat}. One can also measure a noise-biased qubit in the $X$-basis by dispersively coupling ($-\chi a^{\dagger} a Z/2$) it to an ancilla qubit to map the photon parity onto the state of the ancilla qubit which is subsequently measured. To realize a (nondestructive) Pauli $Z$ measurement, distinguishing $\pm \alpha$, Ref.~\cite{grimm+:Kerr-cat} had applied, besides the squeezing drive, a drive at the difference frequency of the SNAIL mode and the read-out cavity mode (b) to get a resonant beam-splitting interaction $ \propto a^{\dagger} b + a b^{\dagger}$. The upshot is that the coherent states $\ket{\pm \alpha}$ are mapped to corresponding coherent states in the cavity mode which are heterodyne-measured when leaking out of the cavity. 

Given that the noise-biased cat qubit is designed to have a low bit-flip error rate, it can function as an ancilla control qubit in the error correction circuit for another code \cite{puri+:Kerr-cat} inducing low feedback noise. Assume we have a code which is an eigenspace of a stabilizer $S=e^{i A}$ and $S$ is to be measured using the noise-biased cat qubit to detect or correct errors. This requires an interaction of the noise-biased cat qubit and the code of the form $H_{\rm int} \propto (a+a^{\dagger}) \otimes A$ since $a+a^{\dagger} \approx Z$ on the noise-biased cat code space (besides some leakage), allowing for a qubit controlled-$S$ operation.  For example, for the cat code, $S=e^{i\pi b^{\dagger}b}=\Pi_{\rm photon}$, requiring a tunable photon-pressure coupling between the two modes of the form $H_{\rm int} \propto (a+a^{\dagger}) b^{\dagger} b$. For the GKP code, see Section \ref{sec:GKP}, $S$ is a displacement so that $H_{\rm int}$ can be chosen to be a tunable beam-splitting interaction of the form $a^{\dagger} b + a b^{\dagger}$.

It has been argued that, if the noise-bias of this qubit is sufficiently strong, only a classical repetition code \cite{GM:repetition} might suffice to correct for the dominant phase-flip ($Z$) errors due to photon loss. Crucial in this idea is that the CNOT gate which is needed to measure the XX checks of this code preserves the noise-bias, that is, $Z$ errors during the gate do not propagate to become $X$ errors after the gate. For the Kerr-cat qubit a noise-bias preserving CNOT gate has been proposed in \cite{puri+:bias}. A similar idea is to use this Kerr-cat qubit as a basic qubit in a surface code architecture in which the XXXX and YYYY checks are measured \cite{puri+:bias, tuckett+:bias}. In this modified form of surface code one gains much more information about $Z$ errors. It has been shown that when the probability for phase-flip errors and measurement errors is a factor 100 more than that of bit-flip errors within a phenomenological error model, the threshold against $Z$ errors can be as high as $5\%$ \cite{tuckett+:bias}. It is an open question whether such high bias will be feasible in practice as experiments for doing the CZ gate and the noise-bias preserving CNOT gate on these noise-biased qubits are still to come. 

% BMT leakage of noise biased qubit weakness, not mentioned 

\subsection{The GKP Qubit}
\label{sec:GKP}

% organize sections and what should not be said

The (square) Gottesman-Kitaev-Preskill (GKP) qubit introduced in Ref.~\cite{GKP} is defined through two commuting displacement operators, acting as translations in phase space, i.e. $S_q=\exp(i 2\sqrt{\pi}\hat{q})$ and $S_p=\exp(-i 2 \sqrt{\pi} \hat{p})$ \footnote{The commutation of $S_p$ and $S_q$ can be verified by using the identity $\exp(A)\exp(B)=\exp(B) \exp(A) \exp([A,B])$ for operators $A$ and $B$ whose commutator is proportional to $I$.}. The ideal GKP code is the space invariant under these two phase-space translations. As a result, any wave function in $q$ (resp. $p$) in this space has support on $q= k \sqrt{\pi}$ (resp. $p=l \sqrt{\pi}$) for integers $k, l \in \mathbb{Z}$.
The logical operators of the qubit are $\overline{Z}=\exp(i \sqrt{\pi}\hat{q})$ and $\overline{X}=\exp(-i \sqrt{\pi} \hat{p})$ with $\overline{X}\overline{Z}=-\overline{Z}\overline{X}$. In addition, $\overline{Y}=i\overline{X}\overline{Z}=\exp(i \pi/2)\exp(-i \sqrt{\pi} \hat{p})\exp(i \sqrt{\pi}\hat{q})=\exp(i \sqrt{\pi} (-\hat{p}+\hat{q}))$. This choice makes the wave function in $q$ of $\ket{\overline{0}}$ a sum of delta functions at values of $q$ which are even multiples of $\sqrt{\pi}$, while $\ket{\overline{1}}$ has uniform support on values of $q$ which are odd multiples of $\sqrt{\pi}$.
The ideal code meets the quantum error correction conditions for a continuous set of `at most half-logical' displacements ${\sf E}=\{ e^{i u \hat{p}}, e^{i v \hat{q}} \colon |u|, |v| \leq \sqrt{\pi}/2  \}$, since any products of these shifts maps a $\ket{\overline{0}}$ onto a state orthogonal to both $\ket{\overline{1}}$ and $\ket{\overline{0}}$ (and vice-versa). The set of correctable displacements forms a square Wigner-Seitz or Voronoi cell (containing only one lattice point such that all points in the cell are closer to this point than to another lattice point) in the code lattice generated by the logical phase-space translations.

Naturally, an asymmetric version of the GKP code which corrects more shift errors in $\hat{q}$ than shift errors in $\hat{p}$ can also be defined. However, when there is no hardware-based noise asymmetry between $\hat{p}$ and $\hat{q}$ this does not seem immediately useful. 

%An appealing feature of the GKP qubit is that it has no leakage. Qubits encoded in two-level subspaces of a larger Hilbert space can leak outside of the two-level space and the formalism of qubit quantum error correction is not designed to detect or correct such leakage errors. For a perfect GKP qubit almost all elements in the continuous displacement error basis can be detected (except for the stabilizer and logical displacements), while the subclass of small displacements can be corrected. Hence there is no need to put the GKP qubit back in the computational subspace in order for some error detection to work. Since error detection has finite precision and noise in practice, the continuous errors on a GKP qubit have to be also dealt with when the qubit is used as a building block in a surface or other code.

In principle, and in theory, to perform quantum error correction the eigenvalues (phases) of the unitary operators $S_p$ and $S_q$ are to be measured.  Performing such measurements projects the continuum of errors onto (superpositions of) possible displacements, and we perform error correction by choosing a displacement of minimal amplitude which resets these eigenvalues to $+1$, corresponding to the code space. In Section \ref{sec:decoding} we will analyze GKP quantum error correction using encoded GKP ancilla qubits, see Fig.~\ref{fig:EC_GKP}. The advantage of this form of error correction is that it does not suffer from feedback errors induced by a poor ancilla qubit (instead, it suffers feedback errors from a GKP ancilla qubit) and the information gained through measuring the GKP ancilla states is analog rather than binary. The disadvantage is that one needs to prepare GKP ancilla states themselves first. 

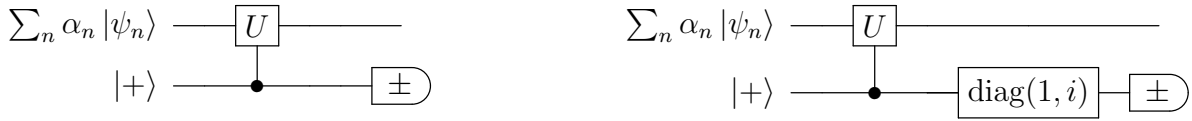
\begin{figure}[htbp]
\centering
   \hspace{2.2cm}\scalebox{1}{\Qcircuit @C=1em @R=.7em {
      \lstick{\sum_n \alpha_n \ket{\psi_n}} & \qw & \gate{U} & \qw & \qw & \qw   \\
      \lstick{\ket{+}}          & \qw & \ctrl{-1} & \qw &  \qw & \measureD{\pm}  }}
\hfill
  \scalebox{1}{\Qcircuit @C=1em @R=.7em {
      \lstick{\sum_n \alpha_n \ket{\psi_n}} & \qw & \gate{U} & \qw & \qw & \qw  \\
      \lstick{\ket{+}}          & \qw & \ctrl{-1} & \qw &  \gate{{\rm diag}(1,i)} \qw & \measureD{\pm}  }}
\caption{Single round of phase estimation with $U \ket{\psi_n}=e^{i\theta_n} \ket{\psi_n}$ where the probability for ancilla qubit to be measured in the state $\ket{\pm}$ equals $\mathbb{P}(\pm)=\sum_n |\alpha_n|^2 \frac{1}{2}(1\pm \cos(\theta_n))$ (left) and $\mathbb{P}(\pm)=\sum_n |\alpha_n|^2 \frac{1}{2}(1\mp \sin(\theta_n))$ (right). In the applications here $U$ is a displacement $S_p$ or $S_q$.}
	\label{fig:PE}
\end{figure}

For this latter task one can perform some form of phase estimation to measure the eigenvalues of the unitary operators $S_p$ and $S_q$. Since the eigenvalues take continuous values, one only ever realizes an approximate estimation of these phases. Phase estimation can readily be executed by coupling the GKP mode repeatedly to a single ancilla qubit via controlled-displacement gates as was proposed and discussed in great detail in Ref.~\cite{TW:GKP}, focusing on a circuit-QED implementation. The idea behind this is simple. To measure the eigenvalue of a unitary operator $U$ such as the displacements $S_p$ or $S_q$, one can use ancilla qubits applying qubit controlled-$U^k$ gates for $k=1, 2 \ldots$. For example, when $k=1$, the circuit on the left in Fig.~\ref{fig:PE} has outcome probabilities $\mathbb{P}(\pm)=\frac{1}{2}(1\pm \langle {\rm Re}(U) \rangle$, while the circuit on the right has probabilities $\mathbb{P}(\pm)=\frac{1}{2}(1\mp \langle {\rm Im}(U) \rangle$.

In phase-estimation schemes, higher powers $k > 1$ of $U^k$ are often used, but applying $U^k$, a displacement of strength $\sim k$, increases the number of photons in the state by $\sim k^2$ and does not provide a good approximation of an approximate GKP state \cite{TW:GKP}. Instead of repeating the phase estimation to collect bits of phase and then do a final corrective displacement, it is experimentally simpler to opt for immediate feedback on the code state based on each new bit obtained in a round of phase estimation. This is the route taken in the experimental realization of the GKP code in \cite{GKP:exp}, where a small conditional displacement on the GKP qubit is executed depending on the ancilla qubit measurement outcome.
In fact, using such immediate feedback the state of the ancilla qubit does not even need to be measured, as the feedback can be done depending on the qubit state itself, followed by an approximate disentangling step \cite{hastrup+:meas-free} or alternatively a qubit reset step (to avoid entropy build-up).  

In addition, in Ref. \cite{GKP:exp} only the right circuit in Fig.~\ref{fig:PE} measuring $ {\rm Im}(U)$ is used (instead of measuring both ${\rm Re}(U)$ and ${\rm Im}(U)$). If the state to be measured is (approximately) symmetrically centered around the vacuum so that its wavefunction is symmetric under $q \rightarrow -q$ and $p \rightarrow -p$, we have $\int dp |\psi(p)|^2{\rm Im}(S_p)=0$ and $\int dq |\psi(q)|^2{\rm Im}(S_q)=0$. This implies that $\mathbb{P}(\pm)=\frac{1}{2}(1 \pm \langle {\rm Im}(U) \rangle)=\frac{1}{2}$, suggesting that the measurement outcome $\pm$ can gain a maximal amount of information by weakly projecting onto $\sin(\theta)\gtrless 0$, and subsequently shifting the state to the point $\theta=0$. These feedback shifts are realized in \cite{GKP:exp} by small displacements.
Note that if the input state has eigenvalue phase $\theta$ close to 0, then ${\rm Re}(U)$ is close to 1, implying that not much is learned by doing the measurement with outcomes $\mathbb{P}(\pm)=\frac{1}{2}(1 \pm \langle {\rm Re}(U) \rangle)$.

% BMT made some changes adding to the view of the benefits of learning im(U) and re(U)

We remark that the length of the displacement of the logical $\overline{Y}$ is $\sqrt{2}$ larger than that of $\overline{X}$ and $\overline{Z}$. This implies some asymmetry in error correction. Namely, if we correct by measuring $S_p$ and $S_q$, shifts such as $\exp(i u \hat{p}+ v \hat{q})$ with $u^2\leq \pi/4$ and $v^2\leq \pi/4$ can be corrected which, as displacements, are a factor $\sqrt{2}$ larger than correctable displacements in pure $\hat{p}$ and $\hat{q}$ directions. Given a noise model which is rotationally-symmetric in phase space, this does not seem to be an optimal choice. It also implies that logical $Y$ eigenstates which can flip due to large displacements in pure $\hat{p}$ and $\hat{q}$ directions can have shorter lifetimes \cite{GKP:exp}. 

%Note that in principle, one could measure the eigenvalues of $S_p$, S_q$ as well as those of $S_p S_q \equiv \exp(i 2\sqrt{\pi} (\hat{q}-\hat{p})$ to stabilize the state in all phase-space directions 
% I don't understand the data of the GKP paper in this respect

A `hexagonal' GKP qubit has also been defined in \cite{GKP} by choosing two phase-space lattice translations which are not orthogonal such that all three logical operators $\overline{X}, \overline{Y}$ and $\overline{Z}$ have the same length as phase-space translation vectors. For this choice we take 
as stabilizers $\exp(i \xi (\sqrt{3} \hat{q}-\hat{p})/2)$ and $\exp(i \xi  \hat{p})$ with $\xi=2\sqrt{2\pi/\sqrt{3}}$, generating a hexagonal lattice in phase space. Again the logical operators are half-stabilizers, forming the vectors generating a hexagonal lattice. The correctable displacements now form a hexagonal Wigner-Seitz cell. This cell is larger in volume than the square Wigner-Seitz cell in the square GKP lattice. If we assume that displacement errors occur according to a stochastic Gaussian model as in Eq.~(\ref{eq:GDC}), it implies that the hexagonal code can correct a larger probability volume of errors.

If we were to choose stabilizers $S_q=\exp(i \sqrt{2\pi} \hat{q})$ and $S_p=\exp(-i \sqrt{2\pi} \hat{p})$, there would be no additional commuting displacement operators, implying that the $+1$ eigenspace $S_p$ and $S_q$ is one-dimensional. This eigenstate, also called the sensor state $\ket{\psi_{\rm sensor}}$ in \cite{DTW:sensor}, is a uniform sum of delta function at $q=k \sqrt{2 \pi}$ with $k \in \mathbb{Z}$ (and similarly a uniform sum of delta functions at $p=l \sqrt{2 \pi}$ with $l \in \mathbb{Z}$). The sensor state is interesting in allowing one to simultaneously estimate the complex and real part of the amplitude $\alpha$ of a displacement $D(\alpha)$, by performing phase-estimation for $S_q$ and $S_p$ on $D(\alpha) \ket{\psi_{\rm sensor}}$ \cite{DTW:sensor}.

We will uniquely focus on the square GKP code in the remainder of this review, although most points apply with small variation to the hexagonal code.

%In this case the correctable displacements in the pure $\hat{p}$ and $\hat{q}$ directions are the longest, namely they are lengthened by only a factor $\frac{2}{3} \times \frac{\sqrt{3}}{2}  \times 2$ as compared to the incorrectable displacements in the lattice generator directions.  In addition, the code can at least correct $e^{i u \hat{p}}$ with at most $u \leq \frac{\xi}{4}=\frac{\sqrt{2\pi/\sqrt{3}}}{2} \approx 1.69$ which is more than the square code with $\sqrt{\pi}/2 \approx 0.88$. For an error $e^{i v \hat{p}}$, its projection onto the generator is at most quarter its length when $v \leq \xi \sqrt{3}/8=\sqrt{2\pi \sqrt{3}}/4=0.824$..

% fidelity, quality
\subsubsection{Approximate GKP States}

Any physical GKP code state will occupy a finite volume in phase space and will have a finite number of photons. In principle, an infinite number of approximations to the perfect GKP code states exist, but some are more useful than other's and here we will mention four. Ref.~\cite{GKP} introduced a form of approximate GKP state obtained by applying a Gaussian superposition of displacements, characterized by a `squeezing' parameter $\Delta > 0$ to a perfect state:
\begin{align}
&\ket{\overline{\psi}} = \frak{E} \ket{\overline{\psi}_{\rm ideal}},
&&\frak{E}\equiv \frac{1}{\sqrt{\pi \Delta^2}} \int_{\mathbb{R}^2} du dv\; \exp\left(-\frac{u^2+v^2}{2\Delta^2}\right) \exp(-i u \hat{p}+iv \hat{q}).
\label{eq:model}
\end{align}
For this model wavefunction it holds that $\bar{n} \approx \frac{1}{2\Delta^2}-\frac{1}{2}$ \cite{GKP, TW:GKP}. One can perform the Gaussian phase-space integral in Eq.~(\ref{eq:model}) and, --neglecting contributions $O(\Delta^4) \ll 1$, see e.g. \cite{tzitrin+:GKPtools}--, one gets a different approximation using an operator $\frak{D}$:
\begin{equation}
\frak{E} \approx 2 \sqrt{\pi \Delta^2} \frak{D}, \;\frak{D} \equiv \exp(-\Delta^2 \hat{n}).
\label{eq:approx}
\end{equation}

\begin{figure}[htbp]
	\center
	\includegraphics[scale=0.7]{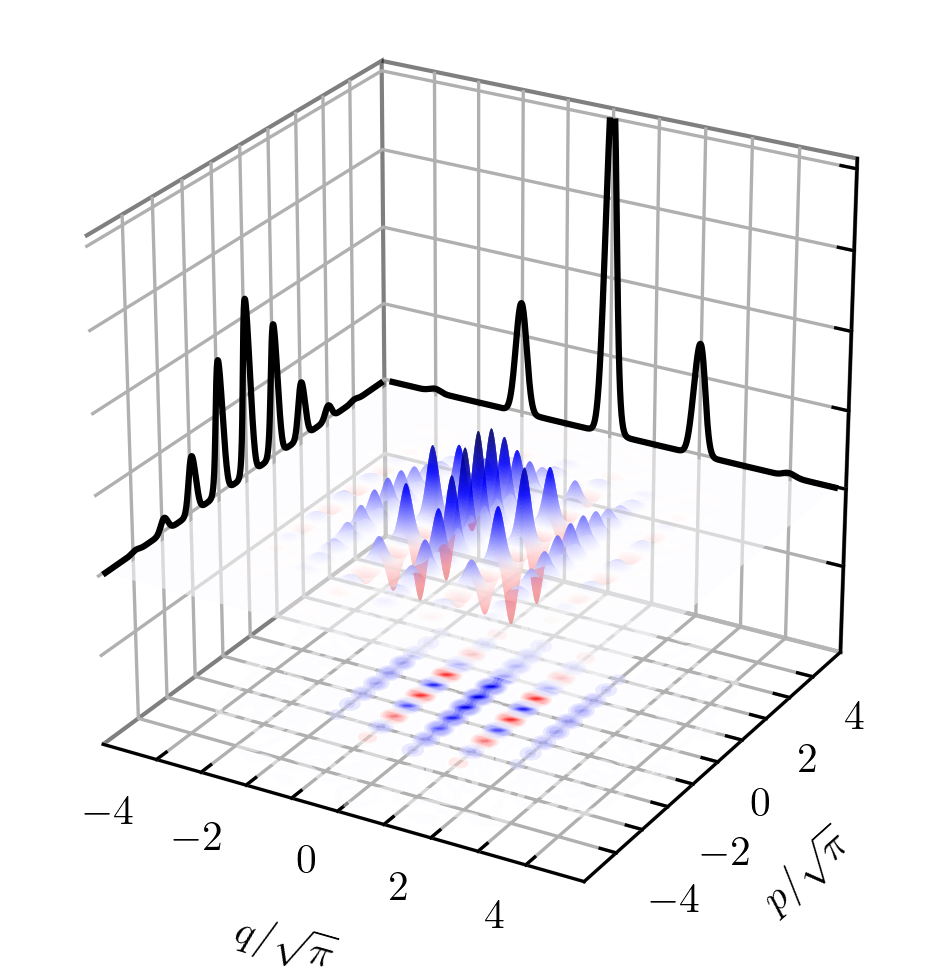}
	\caption{Wigner function of the state $\frak{F} \ket{\overline{0}}$ at $\Delta=0.3$, and the reduced probability distributions over $q$ and $p$ in black. Unlike the $\frak{E}$- and $\frak{D}$-approximation, the $\frak{F}$-approximation has a clear asymmetry with respect to $p$ and $q$. Since the Wigner function has a grid-like periodic structure in phase space, the GKP states are also referred to as grid states.}
	\label{fig:GKP-grid}
\end{figure} 

% useful for tracking effect of gates
The envelope operator $\frak{D}$ has approximately the same effect as the `no loss' Kraus operator of a photon loss channel ${\cal N}_{\gamma}$, Eq.~(\ref{eq:channel-pl}), with $\gamma=2 \Delta^2$. Another approximation, valid for small $\Delta$ is
\begin{align}
\frak{E}\ket{\overline{0}} \approx \frak{F} \ket{\overline{0}} \propto \int_{\mathbb{R}} dq\;  \sum_{k \in \mathbb{Z}} \underbrace{e^{-2\Delta^2 \pi k^2 }}_{envelope} \underbrace{e^{-\frac{1}{2 \Delta^2} (q-2k\sqrt{\pi})^2}}_{comb}\ket{q},\\
\frak{E} \ket{\overline{+}} \approx\frak{F} \ket{\overline{+}} \propto \int_{\mathbb{R}} dq\; \sum_{k \in \mathbb{Z}} \underbrace{e^{-\frac{1}{2}\Delta^2 \pi k^2 }}_{envelope} \underbrace{e^{-\frac{1}{2 \Delta^2} (q-k\sqrt{\pi})^2}}_{comb}\ket{q}
\label{eq:approxE}
\end{align}
The state $\frak{F} \ket{\overline{0}}$ can be interpreted as the result of preparing a squeezed state $\frac{1}{\pi^{1/4}} \int dq \exp(-q^2/2 \Delta^2) \ket{q}$ to which one applies a Gaussian-enveloped coherent sum over stabilizer translations, enacting $\rho \rightarrow \sum_{k,l \in \mathbb{Z}} e^{-2 \Delta^2 \pi (k^2+l^2)} S_p^k \rho S_p^{-l}$. The result is a state which is both an approximate eigenstate of $S_q$ (and $\overline{Z}$) due to squeezing, as well as an approximate eigenstate of the translation $S_p$. Note that unlike $\frak{E}$ and $\frak{D}$, approximation $\frak{F}$ has an asymmetry in $p$ and $q$.
The three approximations $\frak{D},\frak{E},\frak{F}$ have been discussed and shown to fit a standard form in \cite{MYK:approxGKP}. In addition, the normalization of these approximate forms can be computed and expressed in terms of theta functions, see e.g. Appendix \ref{app:GKP-math} for the $\frak{D}$-approximation.

%In Appendix \ref{app:GKP-math} we derive the normalization of this state.

In Eq.~(\ref{eq:ancilla-approx}) we will see a fourth, von-Mises or reverse-Villain, approximation using a cosine function to represent the periodicity in the wave-function comb. This reverse-Villain approximation has been used in \cite{WT:breed} and \cite{vuillot+:GKP}. All these approximate states $\frak{E} \ket{\overline{0}}$ and $\frak{E} \ket{\overline{1}}$ (or $\frak{D}\ket{\overline{0}}$ and $\frak{F}\ket{\overline{0}}$ etc.) are $+1$ eigenstates of the photon parity operator $e^{i \pi a^{\dagger} a}$ as they are invariant under $q\rightarrow -q$ and $p \rightarrow -p$, implying that they only have support on even photon number states. In Appendix \ref{app:GKP-math} we show how to get exact Fock state amplitudes for the approximation $\frak{D} \ket{\overline{0}}$, --which for this purpose has the simplest form--, and this turns out to involve $n$-the order derivatives of theta functions. We show in Appendix \ref{app:GKP-math} that the photon number distribution of these GKP states, as well as the sensor state, is following a thermal distribution \cite{albert+:photon-loss} (see Fig.~\ref{fig:fockcoefgkp} and \ref{fig:fockcoefsensor}), with interesting oscillations on top. 

%This is in line detailing a statement in \cite{albert+:photon-loss}

%\textcolor{red}{The goal had been to show that GKP states are thermal states in the photon number basis (but have zero support on odd photon number states, we have an equation in our long paper with Albert et al which suggests this. However I ran into things not yet converging in photon number representation, either by being too careless about states, delta functions or typos.}

One can propose various measures of state quality or fidelity besides the characterization of the state in terms of $\Delta$. For example, when we measure $\hat{q}$ to infer $\overline{Z}$ on a state, all outcomes in which $q$ is closer to an even multiple of $\sqrt{\pi}$ are interpreted as outcome $\overline{Z}=1$ and vice-versa. For a state $\int dq\, \psi(q)\ket{q}$, the probability for this outcome is then
\begin{equation}
\mathbb{P}(\overline{Z}=(-1)^b)=\int_{I_b} dq  |\psi(q)|^2,  \;\;\; I_b=\left\{q| \;\exists k \in \mathbb{Z},\; -\frac{\sqrt{\pi}}{2} \leq q+ (2k+b) \sqrt{\pi} \leq \frac{\sqrt{\pi}}{2}\right\}.
\label{eq:meas}
\end{equation}
If we apply this to the form $\frak{E} \ket{\overline{0}}$, the error probability $\mathbb{P}(\overline{Z}=-1) < \frac{2\Delta}{\pi}\exp(-\pi/4\Delta^2)$. 
Since a perfect (homodyne) measurement of $\hat{q}$ is practically not possible, $\mathbb{P}(\overline{Z}=-1)$ only provides a lower bound on the logical error probability of an approximate state $\ket{\overline{0}}$. We can also examine the expectation value for $\overline{Z}$ on the approximate form $\frak{F} \ket{\overline{0}}$ (for simplicity) which equals 
\begin{align}
\frac{\bra{\overline{0}} \frak{F}^{\dagger} \overline{Z} \frak{F} \ket{\overline{0}}}{\bra{\overline{0}} \frak{F}^{\dagger} \frak{F} \ket{\overline{0}}} \approx
\frac{\sum_{k \in \mathbb{Z}} e^{-4\Delta^2 \pi k^2 } \int_{\mathbb{R}} dq\;  e^{i \sqrt{\pi} q}  e^{-\frac{1}{\Delta^2} (q-2k\sqrt{\pi})^2}}{\sum_{k \in \mathbb{Z}} e^{-4\Delta^2 \pi k^2 }  \int_{\mathbb{R}} dq\;    e^{-\frac{1}{\Delta^2} (q-2k\sqrt{\pi})^2}}=e^{-\pi \Delta^2/4},
\label{eq:Zexp}
\end{align}
and similarly $\frac{\bra{\overline{1}} \frak{F}^{\dagger} \overline{Z} \frak{F} \ket{\overline{1}}}{\bra{\overline{1}} \frak{F}^{\dagger} \frak{F} \ket{\overline{1}}} \approx -e^{-\pi \Delta^2/4}$, showing that the expectation decays exponentially in $\Delta^2$ towards $0$. In the approximation in Eq.~(\ref{eq:Zexp}) we have assumed that $\Delta$ is small enough so that the peaks at different $k$ do not overlap, giving an easy expression for the probability distribution over $q$ of the approximate GKP state. We further discuss the logical $\overline{Z}$ or $\overline{X}$ measurement of a GKP qubit in Section \ref{sec:FTmeas}.

It has become common to describe the quality of a GKP state in terms of an amount of squeezing expressed in dB. For a regular squeezed state (squeezed along $q$) one has variances ${\rm Var}(q)=\frac{\Delta^2}{2}$, ${\rm Var}(p)=\frac{1}{2 \Delta^2}$ as the vacuum (or coherent state) has ${\rm Var}(q)={\rm Var}(p)=\frac{1}{2}$ with $\Delta=e^{-|\xi|} < 1$. 
%The gain $G$ of the amplifier enacting this squeezing, $a \rightarrow \sqrt{G} a+\sqrt{G-1} a^{\dagger}$, can be expressed in dB as $10 \log_{10}(G)=10 \log_{10} \cosh^2(|\xi|)$. In this vein we can re-express the value of the parameter $0 < \Delta < 1$ in Eq.~(\ref{eq:approxE}) in terms of dB. 
The convention which is used in the literature for denoting the dB of squeezing of an approximate GKP state is $\# {\rm dB}=- 10 \log_{10} \Delta^2$, see e.g. \cite{tzitrin+:GKPtools}.
% dB conventions: Eq. A3 in Tzitrin paper uses this and the Yale paper as well.Japanese paper used -10 log_10 (2sigma^2) 

We can view a GKP state as being `squeezed' in both $p$ and $q$ and interpret this squeezing as the extent in which the state is an eigenstate of a unitary operator such as $S_p$ or $S_q$. Since a quantum state may not fit one of the standard GKP approximations, a measure of the effective squeezing is useful in expressing the quality of the state. Since we are interested in modular values of $\hat{q}$ and $\hat{p}$, it is appropriate to use the Holevo phase variance (or the variance of periodic variables such as phases used in circular statistics) to express this squeezing, i.e. one can define \cite{DTW:sensor, WT:mod}:
\begin{equation}
\Delta_{p/q}=\sqrt{\frac{1}{2\pi}\ln\left(\frac{1}{ |{\rm Tr} S_{p/q} \rho|^2} \right)}.
\label{eq:eff-squeeze}
\end{equation}
Note that this measure does not express a logical error rate, e.g. the completely mixed state inside the perfect code space has $\Delta_p=\Delta_q=0$.

%One can thus additionally consider $\Delta(X)$ or $\Delta(Z)$
%The aim of a GKP qubit preparation is not necessarily to get the highest possible $\Delta$, corresponding to the largest number of photons $\bar{n}$: First of all, photon loss in the storage oscillator will naturally reduce $\Delta$ during preparation and measurement of the ancilla oscillator

% mention circuit-QED earlier

\subsubsection{Logical Gates} 
\label{sec:log-gates}

An appealing feature of the GKP code is that all logical Clifford transformations are Gaussian quantum operations, realizable by optical elements \cite{GKP, tzitrin+:GKPtools} which enact linear transformations on the operators $\hat{p}$ and $\hat{q}$ in the Heisenberg picture. Important gates such as the CNOT and S gate do however involve two-mode, respectively single-mode squeezing: the experimental realization of such squeezing transformations is typical through pumped optical {\em non-linearities}. Such elements are relatively straightforward to obtain for optical fields which travel through nonlinear $\chi^{(2)}$ or $\chi^{(3)}$ materials, while for superconducting devices these elements are engineered through the use of Josephson junctions. In contrast, passive linear optical elements, --beam-splitters and phase-shifters in optics language--,  are readily available in circuit-QED by linear capacitive or inductive (fixed) circuit couplings.

In Section \ref{sec:34-mix} we will discuss the engineered non-linearities in superconducting hardware which can be activated by microwave drives or activated by flux-drives, while here we discuss the logical gates for the GKP code at a formal level.

As unitary displacement operators, $Z$ and $X$ are not self-inverse, i.e. $X \neq X^{\dagger}$. On a perfect, completely shift-invariant code state $X$ acts identically to $X^{\dagger}$, but on a finite-photon number state, see e.g. the wave function in Fig.~\ref{fig:GKP-grid}, it does not: a shift to the left or right moves the envelope away from the center. The Hadamard gate has Heisenberg action $\hat{p} \rightarrow -\hat{q}$ and $\hat{q} \rightarrow \hat{p}$ so that $H^{\dagger} X H=Z$, $H^{\dagger} Z H=X^{\dagger}$ and $H^{\dagger} Y H=-Y$. The Hadamard gate corresponds to a phase-space rotation by an angle $\pi/2$, i.e. we can choose ${\rm Had}\equiv \exp(i \frac{\pi}{2} a^{\dagger} a)$, and note again that ${\rm Had} \neq {\rm Had}^{-1}$. A ${\rm Had}$ gate could be done by a quarter-cycle waiting in the self-evolution of the oscillator (so comes for free). 

A disadvantage of using such quarter-cycle waiting Hadamard gate in a GKP surface code architecture is discussed in Section \ref{sec:scale}. The alternative is to use single-qubit rotations around the logical $X,Y$ or $Z$ axes to compose a Hadamard gate.

\begin{figure}[htbp]
  \centering
\scalebox{1}{
 \Qcircuit @C=1em @R=.7em {  \lstick{\mbox{GKP qubit}} &  \qw & \gate{R_P(\phi)} & \qw  }}
\hspace{1cm}=
\hspace{1cm}
\scalebox{1}{
 \Qcircuit @C=1em @R=.7em {
      & \qw & \gate{P} & \qw & \qw       & \qw &  \gate{P} & \qw   \\
      \lstick{\mbox{reg. qubit } \ket{+}}                 & \qw & \ctrl{-1} & \qw &  \qw & \qw & \measureD{\phi ,\pm} \cwx[-1] & \\}}
\caption{Performing a single-qubit gate $R_P(\phi)$ with $P=X,Y,Z$ on a perfect GKP qubit via a regular ancilla qubit, requiring a qubit controlled-displacement.
The measurement is in the basis $\ket{\phi, \pm}=\frac{1}{\sqrt{2}}(e^{i \phi/2} \ket{+}\pm e^{-i \phi/2} \ket{-})$ and upon outcome $-1$, $P$ is applied.}
	\label{fig:singleU}
\end{figure}
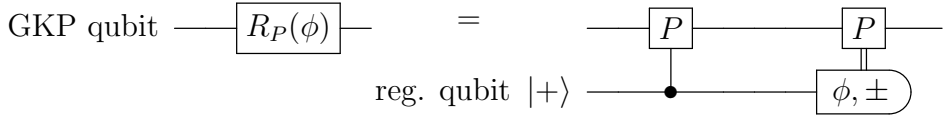

For the GKP code these rotations around logical axes, $R_{P}(\phi)\equiv \exp(-i \phi P/2)$ with logical Pauli $P=X,Y,Z$ are not natural as the logical Pauli, which is a displacement, sits in the exponent. Note also that this gate $R_P(\phi)$ is only unitary when acting on a subspace for which $P^2=I$.
However, one can perform $R_P(\phi)$, using a controlled-displacement coupling with a regular qubit and a regular qubit rotation, as shown in Fig.~\ref{fig:singleU}, and realized in \cite{fluehmann:GKP, GKP:exp}. This circuit applies $R_{P}(\phi)\equiv \exp(-i \phi P/2)$ on the space of states for which $P^2=I$ but we can examine its effect more generally. Imagine applying the circuit in Fig.~\ref{fig:singleU} with $P=Y$ and $\phi=\pi/2$. Upon outcome $\pm$, the Kraus operator action on the GKP qubit equals $A_+=\cos(\phi/2)I - i \sin(\phi/2)P$ resp. $A_-=-i\sin(\phi/2)I+\cos(\phi/2)P$. On the perfect code subspace where $P^2=I$, $A_+$ acts as a unitary and equals $R_P(\phi)$, while $A_-$ can be converted to $R_P(\phi)$ by the additional $\pi$-rotation $P$. However, on a finitely-squeezed GKP state, these Kraus operators are not unitary and their action leads to the envelope of the GKP state to be no longer centered around the vacuum. However, one can apply a displacement $P^{-1/2}$ \cite{GKP:exp} to approximately re-center the GKP state. 

A single-qubit gate such as the $T=R_Z(\pi/4)$ gate can be done in this manner as well. The $S$ gate with action $S^{\dagger} X S=-Y$ and $S^{\dagger} Z S=Z$ can be realized by the transformation $\hat{q} \rightarrow \hat{q}$, $\hat{p} \rightarrow \hat{p}-\hat{q}$ corresponding to $S=\exp(-i \hat{q}^2/2)$ \footnote{Perhaps the simplest way to derive this identity is to calculate $\exp(i \hat{q}^2/2) \hat{p} \exp(-i \hat{q}^2/2)=\sum_{n=0}^{\infty}  \frac{1}{n!}({\rm ad}_{i\hat{q}^2/2})^n(\hat{p}) $, with $({\rm ad}_A)^0(B)=B$, ${\rm ad}_A(B)=[A,B]$, $({\rm ad}_A)^2(B)=[A,[A,B]]$ etc. We can use that $[\hat{q}^2,\hat{p}]=2 i \hat{q}$ and higher-order commutators are zero, leading to $\exp(i \hat{q}^2/2) \hat{p} \exp(-i \hat{q}^2/2)=\hat{p}-\hat{q}$.} The $S$ gate can thus be implemented by means of pump-activated squeezing, see Section \ref{sec:34-mix}, or by using an ancilla qubit as in the circuit in Fig.~\ref{fig:singleU}. Alternative methods for performing a $T$ gate via magic state preparation or using a cubic phase gate $V_{\gamma}=\exp(i \gamma \hat{q}^3)$ exist \cite{GKP}. For example, one can create a $+1$ eigenstate of the Hadamard gate ${\rm Had}=\exp(i \frac{\pi}{2}a^{\dagger} a)$ by starting with a vacuum state, which is already a $+1$ eigenstate of ${\rm Had}$, and measuring $S_p$ and $S_q$ without photon-number changing feedback \cite{baragiola+:all-gauss}.

When using GKP qubits as basic qubits in a surface code, see Section \ref{sec:scale}, we note that $T$ and $S$ gates are not needed for error correction: their only use is to prepare magic GKP ancilla qubits to be grown into the surface code-encoded magic states using GKP CZ and CNOT gates or parity check measurements, see e.g. \cite{CTV:FT} and references therein.

 % check how this works..set theta=pi/2, wrong!
% what is its effect when the code word is not perfect...

The CNOT gate can be realized by the Heisenberg action $\hat{q}_c \rightarrow \hat{q}_c$, $\hat{p}_c \rightarrow \hat{p}_c-\hat{p}_t$, $\hat{q}_t \rightarrow \hat{q}_c+\hat{q}_t$ and $\hat{p}_t \rightarrow \hat{p}_t$. This gate is also called the SUM gate in \cite{GKP} and ${\rm SUM}(g)$ with $g=1$ in \cite{tzitrin+:GKPtools}. We see that ${\rm CNOT}=\exp(-i \hat{p}_t \hat{q}_c)$ by using Eq.~(\ref{eq:id}) with $v=\hat{p}_t$ and $u=\hat{q}_c$.
The inverse CNOT has action $\hat{q}_c \rightarrow \hat{q}_c$, $\hat{p}_c \rightarrow \hat{p}_c+\hat{p}_t$, $\hat{q}_t \rightarrow \hat{q}_t-\hat{q}_c$ and $\hat{p}_t \rightarrow \hat{p}_t$.

%Note that the action of the CNOT or the inverse CNOT are identical on the perfect codespace, but not so on a finite-photon number code space.

We define the action of the ${\rm CZ}$ gate as ${\rm Had}_t\; {\rm CNOT} \;{\rm Had}_t^{\dagger}$ where ${\rm Had}_t$ is a Hadamard gate on the target mode. That is, it enacts the transformation $\hat{q}_t\rightarrow \hat{q}_t$, $\hat{p}_t \rightarrow \hat{p}_t-\hat{q}_c$, $\hat{q}_c \rightarrow \hat{q}_c$, $\hat{p}_c \rightarrow \hat{p}_c-\hat{q}_t$, or ${\rm CZ}=\exp(-i \hat{q}_t \hat{q}_c)$. If either oscillator is a state where $q$ is an even multiple of $\sqrt{\pi}$, then CZ acts as $\exp(-i \pi 2 k)=1$. If both oscillators are in a state where $q$ is an odd multiple of $\sqrt{\pi}$, then CZ acts as $\exp(-i \pi (2n+1) (2k+1))=-1$ for $n,k \in \mathbb{Z}$. 

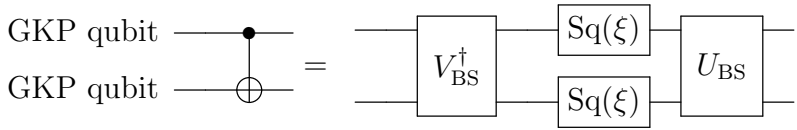
\begin{figure}[htbp]
\centering
\begin{minipage}{0.3\textwidth}
	  \Qcircuit @C=1em @R=1.3em {
      \lstick{\mbox{GKP qubit}} & \qw & \ctrl{1} & \qw  \\ 
     \lstick{\mbox{GKP qubit} }  &  \qw & \targ & \qw \\ }
	\end{minipage}
=\;\;
\begin{minipage}{0.3\textwidth}
 \Qcircuit @C=1em @R=.7em {
      & \qw & \multigate{1}{V^{\dagger}_{\rm BS}} & \qw & \gate{{\rm Sq}(\xi)}   & \multigate{1}{U_{\rm BS}} & \qw  \\
      & \qw & \ghost{V^{\dagger}_{\rm BS}}  & \qw &  \gate{{\rm Sq}(\xi)} & \ghost{U_{\rm BS}} &   \qw \\}
\end{minipage}
\caption{The realization of a CNOT via 50:50 beam-splitters, i.e. $V_{\rm BS}$ and $U_{\rm BS}$ defined in Eq.~(\ref{eq:uv}), and single-mode squeezing ${\rm Sq}(\xi)$ with $\xi \approx -0.4812$.}
	\label{fig:C-BS}
\end{figure}

Sections \ref{sec:GKP-CZ} and \ref{sec:4wave} will discuss how the GKP CZ gate between two GKP modes can be executed using a 3-wave or 4-wave mixing element. There is however another circuit to perform a CNOT gate which uses a sequence of beam-splitters and some single-mode squeezing \cite{TW:GKP, tzitrin+:GKPtools} which can be more useful in some circumstances, see Fig.~\ref{fig:C-BS}. For the CNOT gate the mode transformation on control (c) and target (t) mode equals
\begin{equation}
\left(\begin{array}{c} a_c^{\rm out} \\ a_t^{\rm out} \end{array}\right)=A \left(\begin{array}{c} a_c \\ a_t \end{array}\right)+B \left(\begin{array}{c} a_c^{\dagger} \\ a_t^{\dagger} \end{array}\right),
\end{equation}
with 
\begin{equation}
A=\left(\begin{array}{cc} 1 & -\frac{1}{2} \\ \frac{1}{2} & 1 \end{array}\right), B=\left(\begin{array}{cc} 0& \frac{1}{2} \\ \frac{1}{2} & 0 \end{array}\right).
\end{equation}
By the Bloch-Messiah decomposition \cite{braunstein:squeeze} the singular value decompositions are $A=U D_A V^{\dagger}$ and $B=U D_B V^T$ with unitary matrices $U$ and $V$. For the CNOT gate the singular values are degenerate: $D_A={\rm diag}(\frac{\sqrt{5}}{2},\frac{\sqrt{5}}{2})$ and $D_B={\rm diag}(\frac{1}{2},\frac{1}{2})$, implying that the beam-splitting transformations $U$ and $V$ are not unique. Ref.~\cite{braunstein:squeeze} notes that taking 50:50 beamsplitters with 
\begin{equation} U=U_{\rm BS} \equiv \frac{1}{\sqrt{2}}\left(\begin{array}{cc} i e^{i \theta} & i e^{-i \theta} \\ -e^{i\theta} & e^{-i\theta} \end{array}\right), \;\;V=V_{\rm BS} \equiv \left(\begin{array}{cc} 0 & 1 \\ 1 & 0 \end{array}\right)U^*,
\label{eq:uv}
\end{equation}
with $\theta=\frac{1}{2}\sin^{-1}(2/\sqrt{5})$ can be chosen (while \cite{tzitrin+:GKPtools} makes a different choice). We see that the single-mode squeezing represented by the diagonal matrices corresponds to a squeezer ${\rm Sq}(\xi)$ with $\xi=-\cosh^{-1}(\sqrt{5}/2)$.

It is clear that logical gates are not unique as physical operations as they only have to perform the right action on the code space. 
Ref.~\cite{tzitrin+:GKPtools} has discussed how logical gates propagate or amplify errors on the approximate GKP code states. Keeping the (average) number of photons in an approximate GKP state low by centering the state symmetrically around the vacuum, emerges as a good overall strategy to minimize the propagation of errors and the effect of the inaccurate action of gates.

%squeezing is bad, ie does not commute with the envelope operator 

\subsubsection{Noise on a GKP Qubit}

A simple numerically convenient noise channel, playing the role of depolarizing channel for an oscillator, is the independent Gaussian displacement channel ${\cal N}(\rho)$ with standard deviation $\sigma_0$:
\begin{equation}
{\cal N}(\rho)=\int_{-\infty}^{\infty}{\rm d}u \int_{-\infty}^{\infty}{\rm d}v\;\mathbb{P}_{\sigma_0}(u)\mathbb{P}_{\sigma_0}(v)e^{iu\hat{p}+iv\hat{q}} \rho \; e^{-iu\hat{p}-iv\hat{q}}.
\label{eq:GDC}
\end{equation}
Here $\rho$ is a single-mode density matrix and $\mathbb{P}_{\sigma_0}(x)$ the Gaussian probability density function with mean zero and variance $\sigma_0^2$, i.e. $\mathbb{P}_{\sigma_0}(x)=(2\pi \sigma_0^2)^{-1/2} e^{-x^2/2\sigma_0^2}$. This channel does not naturally correspond to physical sources of noise, but (1) one can convert photon loss via amplification to this channel \cite{albert+:photon-loss}, (2) one can `displacement twirl' noise so that the effective channel is that of probabilistic mixture of displacements \cite{wang:ms-thesis}. The exact displacement twirl is not a physical operation as it requires large displacements, so this type of modeling should be considered less justified than in the qubit Pauli case when we use a depolarizing noise model through a Pauli twirling approximation.

It is thus of interest to study how realistic noise affects the approximate GKP states beyond this toy model. We will explore the question of stochastic Gaussian displacement noise versus coherent finite-squeezing error during quantum error correction in the next Section \ref{sec:decoding}. In this section we describe the interesting effect of photon loss on a GKP qubit using Wigner function dynamics \cite{GKP:exp}, and mention some literature discussing other sources of noise.

\begin{figure}[htbp]
\begin{minipage}[h]{1\linewidth}
  \includegraphics[scale=0.5]{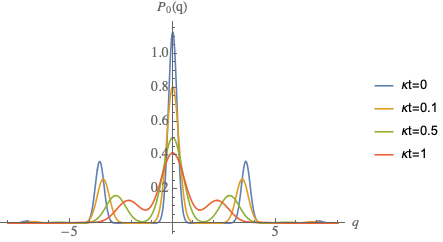}
\includegraphics[scale=0.5]{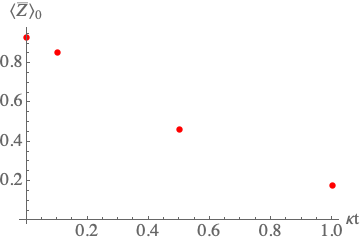}
	\caption{\label{fig:pl-prob-0}The probability distribution $P_0(q)$ of the state $\frak{F} \ket{\overline{0}}$ at $\Delta=0.3$ undergoing photon loss. The squeezed peaks of the initial state $\kappa t=0$ widen and drift inwards. Even though the state has partial support on regions where $q$ is closer to an odd multiple of $\sqrt{\pi}$, $\langle \overline{Z}\rangle_0$ shown on the right, is nonnegative at all times due to the large wave function peak centered at 0.}
     \end{minipage} 
\begin{minipage}[b]{1\linewidth}
\includegraphics[scale=0.5]{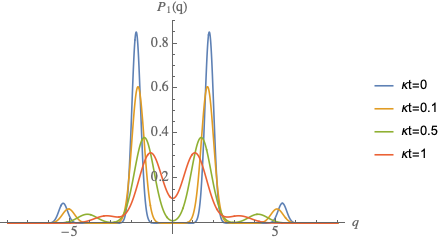}
\includegraphics[scale=0.5]{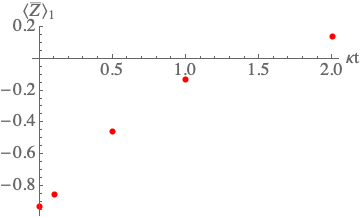}
	\caption{\label{fig:pl-prob-1}The probability distribution $P_1(q)$ of the state $\frak{F} \ket{\overline{1}}$ at $\Delta=0.3$ undergoing photon loss. We observe that $\langle \overline{Z} \rangle_1$ moves from a negative initial value to a final positive value as the state moves to the vacuum state.}
\end{minipage}
\end{figure}

An oscillator state undergoing photon loss at rate $\kappa$ can be described, in a rotating frame at its resonant frequency, using a Lindblad equation $\dot{\rho}=\kappa {\cal D}(a)(\rho)$ using the density matrix $\rho$. Here we assume that the thermal environment which induces this photon loss is at zero temperature, hence there are no photon gain processes. Alternatively, and conveniently, one describes this dynamics through differential equations using phase-space probability distributions such as the Wigner function.
The Wigner function $W(q,p,t)\equiv \frac{1}{2\pi} \int dx e^{- i p x} \bra{q+\frac{x}{2}}\rho(t) \ket{q-\frac{x}{2}}$ for the photon loss dynamics can be shown to obey a two-dimensional Fokker-Planck equation, see \cite{book:carm, GKP:exp, carter:pra}
% BMT Wigner average of displaced parity perhaps
\begin{equation*}
\frac{\partial W(q,p,t)}{\partial t}=\frac{\kappa}{2}\left(\frac{\partial}{\partial q}(q W(q,p))+\frac{\partial}{\partial p}(p W(q,p,t))+\frac{1}{2} \left(\frac{\partial^2 W(q,p,t)}{\partial^2 p} +\frac{\partial^2 W(q,p,t)}{\partial^2 q}\right) \right).
\end{equation*}
This Fokker-Planck equation describes a process of diffusion, --a spread in the variance of the variables $p$ and $q$ to the vacuum noise variance equal to $1/2$--, and drift, i.e. the mean values of $p$ and $q$ flow towards 0. Instead of considering the Wigner function dynamics, we can integrate over, say, $p$ and consider the corresponding Fokker-Planck equation for the probability distribution $P(q,t)=\int_{\mathbb{R}} dp\; W(q,p,t)$, which has the solution:
\begin{align}
P(q,t) =& \int dq' P_{\rm trans}(q,t| q',t') P(q',t=0), \notag \\
 P_{\rm trans}(q,t|q',0) =& \sqrt{\frac{1}{2\pi \sigma^2(t)}}  \exp\left(-\frac{(q-q'e^{-\kappa t/2})^2}{2\sigma^2(t)}\right), \notag \\
\sigma^2(t)=& \frac{1}{2}(1-\exp(-\kappa t)).
\label{eq:OU}
\end{align}
%The solution of this equation in time is
%\begin{align}
%W(q,p,t) =& \iint dp' dq' P(q,p,t| q',p',t') W(q',p',t'), \notag \\
% P(q,p,t|q',p',t') =& \sqrt{\frac{1}{2\pi \sigma^2(t,t')}}  \exp\left(-\frac{(q-q'e^{-\kappa(t-t')/2})^2+(p-p'e^{-\kappa(t-t')/2})^2}{2\sigma^2(t,t')}\right) \notag \\
%\sigma^2(t,t')=& \frac{1}{2}(1-\exp(-\kappa (t-t'))).
%\end{align}
In Fig.~\ref{fig:pl-prob-0} we plot the effect of photon loss of a normalized state $\frak{F} \ket{\overline{0}}$ with $\Delta=0.3$ for $\kappa t=0.1, 0.5$ and 1.

% BMT see mathematica notebook photon-loss.nb logical X?

We can consider the expectation of a stabilizer or logical $\overline{Z}$ over time, i.e. we consider ${\rm Tr}\;e^{i \alpha \hat{q}} \rho(t)=\int dq P(q,t) e^{i \alpha q}$ with $\alpha=\sqrt{\pi}$ or $2 \sqrt{\pi}$. Using Gaussian integration and Eq.~(\ref{eq:OU}) this gives
\begin{align}
{\rm Tr} \;e^{i \alpha \hat{q}} \rho(t)=\left(\sqrt{\frac{1}{2\pi \sigma^2(t)}} \int dq e^{-\frac{q^2}{2\sigma^2(t)}} e^{i \alpha q} \right)
{\rm Tr} \;e^{i \alpha(t) \hat{q}} \rho(0)=e^{-\frac{1}{4}(\alpha^2(0)-\alpha^2(t))}{\rm Tr} \;e^{i \alpha(t) \hat{q}} \rho(0),
\label{eq:time-evolv}
\end{align}
with $\alpha(t)=\alpha e^{-\kappa t/2}$. On the right-hand-side, we see an exponential decrease as well as a direct dependence on the expection value of a displacement operator with exponentially shrinking shift on the initial state. When the initial state $\rho(0)$ is invariant under $q \rightarrow -q$, we can replace ${\rm Tr} \;e^{i \alpha(t) \hat{q}} \rho(0)$ by ${\rm Tr} \cos(\alpha(t) \hat{q}) \rho(0)$. Thus when symmetrically centering the state in phase-space the phases of the stabilizer or logical $\overline{Z}$ never become complex. In addition, when the initial state is an approximate logical $\ket{\overline{0}}$ such as $\frak{F} \ket{\overline{0}}$, the expectation value $\langle \overline{Z}(t) \rangle \geq 0$ at all times as shown for a few points in Fig.~\ref{fig:pl-prob-0} on the right. This is interesting as it shows that $\ket{\overline{0}}$ `never looks more like a $\ket{\overline{1}}$ than a $\ket{\overline{0}}$' under photon loss.  The state $\frak{F} \ket{\overline{1}}$ whose decay is plotted in Fig.~\ref{fig:pl-prob-1} starts at $\overline{Z}(t=0)< 0$ and eventually, for large enough $t$, $\overline{Z}(t) > 0$ as the final state is the vacuum centered around $q=0$. This asymmetry in its effect on $\ket{\overline{0}}$ versus $\ket{\overline{1}}$ is reminiscent of a logical amplitude-damping channel.

Now assume that the initial state is displaced away from its centered location by, say, a stabilizer shift $S_p^m$ which does not affect its initial eigenvalue for $\overline{Z}$. Using Eq.~(\ref{eq:time-evolv}) we get
\begin{equation}
\langle \overline{Z}(t) \rangle =e^{2 \pi i m e^{-\kappa t/2}}e^{-\frac{1}{4}(\alpha^2(0)-\alpha^2(t))}{\rm Tr} \;e^{i \alpha(t) \hat{q}} \rho(0),
\label{eq:pl-loss-displace}
\end{equation}
shows that the expection value can now become complex, {\em but is not faster decaying in its absolute value}. When $m$ is large, we see that the additional phase changes rapidly in time, so that the expectation can rapidly change from positive to negative. However, if we know $m$ and $\kappa$ and it is the only source of noise, this phase change can be treated as a systematic error. Note that if we had applied an arbitrary but known displacement $e^{i u \hat{p}}$ on the initial state, the effect would have been similar.

Going beyond photon loss, other sources of inaccuracy and error could also readily be described using dynamics of the Wigner function. A Lindblad equation dynamics of an $n$-mode system for which the Hamiltonian is quadratic in creation and annihilation operators (beam-splitting, squeezing etc.) or {\em linear} (driving terms $\sim a+a^{\dagger}$ enacting displacements) while the dissipator models photon loss or photon gain, can be mapped to a Fokker-Planck equation of a general solvable form:
\begin{eqnarray*}
\frac{\partial}{\partial t} W(q_1,p_1, \ldots, q_n, p_n)=\left(-\vec{\nabla} \cdot (A \vec{x})+\frac{1}{2} \vec{\nabla} \cdot D \vec{\nabla} \right) W(q_1, p_1,\ldots, q_n, p_n), \; \\
\vec{x}^T=\left(q_1, p_1,\ldots, q_n, p_n \right), \vec{\nabla}^T=\left(\frac{\partial}{\partial q_1}, \frac{\partial}{\partial p_1}, \ldots, \frac{\partial}{\partial q_n}, \frac{\partial}{\partial p_n} \right),
\label{eq:genFP}
\end{eqnarray*} 
with constant $2n \times 2n$ matrices $A$ and $D$. 
%playing the role of generalized drift resp. diffusion coefficients. 
This general behavior follows from the fact that every term in a Lindblad equation which is linear in $a$ or $a^{\dagger}$ (e.g. $a \rho$), gives rise to a first-order derivative in the differential equation for the Wigner function (plus a term which is linear in $\hat{p}$ and $\hat{q}$) \cite{carter:pra,book:carm}, so that terms quadratic in $a$ and $a^{\dagger}$ (e.g. $a^{\dagger} a \rho$) gives second-order derivatives.
The Gaussian Green's function for Eq.~(\ref{eq:genFP}) can be readily given, basically forming a multi-dimensional analog of Eq.~(\ref{eq:OU}), see \cite{book:carm}. All these Gaussian processes keep an initially nonnegative Wigner function nonnegative and hence are simulatable by stochastic means. 

%Since Clifford gates for quantum error correction fall into this class, it implies that the only source of non-simulatability are input ancilla approximate GKP states with Wigner functions.

On the other hand, nonlinear elements such as a self-Kerr nonlinearity $-K {a^{\dagger}}^2 a^2$ lead to third-order derivatives in the differential equation for the Wigner function, as well as terms in which $A$ is not constant (corresponding to a so-called {\rm nonlinear} Fokker-Plank equation): the upshot is that the Wigner function can become negative and non-classical during the dynamics and attempts at classical stochastic simulation will suffer from the sign problem. As an example, Ref.~\cite{SMW:wig} discusses Wigner function dynamics for a single oscillator in the presence of a self-Kerr nonlinearity and dissipation.

%To assess the effect of a Kerr nonlinearity, say of the form $U=\exp(-i K t (a^{\dagger} a)^2)$, we can do a mean-field approximation replacing $a^{\dagger} a$ by the random variable representing the mean photon number $\overline{n}$ of an approximate GKP state.  We then assume that the distribution of photons is approximately thermal as argued before, i.e. $P_{\overline{n}}(n)=\frac{\overline{n}^n}{(1+\overline{n})^{n+1}}$ in the initial state. The effect of this evolution is then \begin{equation}\rho(0) \rightarrow \rho(t) \approx \sum_{n=0}^{\infty} P_{\overline{n}}(n) \exp(-i K t n \,a^{\dagger} a )\rho(0)\exp(i K t n\, a^{\dagger} a \end{equation}This dephasing naturally leads to diagonalization of the initial state $\rho(0)$ to the Fock basis i.e. the matrix elements of the state at time $t$ are\begin{equation}\bra{m} \rho(t)\ket{m'}=\frac{1}{1+\overline{n}(1-e^{i K t (m'-m)})} \bra{m} \rho(0) \ket{m'},\end{equation}hence $|\bra{m} \rho(t)\ket{m'}|$ decreases in magnitude from $|\bra{m} \rho(0) \ket{m'}|$ by a factor $\frac{1}{\sqrt{2+\overline{n}^2 \sin^2(K t (m-m')/2)}}$. An initial GKP state is thus turned into a roughly thermal distribution, --the faster the more average photons the initial state contains--, but with even/odd photon number signature as given by the coefficients $c_n$ in Eq.~(\ref{eq:GKPFock}).

Dephasing, meaning the application of a rotation $e^{i \theta a^{\dagger} a}$ with unknown $\theta$ is a possible error mechanism as it rotates the quadratures $\hat{p}$ and $\hat{q}$ into each other.  Dephasing can come about, for example, from an interplay of photon loss and a Kerr nonlinearity, or a fluctuating mode frequency. In a simple stochastic model the angle $\theta$ is drawn from a distribution $\mathbb{P}(\theta)$ with mean $\langle \theta \rangle=0$ and some moments $\langle \theta^k \rangle$. For small higher-order moments $\langle \theta^k \rangle\ll 1$ for $k>2$, we can expand 
\begin{equation}
{\cal N}_{\rm deph, \langle \theta^2 \rangle}(\rho)=\int d\theta \;\mathbb{P}(\theta) e^{i \theta a^{\dagger} a} \rho e^{-i \theta a^{\dagger} a} \approx 
\rho+ \langle \theta^2 \rangle a^{\dagger} a \rho a^{\dagger} a- \frac{1}{2}\langle \theta^2 \rangle (a^{\dagger} a \rho +\rho a^{\dagger} a \rho) +O(\langle \theta^3 \rangle).
\label{eq:stoch-deph}
\end{equation}
This is a dephasing channel which corresponds to the dynamics of a Lindblad equation $\dot{\rho}=\kappa_{\rm deph}{\cal D}(a^{\dagger} a)\rho$ for a short time with $\kappa_{\rm deph} t=\langle \theta^2 \rangle \ll 1$. The fixed point of this equation is any mixture of Fock states $\ket{n}\bra{n}$; when the initial state is $\sum_n c_n \ket{n}$ the channel maps it onto $\sum_n |c_n|^2 \ket{n}\bra{n}$. In Appendix \ref{app:GKP-math} we evaluate the photon number distribution of such fully-dephased $\frak{D}\ket{\overline{0}}$ and $\frak{D}\ket{\overline{1}}$. We prove that the photon number distribution is asymptotically thermal, independent of the logical state. Hence complete dephasing seems to wash out much of distinction between the two logical GKP states.

% BMT Can dephase be simulated by Wigner method?

%We can examine the change in expectation value for $\overline{Z}$ due to the stochastic channel ${\cal S}_{\rm deph, \langle \theta^2 \rangle}$ in Eq.~(\ref{eq:stoch-deph}), i.e. \begin{align}
%{\rm Tr} \;\overline{Z} {\cal S}_{\rm deph, \langle \theta^2 \rangle}(\rho)={\rm Tr} \;{\cal S}^{\dagger}_{\rm deph, \langle \theta^2 \rangle}(e^{i \sqrt{\pi} \hat{q}}) \,\rho=\int d\theta \;\mathbb{P}(\theta) {\rm Tr} e^{i \sqrt{\pi}(\cos(\theta)\hat{q}+\sin(\theta) \hat{p})} \rho(0) \notag \\\approx \int d\theta \;\mathbb{P}(\theta) {\rm Tr} e^{i \sqrt{\pi}(1-\frac{\theta^2}{2})\hat{q}+\theta \hat{p})} \rho(0)={\rm Tr} E \overline{Z} \rho(0), E=\int d\theta \;\mathbb{P}(\theta)  e^{-i \theta\sqrt{\pi}(\frac{\theta}{2}\hat{q}-\hat{p})}e^{i \pi \theta/2}\end{align}
% BMT JC CV I don't know what to say about the effect on S_p and S_q due to dephasing, see commented out math, not so simple..
% BMT JC CV see also reference to your appendix above

%  dynamics of Wigner function under dephasing? second derivative of phase https://iopscience.iop.org/article/10.1088/1367-2630/15/4/043038#nj463337eqn2.9

Ref.~\cite{albert+:photon-loss} has discussed the detrimental effect of a Kerr nonlinearity on a variety of single-mode bosonic codes. Numerical simulations of several sources of inaccuries on GKP state preparation using an ancilla qubit were also discussed in e.g. \cite{DTW:sensor,fluehmann:GKP, SCC:GKP, GKP:exp,tzitrin+:GKPtools} using Lindblad equation dynamics.

%After $M$ rounds of measurements, we have\begin{align}
%W(q_0,p_0, q_1, p_1, \ldots, q_{2M}, p_{2M})=\int dq_0' dp_0' W_{\rm in}(q'_0,p'_0) \notag \\
%\Pi_{i=0}^{M} \iint dq'_{2i-1} dp'_{2i-1} dq'_{2i} dp'_{2i} P_{\rm trans}(q_0,p_0, q_{2i-1},p_{2i-1}, q_{2i}, p_{2i}|q'_0,p'_0, q'_{2i-1},p'_{2i-1}, q'_{2i}, p'_{2i}) \notag \\
%W_{\rm in}(q'_1, p'_1) \ldots W_{\rm in} (q'_{2M}, p'_{2M})
%\end{align}
%The probability $\mathbb{P}(\frak{q},\frak{p})$ is
%\begin{align}
%P(p_1=\frak{p}_1, q_2=\frak{q}_1, )=\iint dq_0 dp_0 dq_1 dp_2 dq_3 dp_4 \ldots dq_{2M-1} dp_{2M} 
%\int dq_0' dp_0' \ldots dq'_{2M} dp'_{2M} \notag \\
%P_{\rm trans}(q_0,p_0, q_1, p_1, \ldots, q_{2M}, p_{2M})|q'_0,p'_0, q'_1, p'_1, \ldots, q'_{2M}, p'_{2M})) \notag \\
%W_{\rm in}(q'_0,p'_0)W_{\rm in}(q'_1, p'_1) \ldots W_{\rm in} (q'_{2M}, p'_{2M})
%\end{align}

\subsection{Repeated GKP Error Correction and Decoding: Finite Squeezing}
\label{sec:decoding}

In this Section we examine the effect of (coherent) finite-squeezing errors on repeated GKP error correction using GKP ancilla's. This is follow-up work from Ref.~\cite{vuillot+:GKP} in which a similar problem was examined using a stochastic Gaussian displacement error model, Eq.~(\ref{eq:GDC}), applied to GKP ancilla and data qubits as a proxy for finite-squeezing errors. The goal of this Section is to understand whether there are crucial differences between finite-squeezing coherent errors and the Gaussian displacement error model and try to develop a dedicated, computationally-efficient, decoder with good performance. 

\begin{figure}[htbp]
\begin{minipage}{0.2\textwidth}
		\Qcircuit @C=.8em @R=0em {
			& \qw & \gate{\rm EC_{GKP}\left
				(\Delta\right )} & \qw & & \equiv & }
	\end{minipage}
	\begin{minipage}{0.6\textwidth}
		\Qcircuit @C=.8em @R=.7em {
		&& & \qw &\qw & \targ & \qw                          & \qw & \qw & \qw  & \qw & \ctrl{1}     & \gate{D(\frac{-\frak{q}+i\frak{p}}{\sqrt{2}})} & \qw \\
		&&&	 \lstick{\ket{\overline{0}}} & \gate{\frak{F}_V(\Delta)}  & \ctrl{-1}    &  \measureD{\hat{p}=\frak{p}} & & & \lstick{\ket{\overline{+}}}& \gate{\frak{F}_V(\Delta)} & \targ & \measureD{\hat{q}=\frak{q}}  \gategroup{1}{13}{1}{13}{.7em}{--}}    
	\end{minipage}
	\caption{A single round of fault-tolerant GKP error correction for both logical $X$ and $Z$ errors.	
%Here $\ket{\overline{+}}$ is the $+1$ eigenstate of $S_q$ and $X$, and $\ket{\overline{0}}$ is the $+1$ eigenstate of $S_p$ and $Z$.The CNOT gate is the logical CNOT for the GKP code which induces the transformation $q_{\rm target} \rightarrow q_{\rm control}+q_{\rm target}$ (while $p_{\rm control} \rightarrow p_{\rm control}-p_{\rm target}, q_{\rm control}\rightarrow q_{\rm control},p_{\rm target}\rightarrow p_{\rm target}$).
		Each measurement is a perfect homodyne measurement of $\hat{q}$ or $\hat{p}$ with outcomes $\frak{q}$ and $\frak{p}$ respectively. The finitely-squeezed ancilla states are modeled as approximate GKP states, using a slightly-different small-$\Delta$ approximation than $\frak{E}, \frak{D}, \frak{F}$ in Eqs.~(\ref{eq:model}), (\ref{eq:approx}) and Eq.~(\ref{eq:approxE}), given in Eq.~(\ref{eq:ancilla-approx}) and denoted as $\frak{F}_V$. We show that the optional corrective displacement (dashed box) keeps the state at low average photon number $\overline{n}$ in Fig.~\ref{fig:photons}.}
	\label{fig:EC_GKP}
\end{figure}
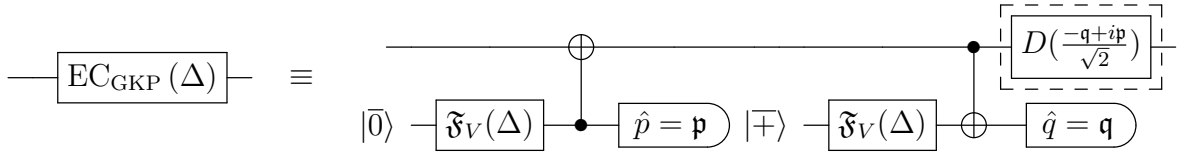

%We also develop a fine-tuned classical decoders to this coherent source of errors, which demonstrate novel ideas about decoding. 
The dynamics to be analyzed is the repeated execution of the quantum circuit in Fig.~\ref{fig:EC_GKP} on a single GKP input state $\frak{F}(\Delta) \ket{\overline{\psi}}$ for $m=1,\ldots, M$ cycles. We remark that a variant of such `Steane error correction' exists: in \cite{GK:error} the authors observed that applying a beam-splitter between GKP ancilla and GKP data qubit followed by squeezing on the GKP data qubit, can also perform error correction. Ref.~\cite{WNK:GKerror} has analyzed the repeated execution of this variant of error correction in more detail.

A clear difference between a stochastic error model and the finite-squeezing model is that in the former entropy build-up is possible, while in the latter the state conditioned on the measurement outcomes in Fig.~\ref{fig:EC_GKP} is pure at all times. One can invoke displacement twirling as a method to convert a coherent noise model in which one applies a superposition of displacements to a stochastic mixture of displacements. For example, displacement twirling a finitely-squeezed state $\frak{E}\ket{\overline{\psi}}$ with some $\Delta$ gives a perfect state $\ket{\overline{\psi}}$ subject to the Gaussian Displacement Channel with $\Delta^2 =2 \sigma_0^2$ ~\cite{vuillot+:GKP}. After such stochastification of the noise on a GKP ancilla, one can then represent the feedback error (a shift in one of the quadratures) induced by the ancilla in the circuit in Fig.~\ref{fig:EC_GKP} effectively as an incoming stochastic shift error on the data qubit. The stochastic shift of the other quadrature of the ancilla then causes a measurement error of the same strength. On this basis, Ref.~\cite{vuillot+:GKP} stochastically modeled finite-squeezing errors as incoming stochastic displacement errors on the GKP data qubit and measurement errors.
Another difference in the models is that in the finite-squeezing error model the measurement outcomes carry non-modular information about the measured quadrature. This can be exploited to recenter the state by choosing a corrective displacement immediately after a single round of error correction, while such corrective displacements would have no effect in the stochastic model.

%Hence we will examine the difference in logical error for the coherent versus stochastic error model using this identification in Fig.~\ref{fig:stoch-coherent}.

%\gate[style={fill=red!20},label style=cyan]

\begin{figure}[htb]
\centering
\includegraphics[width=0.5\textwidth]{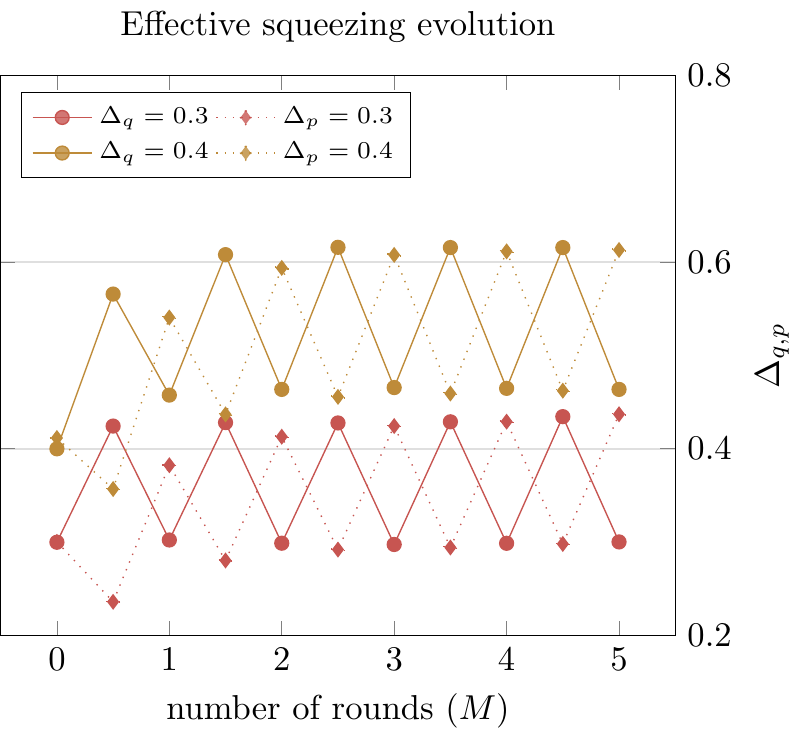}
\caption{Time-evolution of averaged effective squeezing parameters $\Delta_p$ and $\Delta_q$ in Eq.~(\ref{eq:eff-squeeze}) during repeated quantum error correction using Fig.~\ref{fig:EC_GKP} (without corrective displacements). We see an enhancement in $\Delta_q$ (resp. $\Delta_p$) during $Z$-error correction which lowers $\Delta_p$ (resp. $X$-error correction which lowers $\Delta_q$) due to the feedback error induced by the finitely-squeezed GKP ancilla. \label{fig:eff-sq}}
\end{figure}

\begin{figure}[htb]
\centering
\includegraphics[width=0.5\textwidth]{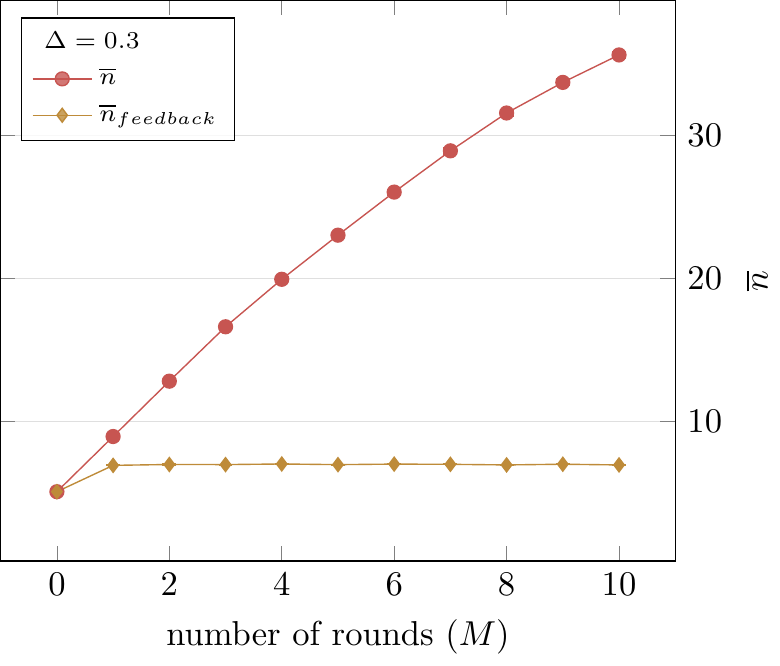}
\caption{Photon number evolution of an initial state $\frak{F}\ket{0}$ state under repeated stabilizer measurement for $\Delta=0.3$. Here, $\overline{n}_{feedback}$ denotes the photon number evolution where each EC-circuit as in Fig.~\ref{fig:EC_GKP} is followed by a corrective displacement $D(\frac{-\frak{q}+i\frak{p}}{\sqrt{2}})$. The data is obtained by averaging over $N=10^4$ trajectories. One observes that the average photon number stays low, corresponding to a centering of the state. 
%We remark that the photon number fluctuations are larger when we do not apply the corrective displacement. 
\label{fig:photons}}
\end{figure}

We will represent the GKP wave-function in the $q$-basis with $\psi_m(q)$, the wavefunction after $m$ rounds of error correction. We will sometimes omit the normalization of states when these normalizations play no role. 

We will now analyze the time evolution without the corrective displacement in the dashed box in Fig.~\ref{fig:photons}. A single round of quantum error correction shown in Fig.~\ref{fig:EC_GKP} with measurement outcomes $\frak{p}_m, \frak{q}_m$ gives $\psi_m(q)=\int dq' G(q \leftarrow q'| \frak{q}_m, \frak{p}_m) \psi_{m-1}(q')$ with Green's function
\begin{align}
G(q \leftarrow q'| \frak{q}_m, \frak{p}_m)& =  \int dq'' G_+( q \leftarrow q''|\frak{q}_m) G_0( q'' \leftarrow q'| \frak{p}_m)\notag \\
& =\psi^+(q-\frak{q}_m) \psi^0(q-q') e^{-i\frak{p}_m(q-q')},
\label{eq:Green}
\end{align}
using $G_0( q'' \leftarrow q'| \frak{p}_m) =\psi^0(q''-q') e^{-i\frak{p}_m(q''-q')}$ ($Z$-error correction) followed by $G_+( q \leftarrow q''|\frak{q}_m) =\delta(q''-q) \psi^+(q-\frak{q}_m)$ ($X$-error correction). To understand this Green's function, observe that in the limit $\Delta \rightarrow 0$, the wavefunction $\psi^+(q)$ has uniform support on $q=k \sqrt{\pi}$, with $k \in \mathbb{Z}$, so that the outgoing wave function is supported solely on $q=\frak{q}_m+k \sqrt{\pi}$, hence the code state sits in the perfect code space with a known shift on top. However, before this, the interaction with the imperfect $\ket{\overline{0}}$ ancilla for the $Z$-error correction applies {\em a convolution} to the incoming wavefunction. If the ancilla is perfect ($\Delta \rightarrow 0$), this convolution amounts to applying superpositions of stabilizer shifts $S_p^k$ with $2k\sqrt{\pi}=q-q'$, each with a phase which depends on $\frak{p}_m$, on the incoming wave-function. If we assume that all wavefunctions are of the form $\frak{F} \ket{\overline{0}}$, i.e. sums of Gaussians, the convolution leads again to a sum of Gaussians and can be exactly evaluated, that is, one has
\begin{align}
\int dq' \psi^0(q-q') e^{-i \frak{p}_m(q-q')}\psi^0(q') \propto \sum_{k,k' \in \mathbb{Z}} e^{-2 \Delta^2 \pi (k'^2+k^2)} \times \notag \\ \int dq' e^{-i \frak{p}_m(q-q')} e^{-\frac{1}{2\Delta^2} \left[(q'-2k'\sqrt{\pi})^2+(q-q'-2k\sqrt{\pi})^2\right]} = \notag \\
\sqrt{\pi \Delta^2}e^{-\Delta^2 \frak{p}_m^2/4} \sum_{k,k' \in \mathbb{Z}} e^{-2 \Delta^2 \pi (k'^2+k^2)} e^{-\frac{(q-2\sqrt{\pi}(k+k'))^2}{4\Delta^2}}e^{-i \frac{\frak{p}_m}{2} (q+2\sqrt{\pi}(k-k'))}.
\label{eq:green-gauss}
\end{align}

%We have supp(fg)=supp(f) \cap supp(g) \subseteq supp(f).

What we observe is that the convolution broadens the peaks and they acquire phases which depend on the location of the peaks and the outcome $\frak{p}_m$. The convolution step, which corrects shifts in $p$, thus introduces a feedback error in the form of peak broadening for the $q$ variable. 

In Fig.~\ref{fig:eff-sq} we plot the effective squeezing parameters $\Delta_p,\Delta_q$ of the state, Eq.~(\ref{eq:eff-squeeze}), after rounds of $Z$ and $X$-error correction. 

% BMT JC with or without correction
%JC BMT The Delta data shows the results without a correction, but note that the Delta is computed via  |Tr[S_{q/p} \rho]|. Adding a sinmple displacement on \rho only results in a phase which is killed by the absolute value. So the finite squeezing paramter is in variant unter displacements, it only quantifies how periodic the state is.

Represented as a state evolution, the stabilizer measurements of $S_p$ and $S_q$with outcomes $\frak{q}_m,\, \frak{p}_m$ effectively map an incoming state $\ket{\phi}$ to $\psi^+(\hat{p}+\frak{p}_m)\ket{\phi}$ and $\psi^+(\hat{q}-\frak{q}_m)\ket{\phi}$. The finite envelope of the approximate GKP states has the effect that the outgoing states are dominantly supported around $\hat{p}=-\frak{p}_m,\;\hat{q}=\frak{q}_m$. We understand this gain of non-modular information in one quadrature as a reflection of the loss of modular information (i.e. in terms of the increase in $\Delta_{q/p}$) in its conjugate. In Fig.~\ref{fig:photons} we observe, that a displacement about $\alpha=\frac{-\frak{q}_m+i\frak{p}_m}{\sqrt{2}}$ indeed contributes to a recentering of the state.

% BMT JC skip para?
%JC BMT I  think we should keep it. This argument is very central to building an intuition for how different the Finite squeezing error is to its stochastic approximation.

What is noteworthy about the error correcting dynamics in Eq.~(\ref{eq:Green}) and Eq.~(\ref{eq:green-gauss}) is that the support of the outgoing wavefunction lies within the support of the incoming wavefunction ($\psi^0(q)$) plus the support of the ancilla wavefunction (here also $\psi^0(q)$) since ${\rm Supp}(f \star g) \subset {\rm Supp}(f) +{\rm Supp}(g)$ for a convolution of two functions $f$ and $g$. The $X$-error correction step {\em multiplies} the convoluted wavefunction by $\psi^+(q-\frak{q}_m)$ which cannot extend its support. Within its support the outgoing wavefunction can have changed amplitudes, depending on the outcomes $\frak{p}_m$ and $\frak{q}_m$. These arguments are relevant as the GKP state $\frak{F} \ket{\overline{0}}$ has support which is concentrated around even multiples of $\sqrt{\pi}$ overlapping by an amount exponentially suppressed in $\Delta^2$ with the support of $\frak{F} \ket{\overline{1}}$. When one uses code states $\psi^0(q)$ and $\psi^1(q)$ whose supports have negligible overlap, --which is the case for $\frak{F}\ket{\overline{0}}$ and $\frak{F}\ket{\overline{1}}$ for sufficiently small $\Delta$--, it implies that, no matter what the measurement outcomes, 0 will largely remain a 0 and 1 will largely remain a 1 in the error correction rounds.
 %\footnote{To argue about the support of $\psi^1(q)$, we can see it as the translate (by $\sqrt{\pi}$ or $\overline{X}$) of some wavefunction $\psi^0(q)$ and the translation $\overline{X}$ can be propagated through the circuit in Fig.~\ref{fig:EC_GKP} so as to only affect the outcome $\frak{q}_m$ which plays no role in the support issue, hence the same arguments apply. Alternatively, using the wavefunction $\frak{F} \ket{\overline{1}}$ directly, we can use convolution of Gaussian distributions to note that the convolution with $\frak{F} \ket{\overline{0}}$ gives a wavefunction with peaks at odd integer multiples of $\sqrt{\pi}$ as $\frak{F}\ket{\overline{1}}$ itself.}. 

\begin{figure}[htb]
\centering
\includegraphics[width=0.5\textwidth]{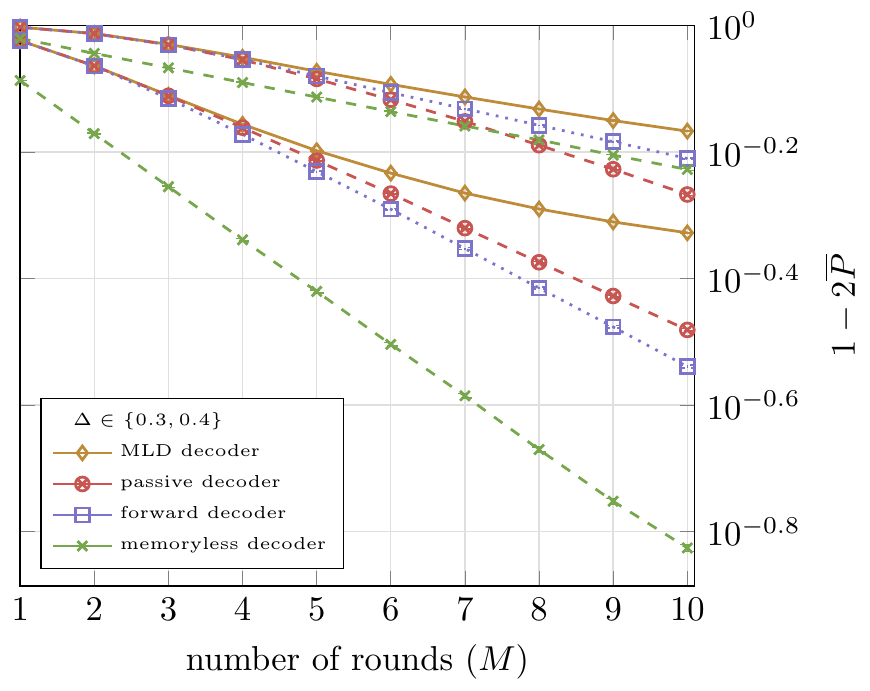}
\caption{Logical error rate $\overline{P}$ for decoding the circuit in Fig. \ref{fig:EC_GKP} ({\em without} the corrective displacement) for different decoders and squeezing parameters $\Delta=0.3$ (top data) and $0.4$ (bottom data) with the number of stabilizer measurement rounds $M=1,..,10$. For $\Delta=0.3$ the mean performance of the forward decoder outperforms passive decoding, while the opposite is true for $\Delta=0.4$.  We have observed that the single sample logical error rates strongly fluctuate at $O(10^{-2})-O(10^{-1})$ around the averages, suggesting rather chaotic behavior. The data are obtained by sampling over $N=5\cdot 10^4$ trajectories of measurement outcomes. Error bars denote the standard deviations of the averaged logical error rates. The memoryless decoder applies a different corrective displacement after each QEC round described in Appendix \ref{app:decode}. \label{fig:sim}}
\end{figure}

It implies that the picture of stochastic error feedback by using a finite-squeezed $\ket{\overline{+}}$ ancilla is inadequate as in this picture the support of the wavefunction gets shifted around by such feedback error, while here we observe that instead amplitudes get changed {\em within the support}. We see some of this behavior in the performance of the maximum likelihood decoder versus a passive decoder in Fig.~\ref{fig:sim}: a passive decoder which decides that 0 stays a 0 does surprisingly well, which may be understandable if we consider that the support of the $\psi_0(q)$ wavefunction can never grow by QEC (of course, in principle, the wave functions have support everywhere albeit exponentially-suppressed with $1/\Delta^2$).
% BMT JC CV agree?
%JC BMT Yes I agree.

% BMT skip para end?

We have formulated a classical `forward' decoder, see details in Appendix \ref{app:decode}, and compared its performance with an optimal, density matrix decoding method (maximum likelihood decoding) as well as a so-called passive decoder, see the numerical results without active displacement feedback in Fig.~\ref{fig:sim}. The passive decoder functions as an important sanity check: this decoder throws away all the measurement data, including the final measurement and {\em simply always decides that the outcome is 0} when the input state to the rounds of error correction is an approximate GKP 0 state. Since we don't necessarily know the input state, we note that this decoder is of little practical value. By comparing the performance of this decoder with the MLD decoder we learn to what extent the quantum error correction circuits are preserving quantum information irrespective of the measurement outcomes and to what extent the measurement outcomes provide the proper logical information for correction. This decoder is similar to the passive decoder in the stochastic model of \cite{vuillot+:GKP} in which none of the error information in each round is used and only the last perfect measurement determines whether the state is identified as 0 or 1. However, in \cite{vuillot+:GKP} this passive decoder clearly performed worse for these values for $\Delta$.

\begin{figure}[htb]
\centering
\includegraphics[width=0.5\textwidth]{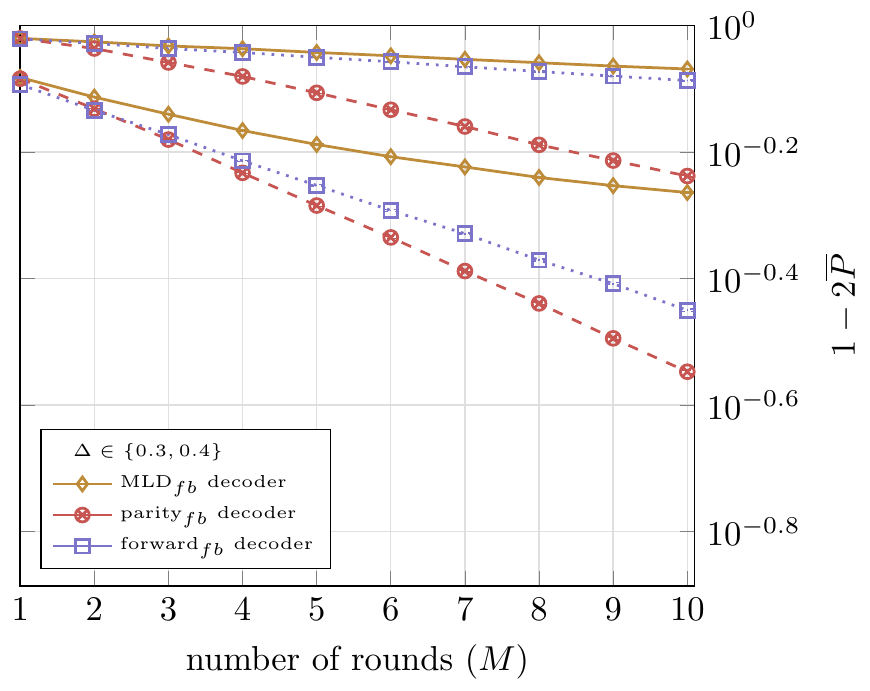}
\caption{Logical error rate $\overline{P}$ for decoding the circuit with active feedback as in Fig. \ref{fig:EC_GKP}. We observe an overall improvement in the logical error rate as compared to the performance of the circuits without immediate feedback. \label{fig:simdisp}}
\end{figure}

\begin{figure}[htb]
\includegraphics[width=\textwidth]{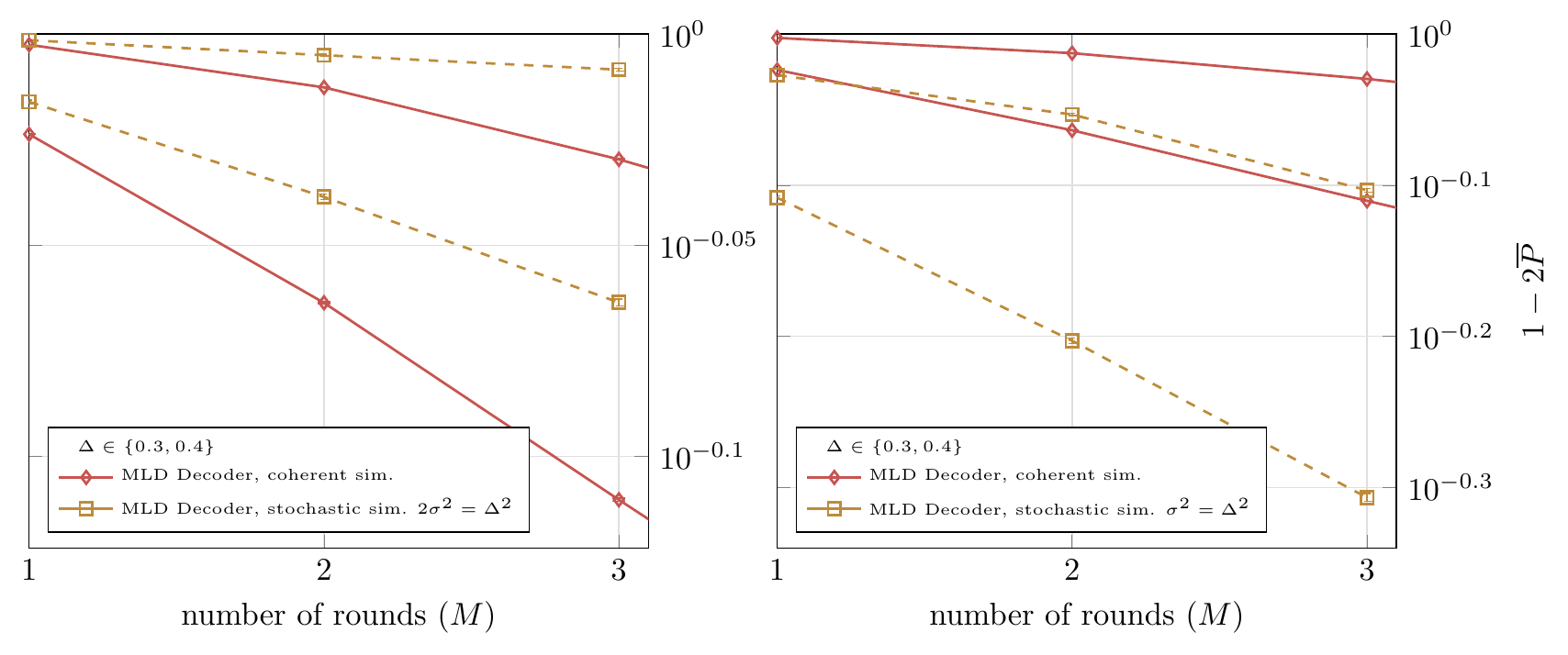}
\caption{Logical error rate for MLD decoders for finite-squeezing (coherent) errors characterized by $\Delta$ versus the stochastic error model characterized by $\sigma_0$ in \cite{vuillot+:GKP}. We compare the logical error rates using two different mappings from $\Delta$ to $\sigma_0$, neither is entirely satisfactory. \label{fig:stoch-coherent}}
\end{figure}

We compare the decoder performance in Fig.~\ref{fig:simdisp} to a `memoryless' decoder using the measurement of the current QEC round for immediate logical feedback, see Appendix \ref{app:decode}, clearly showing worse performance with this strategy. In Fig.~\ref{fig:simdisp} we show the better performance of the MLD and the (feedback-adapted) forward decoder using active feedback which minimizes photon number. We compare their performance with a `parity' decoder which similarly applies the corrective displacement, but then applies a final logical correction (or not) when the sum of all applied shifts is closer to an odd (or even) multiple of $\sqrt{\pi}$.

We observe that for small $\Delta$ and number of EC rounds $M \leq 3$, the passive decoder performs comparably to the MLD decoder, consistent with the intuition given earlier in this section. At larger $M$ it performs worse at $\Delta=0.3$. 
%disp results
Figures \ref{fig:sim} and \ref{fig:simdisp} show the average logical error rates obtained from $N=5\cdot 10^4$ samples for all decoders. We note (not shown in the Figures) that there are large fluctuations of $O(10^{-2})-O(10^{-1})$ per run around the average logical error rate.
% BMT-post JC changed it, check

Last, to study the difference between coherent and stochastic errors, we plot $\overline{P}_{\rm MLD}^{\rm stoch}(\sigma_0)$ and $\overline{P}_{\rm MLD}(\Delta)$ using the identification $\Delta^2 =2 \sigma_0^2$ and $\Delta^2 = \sigma_0^2$ as a function of $M$ and $\Delta$, see Fig.~\ref{fig:stoch-coherent}. The simulation for the MLD decoder based on a stochastic error model is implemented following \cite{vuillot+:GKP}. We observe, that the conversion $\Delta^2 =2 \sigma_0^2$ underestimates the logical error, while it is overestimated for $\Delta^2 =\sigma_0^2$. 

The simulation and data are accessible at \url{https://github.com/JonCYeh/GKP_EC_Sim}.

\section{Circuit-QED Realizations of GKP Qubit Components}
\label{sec:34-mix}

%$\chi^{(2)}$ second-order susceptibility relating polarization ${\bf P}$ to electric field squared.$\chi^{(3)}$ third-order susceptibility relating polarization to electric field cubed.So if we have dipolar coupling ${\bf P} \cdot {\bf E}$, $\chi^{(2)}$ interactions contains 3 creation and/or annihilation operators, hence 3-wave mixing and $\chi^{(3)}$ contains 4 creation and/or annihilation operators, hence 4-wave mixing.

% wustmann & shumeiko stuff? https://arxiv.org/abs/1302.3484

In this Section we review and discuss schemes for state preparation, logical gates and quantum error correction for GKP qubits in circuit-QED. In circuit-QED a natural candidate for a bosonic GKP encoding is a resonant mode of a 3D microwave cavity, having low loss rate. Multiple GKP qubits are then stored in multiple low-loss 3D cavities: an engineering platform for multiple coupled cavities,-- multi-layer microwave-integrated quantum circuits (MMIQC) \cite{brecht+:MMIQC}--, is under development.

\begin{figure}[htbp]
	\center
	\includegraphics[scale=0.08]{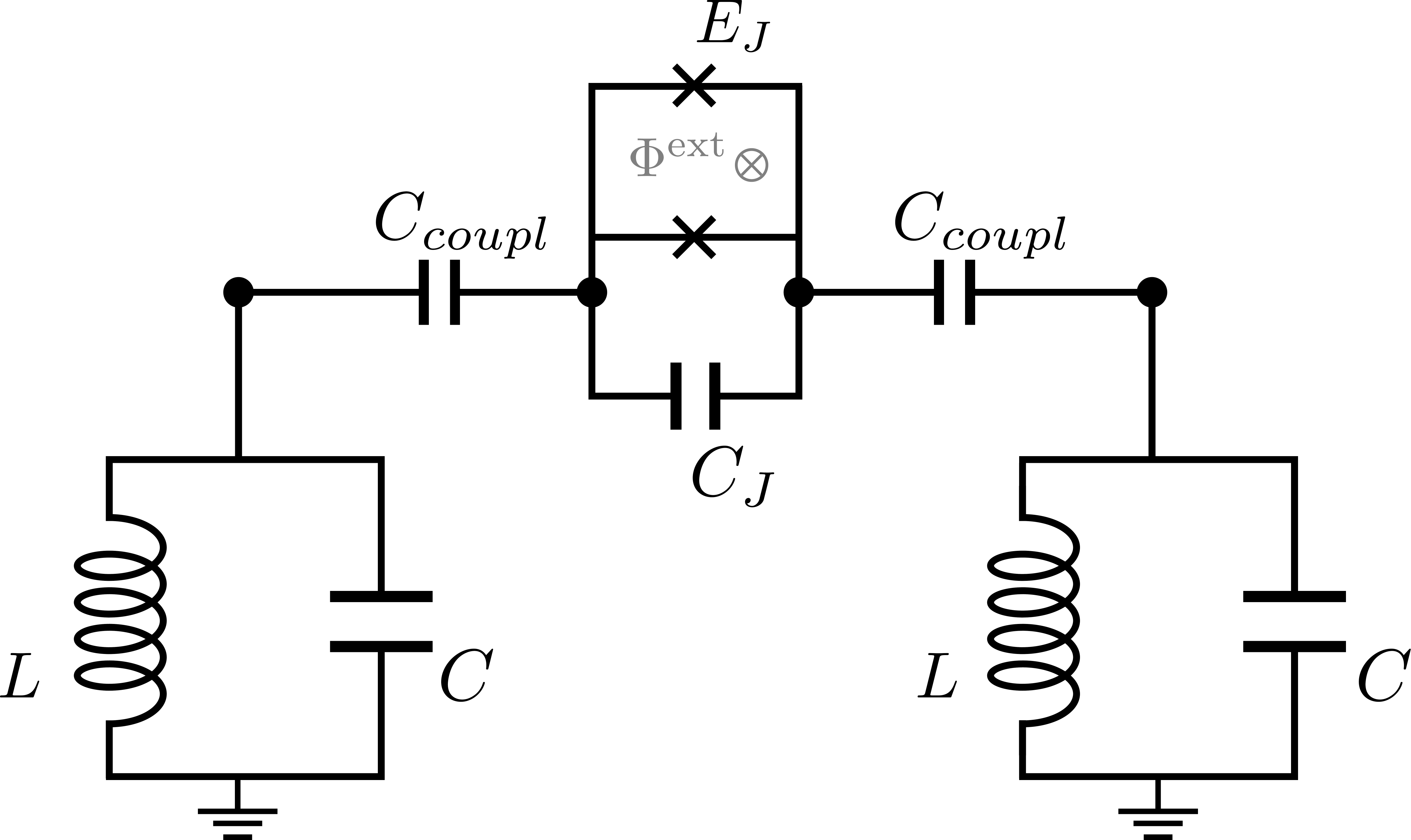}
	\caption{Electric circuit representing a typical capacitive coupling between two LC oscillators, modeling cavity modes, and a (possibly flux-tunable) transmon qubit.}
	\label{fig:circuit}
\end{figure}

The coupling between cavity modes of different cavities can be mediated by dipolar `antenna' couplings between the electric field of the cavity mode and that of an inserted planar chip in the cavity wall hosting a coupler mode. The idea is then to activate two-mode gates such as the CZ gate by applying microwave drives or flux-modulation outside of the 3D cavities on the coupler mode. As a simple circuit example one can take the electric circuit in Fig.~\ref{fig:circuit} in which the two LC oscillators correspond to the cavity modes: two superconducting islands, each protruding into one cavity and coupling to the electric field of the resonant mode, are connected by a Josephson junction (so-called bridge configuration): such setup can generate the $\Phi^4$-interaction for a CZ gate as will be described in Section \ref{sec:4wave}, while more involved circuits could be used to engineer an effective $\Phi^3$-interaction for the same purpose, see Section \ref{sec:GKP-CZ}.

\subsection{Coupling with Regular Qubits}
\label{sec:conD}

To prepare a logical GKP state or to realize single-qubit gates, one can employ an interaction with a regular qubit in which the state of regular qubit controls the application of a displacement on the GKP mode as in Fig.~\ref{fig:PE}. If the regular qubit is realized by an anharmonic oscillator such as a transmon qubit, this then requires the engineering of a tunable qubit controlled-displacement interaction of the form $b^{\dagger} b \,\hat{q}_{\theta}$ with $\hat{q}_{\theta}=\cos(\theta) \hat{q}+ \sin(\theta) \hat{p}$ so that $b^{\dagger} b$ acts as Pauli $Z$ in the regular qubit subspace. 

As mentioned earlier, a common interaction between an off-resonantly coupled transmon qubit and cavity mode is the dispersive or cross-Kerr interaction of the form $-\chi a^{\dagger} a b^{\dagger} b \rightarrow -\chi a^{\dagger} a Z/2$. This interaction realizes a qubit controlled-rotation which can be converted, in principle, to a qubit controlled-displacement, using additional displacements and qubit-flips as follows. Since $e^{i \theta a^{\dagger} a Z} D(\alpha) e^{-i \theta a^{\dagger} a Z}=D(\alpha e^{i \theta Z})$, choosing $\theta=\pi/2$ gives the displacement $D(\alpha)$ when $Z=1$ and the displacement $D(-\alpha)$ when $Z=-1$. Thus this sequence of gates does what is needed. To realize $e^{i \theta a^{\dagger} a Z}$, one can simply conjugate the interaction $e^{-i \theta a^{\dagger} a Z}$ by $\pi$-bit-flips on the qubit, so that we can do all 3 gates in the decomposition of $D(\alpha e^{i \theta Z})$.

Note that the strength of the controlled-displacement $\alpha$ only depends on the strength of the uncontrolled displacement (which can easily be made very strong in O(10) ns). If the entire controlled-displacement is to be done in, say 50 ns, it requires two rotations each with time $t=\pi/\chi=20$ ns, or $\frac{\chi}{2\pi}=25$ Mhz. This realization thus requires making $\chi$ tunable, i.e. when the transmon qubit is to be measured or prepared, it is important that the cross-Kerr interaction be `off', as it induces rotations on the GKP grid state which dephase the state in the Fock basis. However, there is a limit to the on-off ratio of $\chi$ obtained by flux-tuning of the transmon qubit, in particular if $\chi$ is flux-tuned to be stronger, then the resonator becomes more anharmonic as well, see \cite{nigg+:bb} and Eq.~(\ref{eq:off}).

%For example, one can consider just one of the LC oscillators in Fig.~\ref{fig:circuit} and examine the effect of flux-tuning the qubit. The effect of flux-tuning can in its simplest form be modeled as a change in value of $E_J$ to $E_J(\Phi^{\rm ext})=E_J |\cos(\pi \Phi^{\rm ext}/\Phi_0)|$ \cite{koch+:transmon} so that the effective inductance $L_J(\Phi^{\rm ext})=\frac{L_J}{|\cos(\pi \Phi^{\rm ext}/\Phi_0)|}$. In the usual RWA and quartic expansion of the cosine approximation (as in Eq.~\ref{eq:off}) we have
%\begin{equation} H(\Phi^{\rm ext})=\omega_r a^{\dagger} a + \frac{\alpha_r}{2} (a^{\dagger}a)^2 +\omega_q b^{\dagger} b+\frac{\alpha_{q}}{2} (b^{\dagger} b)^2 + \chi_{rq}(\Phi^{\rm ext}) a^{\dagger} a b^{\dagger} b.\end{equation}with $\chi_{rq}(\Phi^{\rm ext})=-2\frac{E_C C_J }{L_J(\Phi^{\rm ext})} \sqrt{\frac{L_J(\Phi^{\rm ext}) L}{C C_J}$, and where in principle all parameters besides $\chi_{rq}(\Phi^{\rm ext})$ such as $\omega_q, \omega_r$ and the mode operators $a, b$ have some dependence on $\Phi^{\rm ext}$. Note however that if $\chi_{rq}$ were negligible, flux-tuning could never create any entanglement between the two modes as the system is uncoupled at all times. It is thus the variability of $\chi$ with flux 

%$Z[\omega]=[i \omega C_{\rm coupl}+i \omega C+\frac{1}{i \omega L}]^{-1}$.

Instead of needing a tunable interaction, Ref.~\cite{GKP:exp} realized a qubit controlled-displacement in $1.2\,\mu$s, using a very weak dispersive coupling $\frac{\chi}{2\pi}=28$ kHz between transmon qubit and cavity mode. This weak coupling obviates the need for a tunable interaction, but would make for a very slow gate when using the method described in the previous paragraph. The idea in \cite{GKP:exp} is to realize a qubit controlled-displacement by temporally displacing the cavity mode to states with $|\beta|^2=320$ photons, so that even a small qubit-induced cavity rotation can have a large effect. The scheme is best understood by using a displacement frame, see Appendix \ref{sec:displace}, on a cavity-driven dispersive shift Hamiltonian $H=-\frac{\chi}{2} a^{\dagger} a Z+i{\cal E}(t)a^{\dagger}-i {\cal E}^*(t)a$. The displacement frame shift shows that the dynamics is due to an effective Hamiltonian
\begin{equation}
\tilde{H}(t)=-\frac{\chi}{2} a^{\dagger} a Z-\frac{\chi}{2} |\beta(t)|^2 Z-\frac{\chi}{2} (a \beta^*(t)+a^{\dagger} \beta(t))Z,
\end{equation}
where $\beta(t)=\int_0^t dt' {\cal E}(t')$. The effect of the last term in this Hamiltonian after a time $T$ (taking $\beta=\beta^*$ for simplicity) is the qubit controlled-displacement $\frac{\chi}{\sqrt{2}}  \int_0^T dt |\int_0^t dt' {\cal E}(t')| \hat{q} Z$. In order to cancel the qubit controlled-rotation (first term) the qubit state is flipped midway in the interval $T$, requiring that the cavity displacement direction is also inverted midway, i.e. $\beta \rightarrow -\beta$. We see that in this realization the applied displacement power and the dispersive shift $\chi$ together determine the strength of the qubit controlled-displacement.

We can ask how to improve on the execution of the qubit controlled-displacement gate and the subsequent qubit measurement, where improvement means a faster as well as more reliable execution. As for the realization in \cite{GKP:exp} one may worry that the large displacements of the state in phase-space during the execution of the gate lead to errors on the GKP state. Even though the cavity has a long lifetime (single-photon life-time $T=245\,\mu$s in \cite{GKP:exp}), the logical operator of a displaced GKP state takes on a complex oscillatory value in time due to photon loss, see Eq.~(\ref{eq:pl-loss-displace}). It is desirable to shorten the duration of the transmon qubit measurement (700 ns in \cite{GKP:exp}), but it is hard to make the dispersive read-out of the qubit via a read-out resonator very fast. For example, the measurement pulse followed by active read-out resonator depletion is O(600) ns in \cite{bultink+:error-detect} and O(250) ns (including resonator occupancy) in \cite{heinsoo:multiplex}. Replacing the qubit measurement by feedback and disentangling \cite{hastrup+:meas-free} requiring other controlled-displacements can only lead to a shorter overall preparation time if the duration of such controlled-displacement can be shortened from what was achieved in \cite{GKP:exp}. Note that the replacement of measurement by coherent interactions could also be done for the GKP qubit rotation in Fig.~\ref{fig:singleU}.

Instead of a transmon qubit as ancilla, one may consider a different qubit such as fluxonium \cite{flux+:manu}, again dispersively coupled to the 3D cavity mode. Advantages of a fluxonium qubit are its long coherence and larger anharmonicity leading to lower leakage \cite{nguyen+:flux}, equally fast-single qubit gate operations ($O(10)$ ns) as well as potentially very fast and powerful qubit measurement, see e.g. the GrAl-based fluxonium qubit in \cite{pop+:GrAl, winkel+:flux, gebauer+:fluxonium}. In addition, flux-tuning fluxonium may give a strongly-tunable dispersive shift $\chi$ \cite{flux+:manu, zhu+:flux}, without the unwanted side-effect of strengthening the cavity anharmonicity.

% BMT not found these points so clearly in literature

Another proposal is to use a noise-biased cat qubit to measure the stabilizer displacements of a GKP qubit \cite{puri+:Kerr-cat} using a tunable beam-splitter interaction between the two cavity modes of the form $H=g(t) a b^{\dagger}+g^*(t) a^{\dagger} b$, as argued in Section \ref{sec:bias-cat}. To use the interaction, we thus imagine first preparing the cat qubit in $\ket{C_{\alpha}^+}$ (by starting in the vacuum state $\ket{0}$ and turning on the pump), then activate the tunable beam-splitter, and measure the noise-biased cat in the $\ket{C_{\alpha}^{\pm}}$ basis or employ feedback and disentangling via qubit controlled-displacements \cite{hastrup+:meas-free}.

\subsection{Logical GKP Measurement}
\label{sec:FTmeas}

How does one determine whether a GKP state is $\ket{\overline{0}}$ or $\ket{\overline{1}}$, that is, realize a logical $\overline{Z}$-measurement?  Such logical measurement may be completely destructive, but is desired to have high-fidelity, hence be fault-tolerant in its implementation, meaning that the outcome is insensitive to imperfections in the state. 
Even though one can measure a logical displacement, i.e. $\overline{X},\overline{Z}$ or $\overline{Y}$, using a single ancilla qubit as was done in Refs.~\cite{fluehmann:GKP, GKP:exp}, such measurement has an intrinsic probability of error on an approximate GKP state. For example, measuring $\overline{Z}$ on $\ket{\overline{0}}$ does not give outcome $1$, since the state is not a perfect eigenstate of $\overline{Z}$ but obeys Eq.~(\ref{eq:Zexp}). If we assume that this measurement circuit is otherwise perfect and is applied to $\frak{F} \ket{\overline{b}}$ with $b=0,1$, the probability for the ancilla qubit to be measured as $\pm$ equals $\mathbb{P}(\pm|\overline{b})=\frac{1}{4}\frac{\bra{\overline{b}}\frak{F}^{\dagger} (I\pm \overline{Z}^{\dagger})(I \pm  \overline{Z})\frak{F} \ket{\overline{b}}}{\bra{\overline{b}} \frak{F}^{\dagger} \frak{F} \ket{\overline{b}}} \approx \frac{1}{2}(1\pm (-1)^{b} e^{-\pi \Delta^2/4)})$, using Eq.~(\ref{eq:Zexp}). The upshot is that the ancilla measurement is flipped with symmetric error probability $q=\frac{1}{2}(1-e^{-\pi \Delta^2/4)})$ which goes to 0 when $\Delta \rightarrow 0$. 
At, say, $\Delta=0.3$, this readout error probability $q$ is about $3.4\%$ and much larger than 
the probability for an incorrect $\overline{Z}$-outcome through the ideal homodyne measurement given in Eq.~(\ref{eq:meas}). Some repetition of the controlled-displacement circuit with the ancilla qubit and taking a majority vote of the answers could bring down the error probability $q$ at the price of more time and possibly additional feedback error.

A target for future work could be to achieve an improved logical GKP qubit measurement by releasing the GKP state from a superconducting cavity via a switch-release mechanism \cite{pfaff+:controlled-release} (taking $O(1)\; \mu$s in time in \cite{pfaff+:controlled-release}) into a transmission line and then enact phase-sensitive amplification (e.g. squeezing) so as to measure one quadrature, say $\hat{q}$, with no further added noise.
After calibration of the measurement using $\Delta$-squeezed displaced states and their targeted measurement outcomes, the measurement could proceed by determining whether the amplified signal corresponds to a $\hat{q}$ which is closer to an even ( $\rightarrow$ outcome 0) or odd multiple ($\rightarrow$ outcome 1) of $\sqrt{\pi}$.  
Photon loss in this process may be expected to be a dominant source of noise. To get an estimate of the error rate in the presence of photon loss, we can compute Eq.~(\ref{eq:meas}) for a state at $\Delta=0.3$ undergoing photon loss as in Eq.~\ref{eq:OU} with $\kappa t=0.1$ (so that a coherent state loses $1-\exp(-\kappa t) \approx 10\%$ of its intensity), giving $\mathbb{P}(\overline{Z}=1|\frak{F}\ket{\overline{0}}, \kappa t=0.1)=99.5\%$. For $\kappa t=0.5$ this measurement success probability is already down to $80\%$. 
%To model the effect of adding noise in the actual measurement, one could assume that Gaussian noise with standard deviation $\sigma_M$ gets added to the outgoing probability distribution $P(q)$, that is, $P_{\rm meas}(q)=\int_{\mathbb{R}} dq' \mathbb{P}_{\sigma_M}(q-q') P(q')$, further decreasing the quality of the measurement.

% BMT see numerics in photon-loss.nb

\subsection{GKP CZ Gate via Three-Wave Mixing}
\label{sec:GKP-CZ}

In this Section we describe how one could realize the CZ interaction between two GKP modes via a 3-wave mixing element which is activated by applying a (strong) microwave pump tone to a coupler mode, see e.g. \cite{RD:3-wave}. An example of pure 3-wave mixing used for broadband parametric amplification is the Josephson-ring modulator circuit \cite{3-wave:ring-mod, bergeal+:ring}.

% BMT extra Krastanov reference?

An example of the use of parametrically-activated 3-wave mixing is the experiment in \cite{vraj+:stimul}: flux-modulation through a coupling Josephson junction (instead of microwave driving) is used to activate a ${a^{\dagger}}^2 b$ coupling between a logical (co-planar microwave) resonator whose state is to be manipulated and an ancilla (co-planar microwave) resonator.

For simplicity, we here assume that the following non-degenerate 3-wave mixing Hamiltonian is available:
\begin{equation}
H_{\chi^{(2)}}=\omega_a (a^{\dagger} a+\frac{1}{2})+ \omega_b (b^{\dagger} b+\frac{1}{2}) +\omega_c (c^{\dagger} c+\frac{1}{2})+ \chi^{(2)} (a+a^{\dagger})(b+b^{\dagger})(c+c^{\dagger}).
\label{eq:H-3wave}
\end{equation}
Here $a, b$ are annihilation operators for two GKP oscillators while $c$ is the annhilation of the pump oscillator. We assume that all frequencies  $\omega_a,\omega_b$ and $\omega_c$ are sufficiently detuned, so that the $\chi^{(2)}$ interaction between the modes will approximately time-average away in the rotating wave approximation (RWA) in the absence of any active driving, see the discussion in Appendix \ref{sec:RWA}. Moving to the rotating frame of the GKP oscillators we have
\begin{equation}
\tilde{H}_{\chi^{(2)}}= \omega_c (c^{\dagger} c+\frac{1}{2})+\chi^{(2)} \left(a b e^{-i (\omega_a+\omega_b)t} + a b^{\dagger} e^{-i (\omega_a-\omega_b)t}+h.c.\right)(c+c^{\dagger}).
\end{equation}
Since there are two time-dependencies involved, we can make all $\chi^{(2)}$-interactions resonant by driving the pump mode $c$ with a two-tone drive, namely at $\omega_p=\omega_a+\omega_b$ and $\omega_p=\omega_a-\omega_b$. Both pump tones will need to be of equal amplitude to get equal contributions from beam-splitting ($a^{\dagger} b +ab^{\dagger}$) as well as two-mode squeezing ($a b +a^{\dagger} b^{\dagger}$). Assuming that the pump mode is a (fairly) harmonic mode which can be strongly driven, we replace the operator $c$ by its classical time-dependent expectation value $\langle c(t) \rangle={\cal E} \left(e^{-i (\omega_a+\omega_b)t}+e^{-i (\omega_a-\omega_b)t}\right)$ with, say, ${\cal E} \in \mathbb{R}$. Making a rotating-wave-approximation, Appendix \ref{sec:RWA}, gives the generating interaction of the CZ gate between modes a and b:
\begin{equation}
H_{\rm CZ}=2 \chi^{(2)} {\cal E} \hat{q}_a \hat{q}_b.
\label{eq:CZ-interact}
\end{equation}
Changing the phase of the pump tone (${\cal E}$) allows one to realize CZ$^{-1}$. If we want to do a GKP CNOT gate via two-tone pump, we cannot start with the interaction in Eq.~(\ref{eq:H-3wave}), but a Hadamard or single-qubit $R_Y(\pi/2)$ would be required to convert $\hat{q} \rightarrow \hat{p}$.

Instead of applying two simultaneous pump tones to get a CZ (and with extra rotations, a CNOT), we could also decompose the CNOT circuit as in Fig.~\ref{fig:C-BS}, i.e. a sequence of beam-splitters and single-mode squeezers. When we drive the pump mode at the difference frequency of the modes, $\omega_p=\omega_a-\omega_b$ in the nondegenerate 3-wave mixing Hamiltonian in Eq.~(\ref{eq:H-3wave}), we realize a beam-splitter interaction as the pump photon assists in converting one mode-$a$ photon to a mode-$b$ photon. 

Single-mode squeezing can be activated by using a degenerate version of the 3-wave mixing element $\hat{q}_a \hat{q}_b \hat{q}_c$ in Eq.~(\ref{eq:H-3wave}) with a Hamiltonian proportional to $\hat{q}_a^2 \hat{q}_c$ (or similarly $\hat{q}_b^2 \hat{q}_c$). By applying a pump tone at frequency $\omega_p=2\omega_a$, one activates a squeezing Hamiltonian $H_{\rm sq}$ on mode $a$: we down-convert one pump photon into two mode-a photons and vice-versa.\\

For superconducting devices the only native non-linear circuit element that we have at our disposal are Josephson junctions which, --in their simplest use, without externally applied fluxes--, realize a $U(\Phi)=-E_J \cos(2 \pi \Phi/\Phi_0)$ potential interaction. Here the flux variable $\Phi$ can be expanded as a linear combination of the $q$ quadratures of the bosonic modes which participate in the junction \footnote{A side comment: in circuit-QED we cannot passively get interactions where some cosine potential depends on the $\hat{q}$s of some subset of modes and another cosine potential depends on the $\hat{p}$ of a subset of modes, since all the variables in the circuit Lagrangian which enter the potential energy, --such as the Josephson junction cosine potential energy--, commute when promoted to quantum operators, i.e. they will never be conjugated variables. The upshot of this is that it is very hard (see an attempt at \cite{sameti:toric}) to entirely passively Hamiltonian engineer, say, the toric code checks on a collection of bosonic modes as the essential property of such stabilizer checks is that they either act as $\hat{q}$ and $\hat{p}$ on a single mode.  An exception would be the simultaneous use of Josephson junctions elements ($\cos(2 \pi \Phi/\Phi_0)$) and so-called phase-slip elements ($\cos(\pi Q/e)$) for conjugate variables flux $\Phi$ and charge $Q$. Ref.~\cite{gyr+:GKP} shows that one can passively engineer an effective GKP Hamiltonian using a gyrator element $\Phi Q$.}. Usually, if $E_J$ is large ($\gg E_C$), we expand this cosine potential around its potential minimum $\Phi=0$, obtaining only interactions which are symmetric under $\Phi \rightarrow -\Phi$ such as $\Phi^4$ (while absorbing the $\Phi^2$-terms in the quadratic part of the Hamiltonian).

It is clear from the discussion above that it would be desirable to engineer a $\Phi^3$-interaction where $2 \pi \Phi/\Phi_0=\alpha q_a+\beta q_b+\gamma q_c$. When the three modes have sufficiently different frequencies, we can observe that {\rm all} terms in this $\Phi^3$-interaction, except those proportional to $a^{\dagger} a, b^{\dagger} b$ or $c^{\dagger} c$, average out in time, hence the interaction is `off' in the absence of active driving of one of the modes, not inducing any nonlinearity on the modes in this off-state. At the same time, by choosing the pump mode drive frequencies appropriately, we can activate, with the same interaction element, either a squeezer for mode a, a squeezer for mode b, a beam-splitter between modes a and b or/and a two-mode squeezer between modes a and b.

Besides the Josephson ring modulator, another 3-wave mixing element, called a SNAIL, has been proposed in \cite{frattini+:snail}: it uses a superconducting (SQUID-like) loop containing an asymmetric array of a few Josephson junctions and external flux is applied through the loop. The effective potential induced by this SNAIL is of the form $U(\Phi)=c_2 \Phi^2+c_3 \Phi^3+c_4 \Phi^4$ with $c_3 \gg c_4 \neq 0$ where $\Phi$ is the flux variable expanded around its potential minimum, determined by the external flux $\Phi_{\rm ext}$. 

A recent paper \cite{hillmann2020universal} discusses the circuit-QED engineering required to realize a universal set of gates for continuous-variable computation using GKP states. By flux-modulating the SNAIL, one can activate some of the terms in the 3-wave mixing Hamiltonian in Eq.~(\ref{eq:H-3wave}), mimicking the effect of microwave driving of the pump mode. The authors in \cite{hillmann2020universal} then use this activation to show for example how to realize an interaction $\hat{q}^3$, required to enact the cubic phase gate $V_{\gamma}=e^{i \gamma \hat{q}^3}$.

%Besides a SNAIL, another circuit element which could be useful is a current-biased junction used in the formulation of a phase qubits, bringing in a washboard potential energy of the form $U(\Phi)=-\Phi I -\cos(2 \pi \Phi/\Phi_0)$. (ok around phi neq 0 point with inductance, given a linear term..Need many levels in the well. 
% current-bias element
% Fluxonium has also 3-wave mixing capability: no, make a flux-loop through the GrAl circuit, not clear how to use current fluxonium and its landscape as a coupler one needs to use many levels in the pump mode

Another use of a $\Phi^3$-coupling for GKP state preparation has been proposed in \cite{WT:mod}. In this paper the aim is to produce a tunable opto-mechanical coupling of the form $b^{\dagger} b \,\hat{q}_{\rm GKP}$ between a GKP mode and (harmonic) ancilla mode (b) which is initially prepared in a coherent state. Such coupling can be used to prepare the GKP mode into a logical state starting from a vacuum state, similar as the preparation via regular qubits discussed in Section \ref{sec:conD}. The idea here is that the frequency of the ancilla oscillator is shifted depending on the value for $\hat{q}$ of the GKP mode, leading to a $\hat{q}$-dependent rotation of the coherent state of the ancilla mode. When the interaction time is chosen so that all $q=k \sqrt{\pi}$ for $k \in \mathbb{Z}$ lead to the same rotation of the coherent state, measuring the coherent amplitude realizes an approximate modular measurement of $\hat{q}$, resolving the value of $q \mod \sqrt{\pi}$. Such modular measurement of $\hat{q}$ is equivalent to measuring the eigenvalue phases of $S_q=\exp(2i\sqrt{\pi} \hat{q})$. A possible advantage of this method over the coupling with regular qubits is that one gets more information per ancilla mode measurement than 1 bit. In this proposal an externally-applied flux is modulated around a value for which there is an effective third-order $\Phi^3$-coupling between the two oscillators while the $\Phi^4$-coupling vanishes at this flux setting. Choosing the GKP oscillator at much lower frequency ($\sim$ 0.5Ghz) than the ancilla oscillator ($\sim$ 10Ghz) creates an asymmetry so that a term like $b^{\dagger} b\, \hat{q}_{\rm GKP}$ dominates in the $\Phi^3$-interaction and the term is made resonant via flux modulation.

\subsection{Use of Four-Wave Mixing?}
\label{sec:4wave}

%Alternatively, such coupler could exist between a 3D cavity mode and a qubit mode, e.g. we replace one of the inductances in the LC oscillator in Fig.~\ref{fig:circuit} by a Josephson junction. We will ignore the additional non-linear terms due to this Josesphon junction which inevitably also lead to some additional non-linearities between modes $a,b$ and $c$ due to the hybridization. 

We comment on the use of a $\Phi^4$-interaction for realizing the GKP CZ gate.
The set-up we have in mind is modeled by the electric circuit in Fig.~\ref{fig:circuit}.
Applying circuit-quantization to this circuit leads to a Hamiltonian with three active modes.
Due to the coupling between the LC oscillators, each described as a single mode, some hybridization will happen between the bare cavity modes and the transmon coupler mode, and so we will associate annihilation operators $a$, $b$ and $c$ with these dressed modes. Due to this hybridization the three dressed modes with annihilation operators $a$, $b$ and $c$ will partake in the Josephson junction. This means that for the flux-variable operator $\hat{\Phi}$ across the Josephson-junction branch, we can write $2\pi \hat{\Phi}/\Phi_0=\alpha \hat{q}_a+ \beta \hat{q}_b+\gamma \hat{q}_c$ with dimensionless $\alpha,\beta,\gamma$ modeling  the participation of the effective modes in the Josephson junction \cite{nigg+:bb}.  Expanding the cosine potential up to fourth-order, and diagonalizing the linear interactions of the Hamiltonian (quadratic in creation and annihilation operators) thus gives rise to three dressed eigenmodes at frequencies $\tilde{\omega}_a$, $\tilde{\omega}_b$ and $\tilde{\omega}_c$, and we have the Hamiltonian:
\begin{equation}
H=\tilde{\omega}_a \left(a^{\dagger} a+\frac{1}{2}\right)+ \tilde{\omega}_b \left(b^{\dagger} b+\frac{1}{2}\right)+ 
\tilde{\omega}_c\left (c^{\dagger} c+\frac{1}{2}\right)-\frac{E_J}{4!} \left(\alpha \hat{q}_a+\beta \hat{q}_b+\gamma \hat{q}_c \right)^4.
\label{eq:phi4}
\end{equation}

% $2 \pi \hat{\Phi}{\Phi_0}$ with $\Phi_0=h/2e$.
%There is a constraint on the alpha, beta and gamma...not arbitrary! Maybe this depends on the characteristic impedance $Z=\sqrt{L/C}$.
% dimensionless constant..
%For an LC oscillator at frequency $\omega=1/\sqrt{LC}$, we can keep its frequency the same but change its characteristic impedance $Z=\sqrt{L/C}$. Flux variable $\hat{\Phi}=\Phi_{\rm ZPF} (b+b^{\dagger})$ and $2\pi \Phi_{\rm ZPF}/\Phi_0$ is dimensionless, so one can consider whether it is large or small. 
% numbers for EJ?

As in the discussion on 3-wave mixing we assume that all frequencies $\tilde{\omega}_a, \tilde{\omega}_b, \tilde{\omega}_c$ are sufficiently different (detuned).
%, $\Delta_{ij}=|\omega_i-\omega_j|$ is large compared to the coupling strength $E_J$. 
If there is no active driving (or flux-modulation), a full RWA approximation, whose accuracy depends on the amount of detuning, will leave only energy-conserving self-Kerr and cross-Kerr terms. In other words, in the off-state, the Hamiltonian is approximately
\begin{align}
H_{\rm off} \approx \omega_a \left(a^{\dagger} a+\frac{1}{2}\right)+ \omega_b \left(b^{\dagger} b+\frac{1}{2}\right)+ \omega_c\left (c^{\dagger} c+\frac{1}{2}\right) +\notag \\ 
-\frac{1}{2}(\chi_{aa} (a^{\dagger} a)^2+\chi_{bb} (b^{\dagger} b)^2+\chi_{cc} (c^{\dagger} c)^2)-\chi_{ab} a^{\dagger}a b^{\dagger} b-
\chi_{ac} a^{\dagger} a c^{\dagger} c-\chi_{bc} b^{\dagger} b c^{\dagger} c,
\label{eq:off}
\end{align}
where $\chi_{ii'}=2\sqrt{\chi_{ii}\chi_{i'i'}}$  \cite{nigg+:bb}. Here $\omega_a=\tilde{\omega}_a-\frac{E_J \alpha^4}{24}$ (and similarly for $\omega_b$ and $\omega_c$) due to rewriting the excitation-conserving terms in $\hat{q}_a^4$ as a quadratic term $\propto a^{\dagger} a$ and the self-Kerr term $\propto (a^{\dagger} a)^2$ and $\chi_{aa}=\frac{E_J\alpha^4}{12}$. Here we clearly see the advantage of a pure three-wave mixing element over a four-wave mixing element: in the off-state the four-wave mixing element induces unwanted Kerr and cross-Kerr anharmoniticies on the GKP storage modes $a$ and $b$. 

In the off-state, mode $c$ is (ideally) in its vacuum state, hence the cross-Kerr interaction with this mode does not contribute. However, if this mode were driven these corrections are relevant and they induce additional cavity rotations. Let us now indeed discuss the effect of applying a drive on mode $c$. For this, we expand the fourth-order term in Eq.~(\ref{eq:phi4}) which becomes
\begin{equation}
H_{\chi^{(3)}}=-\frac{E_J}{4!} \left[{4 \choose 2} \times 2 \alpha \beta \gamma^2 \hat{q}_a \hat{q}_b \hat{q}_c^2+{4 \choose 2} \alpha^2 \gamma^2 \hat{q}_a^2 \hat{q}_c^2+\ldots \right].
\label{eq:H-4wave}
\end{equation}
We can apply a two-tone drive on mode $c$ at frequency $\frac{\omega_a+\omega_b}{2}$ and $\frac{\omega_a-\omega_b}{2}$ with equal amplitudes ${\cal E}$. Replacing $\hat{q}_c$ by its time-dependent expectation $\langle \hat{q}_c(t) \rangle=\frac{{\cal E}}{\sqrt{2}}(e^{i t\frac{\omega_a+\omega_b}{2}}+e^{i t\frac{\omega_a-\omega_b}{2}}+h.c)$ in Eq.~(\ref{eq:H-4wave}) and going to the rotating frame of all modes $a$ and $b$, we find that a term like $ \hat{q}_a \hat{q}_b \hat{q}_c^2$ leads to a time-independent resonant term proportional to $\hat{q}_a \hat{q}_b$. This can be seen as follows. First, note that the signal $\langle q_c(t) \rangle^2$ only contains frequencies $\omega_a, \omega_b, \omega_a+\omega_b$ and $\omega_a-\omega_b$, all of equal strength. 
The frequency $\omega_a+\omega_b$ matches two-mode squeezing ($a b+ h.c$), while $\omega_a-\omega_b$ matches beamsplitting ($a b^{\dagger}+h.c$). Besides this, we throw out all time-dependent terms (RWA). In particular we have
\begin{itemize}
\item Terms without any $\hat{q}_c$ are only leading to self-Kerr and cross-Kerr for modes $a$ and $b$.
\item Terms with a single $\hat{q}_c$ or $\hat{q}_c^3$ are not frequency-matched to become time-independent (as $\omega_a$ and $\omega_b$ are sufficiently different).
\item Terms $\hat{q}_c^4$ leads to self-Kerr for mode c.
\item Terms with $\hat{q}_c^2$ lead to cross-Kerr between modes c and a or c and b. Note that from a term such as $\hat{q}_a^2 \hat{q}_c^2$ there is a contribution proportional to ${\cal E}^2 a^{\dagger} a$ (and similarly $\hat{q}_b^2 \hat{q}_c^2$ gives ${\cal E}^2 b^{\dagger} b$).
\end{itemize} 

We could also realize the CNOT gate using the beam-splitter and squeezer sequence in Fig.~\ref{fig:C-BS}, i.e. we chose a single-tone pump at $\omega_p=(\omega_a-\omega_b)/2$ for the beam-splitter to let two pump photons assist in converting one mode-a photon to one mode-b photon. Single-mode squeezing can be realized by using the interaction $\hat{q}_a^2 \hat{q}_c^2$ in Eq.~(\ref{eq:H-4wave}), i.e. we should take $\omega_p=\omega_a$ so that two pump photons are converted into two mode-a photons. However, note that this also make the unwanted interaction $b^{\dagger} b (a^{\dagger} c+a c^{\dagger})$ (which comes from the $\hat{q}_b^2 \hat{q}_a \hat{q}_c$ term) resonant, which makes this scheme unattractive.

% BMT drive at 2 \omega_a-\omega_b to get a^{\dagger}^2 b c resonant etc. why would you..

An important parameter measuring the quality of the CZ gate via this $\Phi^4$- interaction is the relative strength of the unwanted Kerr and cross-Kerr terms versus the strength of the two-tone pump-activated wanted interactions $\hat{q}_a \hat{q}_b$. In part, this relies on the error contributions due to the rotating wave approximation which should be better quantified theoretically (Appendix \ref{sec:RWA}). Another contributing factor is the relative strength of the participation parameters $\alpha, \beta, \gamma$ and the pump strength, namely, without error contribution from the RWA, one has
\begin{equation}
\frac{||H_{\rm CZ}(a,b)||}{||H_{\rm cross-Kerr}(a,b)||} \propto \frac{{\cal E}^2 \gamma^2}{\alpha \beta};  
\end{equation}
Hence, a large $\gamma^2/\alpha \beta$ is desirable but a large $\gamma$ also makes mode c more anharmonic as $\chi_{cc} \propto \gamma^4$ and this again severley restricts the pump power ${\cal E}$. These conflicting constraints may make this scheme less suitable in practice.

\section{Prospects for a GKP-Surface Code Architecture}
\label{sec:scale}

In this section we would like to provide a perspective of what it would take to build a surface code architecture based on GKP qubits, point out the challenges in this approach, as well as contrast it with existing efforts to engineer a similar architecture using transmon qubits \cite{versluis+:scalable}, see Section \ref{sec:transmons}.

We can partially use the results in Ref.~\cite{vuillot+:GKP} as a starting point for such GKP-surface code architecture. In this code architecture, there are two layers of protection. On the one hand, each GKP qubit is either stabilized or error corrected individually, reducing a continuous set of (displacement) errors to a mostly discrete set of GKP qubit Pauli errors. On the other hand, the surface code layer is there to suppress the logical error rate of a GKP qubit to values which decrease exponentially with the side length of the surface code lattice.

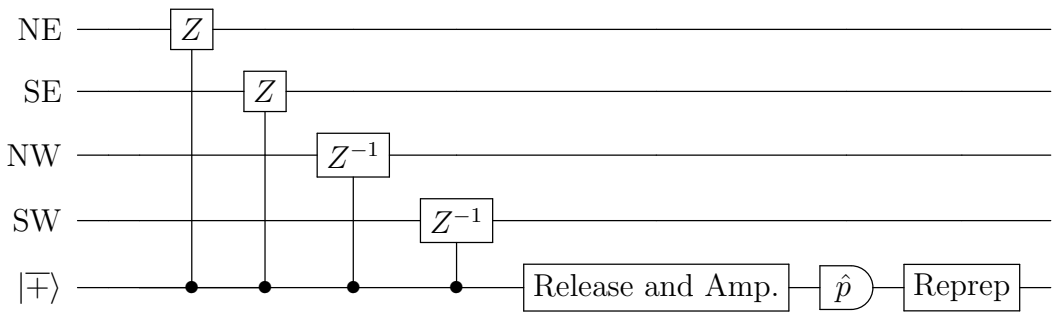
\begin{figure}[htbp]
  \centering
  \scalebox{1}{\Qcircuit @C=1em @R=.7em {
      \lstick{\rm NE} & \qw &  \qw & \gate{Z} & \qw & \qw & \qw & \qw &  \qw & \qw & \qw \\
      \lstick{\rm SE}& \qw & \qw & \qw & \gate{Z} & \qw & \qw & \qw & \qw & \qw & \qw \\
      \lstick{\rm NW}  & \qw & \qw & \qw & \qw & \gate{Z^{-1}} & \qw & \qw & \qw & \qw & \qw \\
      \lstick{\rm SW}  & \qw & \qw & \qw & \qw & \qw & \gate{Z^{-1}} & \qw &  \qw & \qw & \qw \\
      \lstick{\ket{\overline{+}}} & \qw & \qw & \ctrl{-4} \qw & \ctrl{-3} \qw  & \ctrl{-2} & \ctrl{-1} & \gate{\mbox{Release and Amp.}} &  \measureD{\hat{p}} & \gate{\rm Reprep} & \qw
	}}
\caption{A $Z$-check parity measurement for the surface-GKP code in Fig.~\ref{fig:surface-code}, on oscillators labelled NE-SW (Northeast to Southwest of ancilla qubit) using CZ and CZ$^{-1}$ gates as defined in Section \ref{sec:log-gates}. $\ket{\overline{+}}$ is an approximate $+1$ eigenstate of $\overline{X}$ and the GKP stabilizers. Release and amplification (Amp) followed by measurement of the quadrature $\hat{p}$ is a way to do a logical $\overline{X}$ measurement and Reprep is a unit standing for the repreparation of the GKP ancilla state.}
	\label{fig:setup-concat1}
\end{figure}

\begin{figure}[htbp]
	\centering
	\includegraphics[scale=0.5]{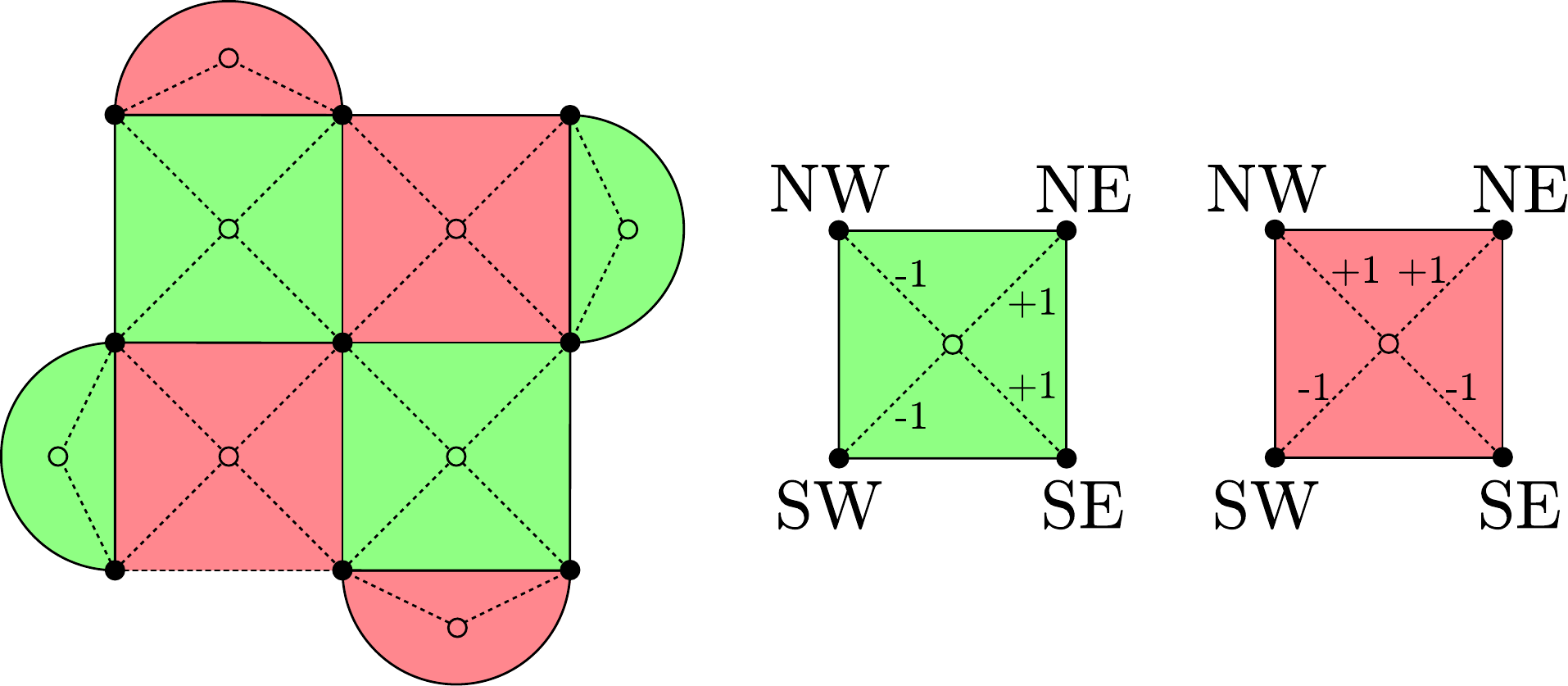}
	\caption{Distance-3 rotated surface code in its standard Surface-17 layout with green $Z$-checks and red $X$-checks: black filled circles are data qubits, open circles are ancilla qubits, dashed lines are two-qubit interactions. The $\pm1 $ patterns on each check denotes the use of inverses as in Eq.~(\ref{eq:check1}), so that all checks commute as displacements.}
	\label{fig:surface-code}
\end{figure}

In Ref.~\cite{vuillot+:GKP} GKP error correction takes place with ancilla GKP modes using the circuits in Fig.~\ref{fig:EC_GKP}. Note that these circuits can also be implemented via CZ gates, but will then require Hadamard or $R_Y(\pi/2)$ rotations on the GKP data mode.
Interspersed with this GKP error correction, parity checks of the surface code, shown in Fig.~\ref{fig:surface-code}, are to be measured in QEC cycles. These circuits are similar as for a regular surface code, except that the underlying qubits are GKP qubits encoded in oscillators, see Fig.~\ref{fig:setup-concat1} for a $Z$-check. We use the fact that GKP logical operators are not self-inverse as displacements, --and as displacements they obey $[\overline{X}_1 \overline{X}_2, \overline{Z}_1^{-1} \overline{Z}_2]=0$--, to measure checks which mutually commute on the entire oscillator space \footnote{One expects worse error behavior when one measures non-commuting checks outside the ideal code space; for example, this also happens when regular qubits leak \cite{varbanov+:leak}. In addition, the maximum likelihood decoding analysis in \cite{vuillot+:GKP} relies on this choice to map onto compact-QED model with proper lattice versions of rotations and divergences. }. In this Figure the measurement of the GKP ancilla is shown as the release and amplification of the cavity state, followed by a quadrature measurement, as discussed in Section \ref{sec:FTmeas}. Such measurement would give useful analog information, but the usefulness of this analog information is challenged by losing photons in the step, nor has it been experimentally realized.

We call this GKP-surface code architecture {\sf All-GKP-Ancilla}. This set-up would require each GKP qubit oscillator to have CZ capability with 5 other GKP oscillators, namely 4 GKP ancilla oscillators for the surface code and 1 GKP ancilla oscillator for its own error correction. 

Ref.~\cite{vuillot+:GKP} used a model of Gaussian stochastic displacement noise, Eq.~(\ref{eq:GDC}), as an effective, numerically-simulatable, error model for this architecture. The noise channel acts in the different locations: (1) on each GKP qubit prior to GKP error correction and a round of surface-code parity check measurements, (2) prior to the homodyne measurement in Fig.~\ref{fig:EC_GKP} and (3) prior to the homodyne measurement of the ancilla GKP in the surface code check. Taking the standard deviations from these Gaussian channels to be equal, a threshold standard deviation ${\sigma_0}_c \approx 0.243$ for the toric code was found. Note that this model includes all sources of errors, including finite squeezing and feedback errors, albeit stochastically. Using the conversion $\Delta^2=2 \sigma_0^2$, this gives a threshold of $\Delta=0.34$ or 9.3 dB, but the data in Fig.~\ref{fig:stoch-coherent} show this conversion is somewhat too optimistic: using squeezed states with $\Delta$ gives an error rate which is somewhat worse than a stochastic model with $\sigma_0=\Delta/\sqrt{2}$, so a worst-case threshold estimate using $\Delta^2=\sigma_0^2$ would be 12.3 dB. Ref.~\cite{NC:surface-GKP} considered a variation on this stochastic noise model, --explictly including error feedback--,  and applied this noise to a concatenation of the GKP qubit with the surface code. Both \cite{vuillot+:GKP} and \cite{NC:surface-GKP} used minimum weight matching decoders to find thresholds. Ref.~\cite{vuillot+:GKP} identified the defects and the distance function between them following the associated compact-QED model closely in order 
to approach exact minimum-weight decoding. Different from this, Ref.~\cite{NC:surface-GKP} identified the positions where the surface code check outcomes change as defects, but altered the distance function between these defects based on GKP error information.

% BMT TD does it include feedback errors really?

%   Indeed, individual GKP error correction information needs to be used and integrated in surface code decoding. Furthermore, if individual GKP error information is not sufficiently reliable, --making quantum information in each oscillator not sufficiently binary--, then no memory threshold for surface coding exists (even when its parity checks are executed perfectly and the incoming error rate goes to zero). 
%The perspective that there is a deep advantage due to the analog nature of the error syndrome information \cite{fukui+:analog} does not seem fully justified as it is a consequence of the fact that the set of possible shift errors on an oscillator is analog/continuous (and not digital as for a qubit or finite-dimensional qudit) which can rather, --when the analog information is not measured with sufficient accuracy--, throw rain on the whole quantum error correction parade as we cannot correct for such continuous error model in a scalable way, at least not in a 2D code. 
% BMT 4D code phase boundary is open..

We add two more observations about this scheme. First, when we use stabilizer error correction, such as surface code error correction, on bosonic codes, we need to implement parity check operators which sometimes act like a logical $X$ on a bosonic qubit, and sometimes like a logical $Z$. For GKP qubits this translates into the ability to perform CZ gates as well as CNOT gates. For standard (transmon) qubits, the switch between CZ and CNOT is easily achieved by applying a layer of Hadamard gates between a parity $X$-cycle and a parity $Z$-cycle. For a GKP qubit encoded into an oscillator with frequency $f$, such Hadamard gate seems simple: it constitutes waiting for time $t=1/(4 f)$. But since all data qubits have to undergo this Hadamard gate, it implies that the resonant frequencies of the data qubit oscillators (resonant modes of identical 3D cavities) should all be identical, which seems like a narrow target to aim at (although the difference between a simulation-based predicted 3D cavity frequency and the measured frequency can be less than 0.1\% \cite{reagor+:3D}).

As alternative to the Hadamard gate one can use $R_Y(\pi/2)$ gate and $R_Y(-\pi/2)$ gate, using a regular qubit as in Fig.~\ref{fig:singleU}, to toggle back and forth between $Z$ and $X$ error corrrection, but it costs a lot more hassle and time than doing a $R_Y(\pi/2)$ on a transmon qubit in O(10) ns.
A second observation is that the use of parametrically-driven 3-wave or 4-wave mixing as discussed in Sections \ref{sec:GKP-CZ} and \ref{sec:4wave} could allow for the simultaneous execution of the CZ and ${\rm CZ}^{-1}$ gates needed to do a surface-code parity check measurement, as the activation of the CZ or (CZ$^{-1}$) gate only requires the application of a pump tone to the coupler between each data oscillator and ancilla oscillator (4 couplers in total). The coupling strength of these CZ couplers may not be equally strong, hence the duration of these four pump drives can vary, but an advantage of only driving the coupler mode (instead of the GKP mode) is to enable the simultaneous execution of these commuting gates. 
%In the $\Phi^4$-interaction scheme of Section \ref{sec:4wave} the pump driving does change the resonant frequency of the GKP mode, via the cross-Kerr of the pump mode and the GKP mode, so this needs to be included in choosing pump frequencies for executing simultaneous CZ gates).
Another way of looking at the simultaneously-executed parity check is to observe that a green $Z$-check in Fig.~\ref{fig:surface-code} on oscillator NE, SE, NW, SW corresponds to 
\begin{equation}
\hat{G}=\hat{q}_{\rm NE}+\hat{q}_{\rm SE}-\hat{q}_{\rm NW}-\hat{q}_{\rm SW}.
\label{eq:check1}
\end{equation}
An interaction Hamiltonian $H_G=-\hat{q}_{A} \hat{G}$ applied for time $t$ has the effect that $\hat{p}_A \rightarrow \hat{p}_A- \hat{G} t$, using Eq.~(\ref{eq:id}) (while for all data oscillators $i={\rm NE}, {\rm SE}, {\rm NW}, {\rm SW}$ participating in the check, we have $\hat{p}_i \rightarrow \hat{p}_i \pm t \hat{q}_A$). Taking $t=1$, we see that by measuring the ancilla quadrature $\hat{p}_A$, we measure $\hat{G}$ modulo even multiples of $\sqrt{\pi}$ (as $\ket{\overline{+}}$ has sharp peaks at $p_A$ being even multiples of $\sqrt{\pi}$). Thus, if one of the oscillators undergoes a $\sqrt{\pi}$ shift in $q$, the measurement will detect this. 

Besides the Steane error correction in Fig.~\ref{fig:EC_GKP}, one can also imagine a more hardware-efficient form of GKP error correction via stabilization using a regular qubit as discussed in Section \ref{sec:conD}, regularly interspersed with parity check measurements for the surface code which, for example, do use a GKP ancilla. The advantage here is that one does not need to prepare and couple the ancillary GKP qubit as in Fig.~\ref{fig:EC_GKP} (which again requires a regular qubit). In particular, (ancilla) GKP state preparation is time-consuming ($60\, \mu$s in \cite{GKP:exp}) due to requiring slow controlled-displacement gates and slow qubit measurement, and during this process photon loss is affecting the GKP qubit. At the same time, we keep the GKP ancilla for the possibly-less frequent surface code QEC cycle in order to get still analog error information. We refer to this intermediate scheme as {\sf Only-SurfaceCode-GKP-Ancilla}.

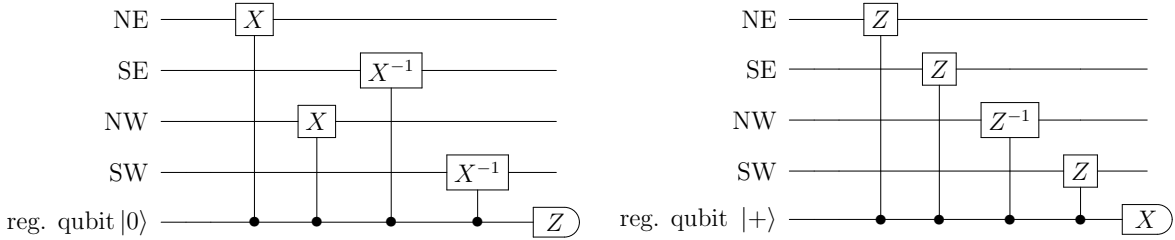
\begin{figure}[htbp]
\centering
\hspace{2cm}
\scalebox{0.8}{
  \Qcircuit @C=1em @R=0.7em{
      \lstick{\rm NE} & \qw &  \qw & \gate{X} & \qw & \qw & \qw & \qw \\
      \lstick{\rm SE}& \qw & \qw & \qw & \qw & \gate{X^{-1}} & \qw & \qw \\
      \lstick{\rm NW}  & \qw & \qw & \qw & \gate{X} & \qw & \qw & \qw \\
      \lstick{\rm SW}  & \qw & \qw & \qw & \qw & \qw & \gate{X^{-1}} & \qw \\
      \lstick{\mbox{reg. qubit} \ket{0}} & \qw & \qw &  \ctrl{-4} &  \ctrl{-2}  & \ctrl{-3} & \ctrl{-1} & \measureD{Z}}}
\hfill
\scalebox{0.8}{
\Qcircuit @C=1em @R=0.7em {
      \lstick{\rm NE} & \qw &  \qw & \gate{Z} & \qw & \qw & \qw & \qw \\
      \lstick{\rm SE}& \qw & \qw & \qw & \gate{Z} & \qw & \qw & \qw \\
      \lstick{\rm NW}  & \qw & \qw & \qw & \qw & \gate{Z^{-1}} & \qw & \qw \\
      \lstick{\rm SW}  & \qw & \qw & \qw & \qw & \qw & \gate{Z} & \qw \\
      \lstick{\mbox{reg. qubit } \ket{+}} & \qw & \qw &  \ctrl{-4} &  \ctrl{-3}  & \ctrl{-2} & \ctrl{-1} & \measureD{X}}}
\caption{A single round of error correction for an $X$-check (left) and a $Z$-check (right) on the GKP data oscillators using regular qubit ancillas. The interactions between ancilla qubit and GKP oscillators are all tunable qubit controlled-displacement interactions.}
	\label{fig:pchecks-reg}
\end{figure}

Another choice is to use regular qubits to extract both GKP and surface code error information, see the circuits in Fig.~\ref{fig:pchecks-reg}. We refer to this scheme as {\sf All-Regular-Qubit-Ancilla}. An advantage of this scheme is that no Hadamards or $R_Y(\pi/2)$ rotations are needed on GKP modes and tunable controlled-displacement gates are used {\em throughout}. 

The {\sf All-Regular-Qubit-Ancilla} architecture can however be less tolerant towards errors: it might be hard to get below threshold for the surface code, when all error information is obtained through qubits, giving 1 bit of information at the time. We can provide arguments for this by using a simple error model in which we assume that GKP error correction generates an effective phenomenological error model in each surface code QEC cycle and we assume that the surface code QEC cycle is otherwise perfect. We model the effect of GKP error correction as stabilizing an approximate GKP qubit of the form, say, $\frak{F} \ket{\overline{b}}$ at some $\Delta$, besides having a logical error $b \rightarrow \neg b$ on top with probability $p$. Effectively then, the approximate GKP code states coming into a perfect surface-code parity check circuit as in Fig.~\ref{fig:pchecks-reg} will flip the regular qubit ancilla with some effective error probability $q$ which depends on $\Delta$. We thus map our error model onto a known (phenonomenological) surface code error model in which there is an incoming error with probability $p$ in each QEC round and a measurement error with probability $q(\Delta)$. For this model, Fig.~3 in Ref.~\cite{andrist+:pq-boundary} shows the numerically-found below-threshold region and for $q=p$ the threshold is optimally $3.3\%$ \cite{ohno+:toric}. Ref.~\cite{andrist+:pq-boundary} does not investigate the below-threshold region and its shape for low $p$, but it certainly lies within the below-threshold region for the repetition code which is conjectured to have a below-threshold region given by $H_2(p)+H_2(q) \leq 1$ with $H_2(p)$ the binary entropy \cite{TSN:threshold}.

To estimate $q(\Delta)$ we can write the probability to measure the ancilla qubit in the $X$-basis as $\mathbb{P}(\pm)=\frac{1}{2}\frac{(\bra{\overline{0}}\frak{F}^{\dagger}\frak{F}\ket{\overline{0}})^4 \pm \frac{1}{2}[(\bra{\overline{0}}\frak{F}^{\dagger}\overline{Z}\frak{F}\ket{\overline{0}})^4 +(\bra{\overline{0}}\frak{F}^{\dagger}\overline{Z}^{\dagger}\frak{F}\ket{\overline{0}})^4]}{(\bra{\overline{0}}\frak{F}^{\dagger} \frak{F}\ket{\overline{0}})^4} $, which, using Eq.~(\ref{eq:Zexp}), approximately gives
$\mathbb{P}(\pm) \approx \frac{1}{2}(1\pm e^{-\pi \Delta^2})$ and thus $q(\Delta)=\frac{1}{2}(1-e^{-\pi \Delta^2})$. At $\Delta=0.3$, we already have 
$q(\Delta)=12\%$ while $\Delta=0.15$ just suffices to get $q(\Delta)=3.4\%$.

The frequency of doing the surface code error correction could also be adapted to the logical decay rate of the stabilized GKP state, e.g. $275\,\mu$s in \cite{GKP:exp}, so that the logical qubit error rate between surface code QEC cycles is at least less than $3.3\%$. It is an open question how to analyze the noise threshold for the {\sf All-Regular-Qubit-Ancilla} architecture for a more elaborate error model. 

%It is also an open question how well we can numerically simulate inaccurate surface code dynamics beyond stochastic shift noise models.

In this {\sf All-Regular-Qubit-Ancilla} architecture the workhorse is the controlled-displacement gate with the regular qubit and the regular qubit preparation and measurement. The desiderata for these regular qubits are clearly (1) ability to enact a fast and accurate tunable controlled-displacement with a 3D cavity mode, (2) low leakage to higher excited states, (3) long $T_1$ and $T_2$, beyond $100\,\mu$s, and (4) fast measurement below $O(100)$ ns, (5) fast preparation of $\ket{+}$ and single-qubit gates ($O(10)$ ns). At first sight, this seems like a wishlist for any good qubit, however it is not necessary to have a high-quality two-qubit gate between these qubits, which is a nontrivial component for the surface code with transmon qubits. Furthermore, the frequency of the 3D cavity GKP modes can be taken to be far different than those of the ancillary qubits and their coupled read-out resonators, possibly leading to easier frequency control and less frequency crowding than in an architecture with only one type of device qubit such as the surface code with transmon qubits \cite{brink+:device, versluis+:scalable}.

% frequency control
%In a scalable surface code architecture, see Section \ref{sec:scale}, these CZ gates will be employed between an ancilla mode and a storage mode, requiring storage and ancilla modes to be at different frequencies. For example, one can choose the frequencies of the storage modes at 3GHz, while the ancilla modes are all at 8Ghz. In the 4-wave mixing picture this requires drive tones at frequencies 3GHz and 6.5GHz.
% table of strength in terms of $k$Hz Kerr, $\chi$ ranging from X..Y.

\subsection{Comparison: Fock Qubit Surface Code and Transmon Qubit Surface Code}
\label{sec:transmons}

Given that we imagine using high-Q 3D cavities for qubit storage, we can ask how to compare a GKP encoding with a simple Fock encoding in a surface code architecture, omitting any additional error correction. CZ gates between a 3D-cavity encoded Fock qubit (mode a) and an ancilla transmon qubit can be realized by a dispersive coupling $-\chi a^{\dagger} a Z/2$, allowing for the execution of the $Z$-check measurement. Similar as for the controlled-displacement in the GKP encoding, tunability of this interaction, for example, by using an intermediate frequency-tunable resonator to vary the coupling strength, is important. This type of parity check, using 1 transmon qubit to read out 4 Fock qubits, is the reverse of using one bosonic mode to read out the parity of four coupled transmon qubits as realized in Ref.~\cite{blumoff+:parity}. For the $X$-check measurement one requires a CNOT gate with transmon qubit as target, which can be realized by performing a CZ followed and preceded by Hadamard or $R_Y(\pi/2)$. Again, similar as in the GKP encoding, these simple single-qubit gates require the use of an ancillary qubit, but arbitrary cavity manipulations through such coupled transmon qubit have been demonstrated in \cite{heeres+:cv-mani}, albeit of rather long, O(1) $\mu$s, duration, and having some, inevitable, leakage towards the state $\ket{2}$ or higher.

An engineering effort for making a surface code architecture using 2D transmon qubits is underway at e.g. Google, IBM, TU Delft and ETH Z\"urich. Besides using an optimized decoder \cite{OTD:dens}, the crucial numbers which determine whether such architecture will be `below threshold' are the quality, leakage \cite{varbanov+:leak}, time-duration and cross-talk of the two-qubit gate and the duration (and cross-talk) of the qubit measurement versus the dephasing and relaxation time of the qubits.  Flux-tunable transmon qubits have recently achieved very good numbers for their two-qubit gates: Ref.~\cite{rol+:netzero} reports on a $99.1\%$ CZ fidelity of 40 ns duration and low leakage $0.1\%$, while Google's supremacy experiments \cite{arute+:supr} have shown the performance of ISWAP-like two-qubit gates on a 54-qubit Sycamore chip with an average error rate of $0.62\%$ and duration $O(10)$ ns. It is an open question how much further transmon performance numbers, including measurement duration, can be pushed beyond their current values. The use of different superconducting materials \cite{houck+:tantalum} can provide new opportunities to lengthen $T_2$ and $T_1$ times. Note however that an enhanced $T_1$ also leads to an enhanced duration of leakage. Frequency crowding and limits on highly-accurate frequency targeting, in particular for non-flux tunable transmons, leading to spurious cross-talk couplings is another challenge in realizing the surface code. 
%Roughly a factor of 10 in life-time improvement and no leakage might be needed to get us below threshold.

We thus believe that there is plenty of room and, in fact, necessity for developing an alternative surface code architecture in which a data qubit, such as a Fock or GKP qubit, is encoded in a very harmonic mode of high-Q (3D) cavity, while transmon qubits or their next-generation versions such as fluxonium or noise-biased cat qubits, are used as ancilla qubits. If a pump-activated CZ or controlled-displacement gate has high on/off ratio, one expects that spurious couplings between the 3D cavity data modes, due to common coupling to the ancilla qubits, would be well suppressed.

\section{Acknowledgements}

We thank Alessandro Ciani, David DiVincenzo, Ioan Pop and Daniel Weigand for useful feedback and discussions and some help with the Figures and the numerics.
We acknowledge support from the European Research Council (EQEC, ERC Consolidator Grant No. 682726). CV and BMT acknowledge support from a QuantERA grant for the QCDA consortium. 
This research was supported in part by Perimeter Institute for Theoretical Physics. Research at Perimeter Institute is supported by the Government of Canada through Industry Canada and by the Province of Ontario through the Ministry of Economic Development $\&$ Innovation.

\appendix

\section{Fock State Representation of GKP Grid States}
\label{app:GKP-math}

%\begin{equation}
%a\geq0,\, b \in \mathbb{R}: \;\int_{\mathbb{R}} dx \exp(-a x^2+i b x)= \sqrt{\frac{\pi}{a}}\exp\left (-\frac{b^2}{4a}\right )
%\end{equation}
%\begin{equation}
%\sum_{n \in \mathbb{Z}} \exp(-\pi n^2/a) \exp(
%\end{equation}
In this Appendix we examine the Fock coefficients of approximate GKP states, using the $\frak{D}$-approximation, and the sensor grid state \cite{DTW:sensor}. We study the asymptotic behavior of these Fock coefficients, showing that the photon number distribution trends along a geometric or thermal distribution. It turns out that these Fock coefficients relate to some interesting nontrivial mathematics.

The theta function with rational parameters $(a,b)\in\mathbb{Q}^2$, adopting the notations of \cite{mumford_tata_2007,MYK:approxGKP}, is given by
\begin{equation}
\vartheta\left [\begin{matrix}
a\\b
\end{matrix}\right ](z,\tau)=\sum_{k \in \mathbb{Z}} \exp\left (\pi i \tau \left (k+a\right )^2 + 2 \pi i \left (k+a\right )\left (z+b\right )\right ),
\end{equation}
where $(z,\tau)\in\mathbb{C}$ and ${\rm Im}(\tau)>0$ ensuring absolute convergence of the series.
%BMT use k instead of n in sum to avoid clash with photon number n in N
Some common shorthands are the following
\begin{align}
\vartheta\left [\begin{matrix}
0\\0
\end{matrix}\right ]\left(0, ix\right) &= \theta_3\left(x\right)=\sum_{k \in \mathbb{Z}} \exp(-\pi x k^2),\\
\vartheta\left [\begin{matrix}
\frac{1}{2}\\0
\end{matrix}\right ]\left(0, ix\right) &= \theta_2\left(x\right)=\sum_{k \in \mathbb{Z}} \exp(-\pi x (k+1/2)^2),\\
\vartheta\left [\begin{matrix}
0\\\frac{1}{2}
\end{matrix}\right ]\left(0, ix\right) &= \theta_4\left(x\right)=\sum_{k \in \mathbb{Z}}\exp(-\pi x k^2+\pi i k).
\label{eq:bound2}
\end{align}
The multidimensional generalization with rational vectors $(\vec{a}, \vec{b})\in\left (\mathbb{Q}^m\right )^2$ is given by
\begin{equation}
\bigvartheta\left [\begin{matrix}
\vec{a}\\\vec{b}
\end{matrix}\right ]\left (\vec{z},\Omega\right ) = \sum_{\vec{k}\in\mathbb{Z}^m}\exp\left [\pi i \left (\vec{k}+\vec{a}\right )^{\rm T}\Omega\left (\vec{k}+\vec{a}\right ) + 2\pi i \left (\vec{k}+\vec{a}\right )^{\rm T}\cdot\left (\vec{z}+\vec{b}\right )\right ],
\end{equation}
where $\vec{z}\in\mathbb{C}^m$ is a complex vector, $\Omega\in\mathbb{C}^{m\times m}$ is a complex matrix and ${\rm Im}\left (\Omega\right )$ is positive definite which ensures the absolute convergence of the series.

First, it is important to note that one can properly normalize the approximate GKP state $\ket{j_{\rm approx}}$, using the $\frak{D}$-approximation defined in Eq.~\eqref{eq:approx}, for the two logical states $j=0,1$:
\begin{align}
\ket{j_{\rm approx}}&=\frac{1}{\sqrt{N(\Delta, j)}}\;\frak{D} \ket{\overline{j}_{\rm ideal}}\\
&= \frac{1}{\sqrt{N(\Delta, j)}}\; \exp\left (-\Delta^2\hat{n}\right )\sum_{k\in\mathbb{Z}}\ket{q=(2k+j)\sqrt{\pi}}
\end{align}
with 
\begin{align}
N(\Delta,j) &= \frac{1}{\sqrt{\pi(1-u^2)}}\,\bigvartheta\left [\begin{matrix}
\frac{j}{2}\vec{1}\\\vec{0}
\end{matrix}\right ]\left (\vec{0}, \Omega\right ),\label{eq:normDj}
\end{align}
where
\begin{align}
u &= \exp(-2\Delta^2),\label{eq:u}\\
\Omega &= \frac{2i}{1-u^2}\begin{pmatrix}
1+u^2 & -2u\\ -2u & 1+u^2
\end{pmatrix},\label{eq:Omega}
\end{align}
and $\vec{0}$ and $\vec{1}$ are the all-zeros and all-ones vectors respectively.
Note that there is a bijection between $\Delta\in(0,\infty)$ and $u\in[0,1)$ so we write either $N(\Delta,j)$ or $N(u,j)$ depending on convenience.
 
In order to obtain this expression one can use the position representation of Fock states in terms of Hermite functions $\Psi_n(q)$ (or Hermite polynomials $H_n(q)$),
\begin{equation}
\braket{n\vert q} = \Psi_n(q) = \frac{1}{\sqrt{2^n\sqrt{\pi}n!}}\exp\left (-\frac{q^2}{2}\right )H_n(q),
\end{equation}
and the so-called Mehler's Hermite polynomial formula:
\begin{equation}
\sum_{n\in\mathbb{N}}u^n\Psi_n(x)\Psi_n(y) = \frac{1}{\sqrt{\pi(1-u^2)}}\exp\left (\frac{(1+u^2)(x^2+y^2) - 4uxy}{2(1-u^2)}\right ).
\end{equation}
It is possible to rewrite Eq.~\eqref{eq:normDj} in the form given in Ref.~\cite{MYK:approxGKP}
\begin{align}
N(u, j) = \frac{1}{2\sqrt{\pi}(1+u)}\Bigg (&\vartheta\left [\begin{matrix}
\frac{j}{2}\\0
\end{matrix}\right ]\left (0,i8\sigma^2\right )\,\vartheta\left [\begin{matrix}
0\\0
\end{matrix}\right ]\left (0,i\sigma^2/2\right ) \nonumber\\
&+ \vartheta\left [\begin{matrix}
\frac{j+1}{2}\\0
\end{matrix}\right ]\left (0,i8\sigma^2\right )\,\vartheta\left [\begin{matrix}
0\\\frac{1}{2}\end{matrix}\right ]\left (0,i\sigma^2/2\right )\Bigg ).\label{eq:normapproxgkp}
\end{align}
%\begin{equation}
%N(\Delta, j) = \frac{1}{2\sqrt{\pi}(1+u)}\left (\theta_{3-j}\left (8\sigma^2\right )\,\theta_3\left (\sigma^2/2\right ) + \theta_{2+j}\left (8\sigma^2\right )\,\theta_4\left (\sigma^2/2\right )\right ).
%\end{equation}
This expression can be recovered using
\begin{align}
\Omega &= i\begin{pmatrix}
1 & 1 \\1 & -1
\end{pmatrix}\begin{pmatrix}
2\sigma^2 & 0 \\ 0 & \frac{1}{2\sigma^2}
\end{pmatrix}\begin{pmatrix}
1 & 1 \\1 & -1
\end{pmatrix}\label{eq:diagOmega}\\
2\sigma^2 &= \tanh\left (\Delta^2\right ) = \frac{1-u}{1+u},
\end{align}
and the modular transformation
\begin{equation}
\vartheta\left [\begin{matrix}
\frac{j}{2}\\0\end{matrix}\right ]\left (0,i/x\right ) = \sqrt{x}\,\vartheta\left [\begin{matrix}
0\\\frac{j}{2}\end{matrix}\right ]\left (0,ix\right ).
\end{equation}
%\begin{equation}
%\theta_{3-j}\left (1/x\right ) = \sqrt{x}\,\theta_{3+j}\left (x\right ).
%\end{equation}

%This implies that $c_n(j)=\langle n \ket{j_{\rm approx}}$ with Fock state $\ket{n}$, has the property that $\sum_{n=0}^{\infty} |c_n(j)|^2=1$. 

%In addition, we will also consider $I_{\rm approx}=\frac{1}{2}\sum_j \ket{j_{\rm approx}}\bra{j_{\rm approx}}$ and $\bra{n} I_{\rm approx} \ket{n}=P_n(I)\geq 0$ which obey $\sum_{n=0}^{\infty} P_n(I)=1$.

%Let $\theta_3(x)=\sum_{k \in \mathbb{Z}}  \exp(-\pi k^2 x)=1+2\sum_{k=1}^{\infty} \exp(-\pi k^2 x)$ for $x > 0$ be the Jacobi theta constant. Furthermore, we have $\bra{n}q\rangle=(2^n \sqrt{\pi} n!)^{-1/2} \exp(-q^2/2) H_n(q)$ with Hermite polynomial $H_n(q)=(-1)^n e^{q^2}\frac{d^n}{dq^n} e^{-q^2}$. We thus have 
% \begin{equation}
%c_n(j)=\frac{1}{\sqrt{N(\Delta, j)}}(2^n \sqrt{\pi} n!)^{-1/2}e^{-\Delta^2 n}   \sum_{k\in \mathbb{Z}}  e^{-((2 k+j)\sqrt{\pi})^2/2} H_n((2 k+j) \sqrt{\pi}).
%\end{equation}

We can turn to the Fock coefficients
\begin{equation}
c_n(j) = \braket{n\vert j_{\rm approx}} = \frac{1}{\sqrt{N(\Delta, j)}}\e^{-\Delta^2 n}   \sum_{k\in \mathbb{Z}}  \Psi_n((2 k+j) \sqrt{\pi}).\label{eq:hermfunccoef}
\end{equation}
As the parity of the Hermite function $\Psi_n(q)$ is that of the parity of $n$, the coefficients $c_{n}(j)$ vanish for odd $n$. We can use that for $q\geq 0$, see \cite{romik:taylor},
\begin{equation}
\Psi_{2n}(q)=\sqrt{\frac{4^n}{\sqrt{\pi}(2n)!}}\, \frac{d^n}{dz^n}\left(\frac{1}{\sqrt{1+z}} \exp\left(-\frac{q^2}{2}\cdot \frac{1-z}{1+z}\right)\right)\Bigg|_{z=0}.
\end{equation}
with $q=(2k+j)\sqrt{\pi}$. This implies that we have
%\begin{equation}
%c_{2n}(j)=\sqrt{\frac{4^n}{N(\Delta, j)\sqrt{\pi}(2n)!}}\,\e^{-2\Delta^2 n} \frac{d^n}{dz^n} \left[\frac{1}{\sqrt{1+z}} \vartheta\left [\begin{matrix}
%\frac{j}{2}\\0
%\end{matrix}\right ]\left(0, i2\frac{1-z}{1+z}\right)\right]\Bigg|_{z=0}.
%\label{eq:bound}
%\end{equation}
\begin{equation}
c_{2n}(j)=\sqrt{\frac{4^n}{N(u, j)\sqrt{\pi}(2n)!}}\,u^{n} \frac{d^n}{dz^n} \left[\frac{1}{\sqrt{1+z}} \theta_{3-j}\left (2\frac{1-z}{1+z}\right)\right]\Bigg|_{z=0}.
\label{eq:bound}
\end{equation}

\begin{figure}[h]
	\centering
	\includegraphics[width=\textwidth]{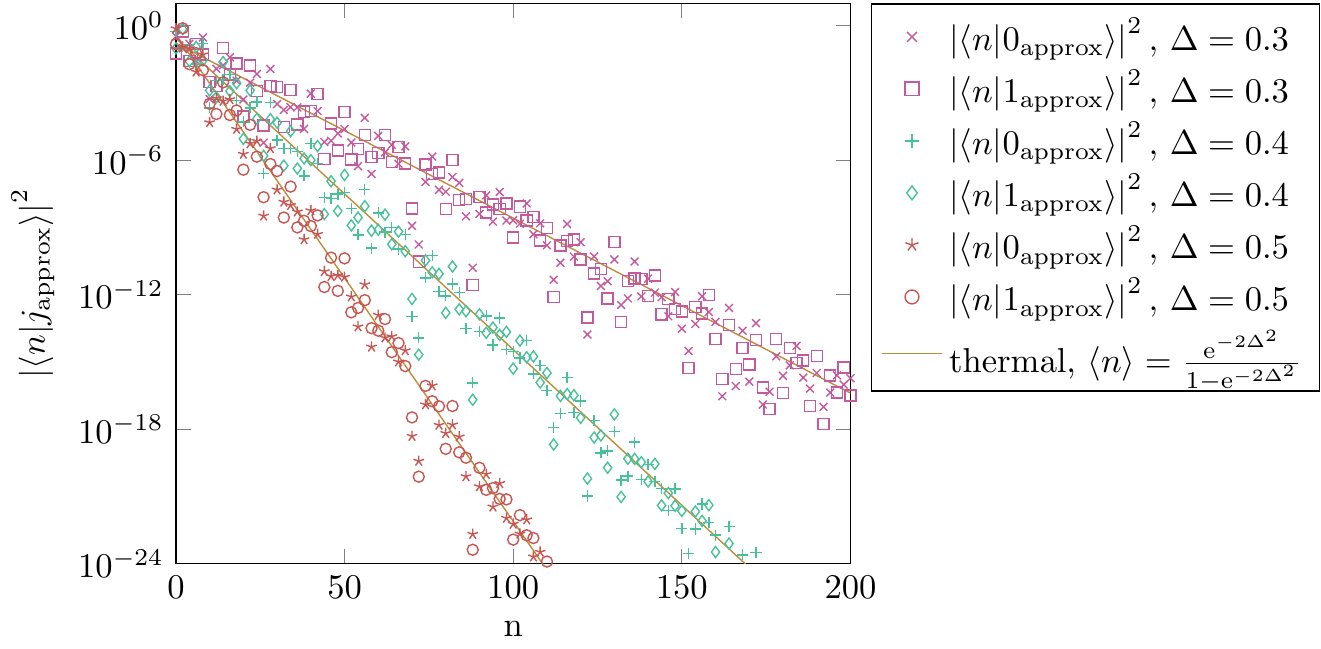}
	\caption{Fock coefficients of the ${\frak D}$-approximate GKP state $\ket{j_{\rm approx}}$ for $\Delta\in\{0.3,0.4,0.5\}$.
		Eqs.~\eqref{eq:hermfunccoef} and \eqref{eq:normapproxgkp} are used to numerically compute the coefficients.
		The photon distributions of a thermal distribution with an average photon number $\overline{n} =\frac{u}{1-u}= \frac{{\rm e}^{-2\Delta^2}}{1-{\rm e}^{-2\Delta^2}}$ are shown for comparison.}
	\label{fig:fockcoefgkp}
\end{figure}

This gives a somewhat concise expression although it is not directly useful.
In order to numerically evaluate the coefficients for example, Eq.~\eqref{eq:hermfunccoef} is more convenient as the Hermite functions, $\Psi_n(q)$,  have support essentially within $\left [-2\sqrt{n},\, 2\sqrt{n}\right ]$ and are easily computed recursively using
\begin{align}
\Psi_0(q) &= \pi^{-1/4}{\rm e}^{-\frac{q^2}{2}}, \\
%	\Psi_1(x) &= x\sqrt{2}\pi^{-1/4}{\rm e}^{-\frac{x^2}{2}},\\
\Psi_n(q) &= q\sqrt{\frac{2}{n}}\Psi_{n-1}(q) - \sqrt{\frac{n-1}{n}}\Psi_{n-2}(q).
\end{align}
This is used together with Eq.~\ref{eq:normapproxgkp} to plot the coefficients in Fig.~\ref{fig:fockcoefgkp}.

The case of the Fock representation of the so-called sensor state \cite{DTW:sensor} which is an approximate  eigenstate of $e^{i \sqrt{2\pi} \hat{q}}$ and $e^{i \sqrt{2\pi} \hat{p}}$ is very similar. For the perfect sensor state $\ket{\psi_{\rm ideal}}$ the wavefunction in $q$ is a sum over $\delta$-functions at integer multiples of $\sqrt{2\pi}$, and similarly the wavefunction in $p$ is a sum over $\delta$-functions at integer multiples of $\sqrt{2\pi}$.

\begin{figure}[h]
	\centering
	\includegraphics[width=\textwidth]{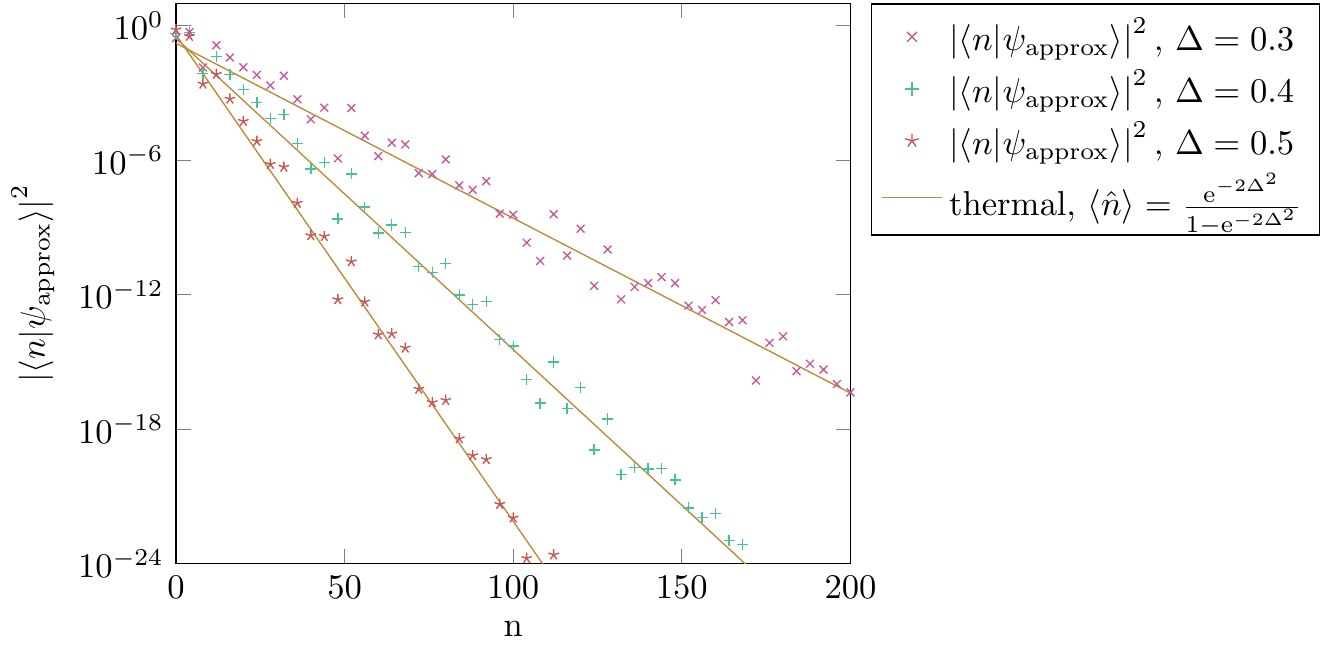}
	\caption{Fock coefficients of the ${\frak D}$-approximate sensor state $\ket{\psi_{\rm approx}}$ for $\Delta\in\{0.3,0.4,0.5\}$.
		Eqs.~\eqref{eq:cnsensor} and \eqref{eq:normapproxsens} are used to numerically compute the coefficients.
		The thermal distributions with average photon number $\overline{n} = \frac{u}{1-u} = \frac{{\rm e}^{-2\Delta^2}}{1-{\rm e}^{-2\Delta^2}}$ are shown for comparison.}
	\label{fig:fockcoefsensor}
\end{figure}

The approximate state is $\ket{\psi_{\rm approx}}=\frac{1}{\sqrt{N(\Delta)}}{\frak D} \ket{\psi_{\rm ideal}}$ with normalization 
\begin{equation}
N(\Delta)= \frac{1}{\sqrt{\pi(1-u^2)}}\,\bigvartheta\left [\begin{matrix}
\vec{0}\\\vec{0}
\end{matrix}\right ]\left (\vec{0}, \frac{1}{2}\Omega\right ),
\label{eq:normapproxsens}
\end{equation}
where $u$ and $\Omega$ are the same as defined in Eqs.~\eqref{eq:u} and \eqref{eq:Omega}.
This can be obtained in a similar way and we can also write $N(u)$ instead of $N(\Delta)$ when convenient. The sensor state $\ket{\psi_{\rm approx}}$ is an eigenstate of $\exp(i \pi a^{\dagger} a/2)$, hence the photon number is $0 \mod 4$ for this state \cite{DTW:sensor}. We now have $c_n=\bra{n} \psi_{\rm approx} \rangle$ with $c_{2n+1}=0$ and 
\begin{equation}
c_{2n}=\frac{1}{\sqrt{N(\Delta)}}\e^{-2\Delta^2 n}   \sum_{k\in \mathbb{Z}} \Psi_{2n}(k \sqrt{2\pi}).\label{eq:cnsensor}
\end{equation}
Ref.~\cite{romik:taylor} derives an expression analogous to Eq.~\eqref{eq:bound} for the sum on the right-hand-side:
\begin{align}
\sum_{k\in \mathbb{Z}}  \Psi_{2n}(k \sqrt{2\pi})=\theta_3(1)\sqrt{\frac{4^n\Phi^{n}}{\sqrt{\pi}(2n)!}}  d(n/2), n\; {\rm even} \label{eq:romik} \\
\sum_{k\in \mathbb{Z}}  \Psi_{2n}(k \sqrt{2\pi})=0, n\; {\rm odd}.
\notag
\end{align}
with $\Phi=\frac{\Gamma(\frac{1}{4})^8}{128 \pi^4}$, with $\Gamma()$ the Euler gamma function, and $\{d(n)\}_{n=0}^{\infty}$ is a particular integer sequence studied in \cite{romik:taylor} which is directly related to the derivatives of $\theta_3(x)$ at $x=1$ by
\begin{equation}
\frac{1}{\sqrt{1+z}}\theta_3\left (\frac{1-z}{1+z}\right ) = \theta_3(1)\sum_{n=0}^\infty\frac{d(n)}{(2n)!}\Phi^nz^{2n}.
\end{equation}
We show the Fock coefficients for the sensor state in Fig.~\ref{fig:fockcoefsensor}. Note that the sign of the integers $d(n)$ is the same as the sign of $c_{4n}$ through Eqs.~\ref{eq:cnsensor} and \ref{eq:romik}. We can write the sensor state $\ket{\psi_{\rm ideal}} \propto \sum_{t,s \in \mathbb{Z}}
e^{-i \sqrt{2 \pi} \hat{p}t} e^{i \sqrt{2\pi} \hat{q}s} \ket{0}$ where $\ket{0}$ is the vacuum state. Using that $e^{i \sqrt{2\pi} \hat{q}}=D(i \sqrt{\pi})$, $e^{-i \sqrt{2\pi} \hat{p}}=D(\sqrt{\pi})$ and $D(\alpha) \ket{\beta}= e^{i {\rm Im}(\alpha \beta^*)} \ket{\alpha+\beta}$, $\bra{n} \alpha \rangle=e^{-\frac{|\alpha|^2}{2}} \frac{\alpha^n}{\sqrt{n!}}$, we have
\begin{equation}
{\rm sign}(d(n))={\rm sign}(c_{4n})={\rm sign}\left(\sum_{t\in \mathbb{Z}, s \in \mathbb{Z}\colon z=t+is} e^{-\frac{1}{2}\pi |z|^2} z^{4n} (-1)^{t s}\right).
\end{equation}

We want now to derive the asymptotic behavior of the coefficients.
This can be done by expressing the normalization of the states by an equality which has to hold for all $\Delta\in(0,\infty)$ or equivalently all $u\in[0,1)$.
In the case of the sensor state, $\sum_{n \in \mathbb{N}} |c_{4n}|^2=1$ implies:
\begin{equation}
N(\Delta) =\frac{\theta_3(1)^2}{\sqrt{\pi}}\sum_{n \in \mathbb{N}}\frac{\left (2\sqrt{\Phi}u\right )^{4n}}{(4n)!} d^2(n),\label{eq:asympteqsensor}
\end{equation}
which is of the form:
\begin{equation}
N(u)= \frac{A}{\sqrt{\pi}}\sum_{n\in\mathbb{N}} a_{4n}u^{4n},
\end{equation}
where the constant $A$ and the sequence $a_{4n}$, both independent of $u$, are defined by Eq.~\eqref{eq:asympteqsensor}.
Similarly for the approximate GKP states we have: 
\begin{equation}
N(u,j) =\frac{1}{\sqrt{\pi}}\sum_{n \in \mathbb{N}}\frac{\left (2u\right )^{2n}}{(2n)!}\left (\frac{d^n}{dz^n} \left[\frac{1}{\sqrt{1+z}} \theta_{3-j}\left (2\frac{1-z}{1+z}\right)\right]\Bigg|_{z=0}\right )^2,\label{eq:asympteqgkp}
\end{equation}
which is of the same form:
\begin{equation}
N(u, j)= \frac{B(j)}{\sqrt{\pi}}\sum_{n\in\mathbb{N}} b_{2n}(j)u^{2n},
\end{equation}
for some other constant $B(j)$ and sequence $b_{2n}(j)$, also both independent of $u$, defined by Eq.~\eqref{eq:asympteqgkp}.

Seen as complex functions of $u\in\mathbb{C}$, $N(u, j)$ and $N(u)$ are both analytic and have a convergence radius of $1$.
Their behavior for $u\rightarrow 1$ can be obtained (see also \cite{MYK:approxGKP}) using Eq.~\eqref{eq:normapproxgkp} and the fact that
\begin{equation}
\theta_{3-j}(x) \;\underset{x\rightarrow 0}{\sim}\; \frac{1}{\sqrt{x}}, \qquad \theta_{4}(x) \;\underset{x\rightarrow 0}{\sim}\; \frac{2{\rm e}^{-\pi/4x}}{\sqrt{x}}.
\end{equation}
This gives
\begin{align}
\frac{A}{\sqrt{\pi}}\sum_{n\in\mathbb{N}} a_{4n}u^{4n} &= N(u) \;\underset{u\rightarrow 1}{\sim} \;\frac{1}{2\sqrt{\pi}(1-u)},\\
\frac{B(j)}{\sqrt{\pi}}\sum_{n\in\mathbb{N}} b_{2n}(j)u^{2n} &= N(u,j) \;\underset{u\rightarrow 1}{\sim}\; \frac{1}{2\sqrt{\pi}(1-u)}.
\end{align}
We then apply a transfer theorem, see \cite{flajolet_analytic_2009}, to deduce the asymptotic behavior of the $a_{4n}$ and $b_{2n}(j)$ sequences. 
More precisely they both converge to some finite value
\begin{equation}
a_{4n} \;\underset{n\rightarrow \infty}{\sim} \; \frac{1}{2A},\qquad b_{2n}(j) \;\underset{n\rightarrow \infty}{\sim} \; \frac{1}{2B(j)}.
\end{equation}
In turn this gives the asymptotic behavior of the Fock coefficients:
\begin{equation}
	\vert\braket{4n\vert\psi_{\rm approx}}\vert^2 \;\underset{n\rightarrow\infty}{\sim}\; \frac{u^{4n}}{2\sqrt{\pi}N(u)},\qquad \vert\braket{2n\vert j_{\rm approx}}\vert^2 \;\underset{n\rightarrow\infty}{\sim}\; \frac{u^{2n}}{2\sqrt{\pi}N(u,j)}.
\end{equation}

In both cases the coefficients are asymptotically equivalent to a geometric or thermal distribution which is usually parametrized by the average photon number $\overline{n}$ as follows
\begin{equation}
p_n^{\rm thermal} = \frac{1}{\overline{n}+1}\left (\frac{\overline{n}}{\overline{n}+1}\right )^n.
\end{equation}
We can therefore deduce the average photon number of the equivalent asymptotic thermal distribution to the approximate GKP and sensor states using
\begin{equation}
 u = \frac{\overline{n}}{\overline{n}+1} \;\Rightarrow\; \overline{n} = \frac{u}{1-u} = \frac{{\rm e}^{-2\Delta^2}}{1 - {\rm e}^{-2\Delta^2}}.
\end{equation}
These thermal distributions for the different $\Delta$s considered are also shown in Figs.~\ref{fig:fockcoefgkp} and \ref{fig:fockcoefsensor}. One can see a good agreement of the general trend although oscillations of the order of the probabilities themselves persist. Note also that for small $\Delta$, $\overline{n}  \sim \frac{1}{2\Delta^2}$, consistent with approximate expressions derived in other literature (see main text). 
 
% BMT TD 1/4 Delta^2

%Note that when you have the Wigner function, see \cite{MYK:approxGKP} you can determine all moments $n^k$ as these can be expressed as polynomials in $p$ and $q$. A thermal state equals $\rho_{\overline{n}}=\sum_{n=0}^{\infty} (\frac{\overline{n}}{1+\overline{n}})^n \ket{n}\bra{n}$.
%When $\Delta^2 \ll 1$, we can approximate $\overline{n}_{\rm GKP} \approx \frac{1}{2 \Delta^2}-\frac{1}{2}$.

%photon loss on a coherent state, i.e. $\frak{E} \ket{\alpha} \propto \ket{\alpha \exp(-\Delta^2)}$, corresponding to photon loss with rate $\gamma=2 \Delta^2$. 

\section{Decoders For Repeated GKP Error Correction With Finite Squeezing}
\label{app:decode}

For computational, simulation efficiency (as well as our formulation of a classical decoder) we use the approximation $\frak{F}_V$ for the GKP ancillas. This form of the states can be viewed as applying a reverse Villain approximation to the approximate $\frak{F}$-states in Eq.~(\ref{eq:approxE}). The reverse Villain approximation, which is tight for $a \geq 4 \pi^2$, reads
\begin{equation}
\sum_{n\in \mathbb{Z}} e^{-a n^2 +b n}\approx e^{-\frac{a}{2 \pi^2}}e^{\frac{b^2}{4a} +\frac{a}{2\pi^2}\cos{(\frac{\pi}{a} b)}}.\label{Vill} 
\end{equation}
Using Eq.~\eqref{Vill} and $\frac{\Delta^2}{\Delta^4+1} \approx \Delta^2$, 
%and taking $\Delta \leq 0.2$, 
we then have
\begin{align}
\frak{F}_V \ket{\overline{0}} =\int_{\mathbb{R}} dq\; \psi^0(q) \ket{q}, \;\; \psi^0(q)\propto e^{-\frac{\Delta^2}{2} q^2}e^{\frac{1}{\pi \Delta^2} \cos(\sqrt{\pi} q)}, \notag \\
\frak{F}_V \ket{\overline{+}}= \int_{\mathbb{R}} dq\;\psi^+(q) \ket{q}, \;\; \psi^+(q)\propto e^{-\frac{\Delta^2}{2}q^2} e^{\frac{1}{4 \pi \Delta^2}\cos(2\sqrt{\pi}q)}.
\label{eq:ancilla-approx}
\end{align}
%where we have made $\Delta_p \neq \Delta_q$ to see the explicit effect of finite-squeezing in either quadrature later on.
Using these ancillas, the Green's function for $M$ rounds of error correction ({\em without} active feedback between rounds), with outcomes denoted as $M$-dimensional vectors $\vec{\frak{p}}$ and $\vec{\frak{q}}$, can be written as
\begin{equation}
G^M(q_{\rm out} \leftarrow q_{\rm in}|\vec{\frak{q}},\vec{\frak{p}})=\int_{\mathbb{R}^{M+1}} d\vec{q}\, \exp(-S[\vec{q}|\vec{\frak{q}},\vec{\frak{p}}]) \delta(q_{\rm out}-q_M)\delta(q_{\rm in}-q_0), \label{GtoM}
\end{equation} 
with one-dimensional `action' $S[\vec{q}|\vec{\frak{q}},\vec{\frak{p}}]$:
\begin{align}S[\vec{q}|\vec{\frak{q}},\vec{\frak{p}}]& = \sum_{m=1}^M \frac{\Delta^2}{2}(q_m-q_{m-1})^2-\frac{1}{\Delta^2 \pi} \cos{\sqrt{\pi}(q_m-q_{m-1})} \nonumber \\ 
& -\sum_{m=1}^{M} \frac{1}{4 \pi \Delta^2} \cos{2\sqrt{\pi}(\frak{q}_m-q_{m}})+ \frac{\Delta^2}{2}(\frak{q}_m-q_{m})^2 \nonumber \\
& +i\sum_{m=1}^M \frak{p}_m (q_m-q_{m-1}). \label{eq:action}
\end{align}
% sum over M for the potential term? compare with our previous formulation, think is ok not a last round of errors..
% maybe specify parameters for the different ancilla to see how they affect the action
We can readily interpret the first line in the last equation as a kinetic energy term $T$, generating dynamics in the position variable, while the second line is a potential energy term $-U$, pinning the position variable to the measured values $\frak{q}_m$. We note that in the potential energy, the $\cos{2\sqrt{\pi}(\frak{q}_m-q_{m})}$ term is dominant when $\Delta$ is small and hence we can omit the wide parabolic potential proportional to $\Delta^2$. Similarly, for the kinetic energy, when $\Delta$ is small, expanding $\cos{\sqrt{\pi}(q_m-q_{m-1})}$ gives heavy-mass terms with mass $\sim \frac{1}{\Delta^2}$ and the light-mass quadratic term $\sim \Delta^2$ contributes little. We see that aside from the imaginary term on the last line, the action in this approximation is the same as in the Hamiltonian developed for stochastic noise in the reverse Villain approximation in Ref.~\cite{vuillot+:GKP} identifying $\sigma_0^2=\Delta^2$ with $\sigma_0$ the standard deviation in the Gaussian displacement model in Eq.~(\ref{eq:GDC}).
Note however that here the dynamics occurs at the level of wave functions, whereas the description in Ref. \cite{vuillot+:GKP} took place at the level of the probabilities (for shift errors).
The interesting difference lies in the pure imaginary term making the action $S$ complex. If the path integral is approximated by taking a single `classical' path which minimized ${\rm Re}(S)$ then this phase factor only gives an additional phase to this path. However, the pure imaginary term contributes a phase to each path so that the total sum of paths $q_{\rm in} \rightarrow q_{\rm out}$ can be different, due ito interference, than the case in which these phases are absent. Note that these terms comes from the $Z$-error correction step with outcome $\frak{p}_m$ which puts a feedback error on the data oscillator.

We formulate a decoder which will provide an approximate tracking of the dynamics of the wave function in $q$ and $p$ of the encoded state which is classical. This approximation can be viewed, to some extent, as making a classical approximation, i.e. selecting a single optimal classical path of a quantum path integral. 
We believe that similar ideas could be applied to simplified tracking of the Wigner function for the purpose of decoding: this may be of interest when we want to study the effects of a fuller noise model, which includes, say, photon loss on the cavity mode during ancilla GKP state preparation and photon loss on the ancilla prior to measurement.

The use of a classical approximation \footnote{One can also define a semi-classical approximation by expanding around the classical path and performing the Gaussian integral, but we have not numerically explored these variations on decoding.} to this path-integral is to determine the outgoing wave function without fully calculating the entire evolution by executing the $M-1$-dimensional integral. In this classical approximation we would have 
\begin{equation}
\psi_{\rm out}(q_{\rm out})\approx \int dq_{\rm in} \exp(- S[\vec{q}_{\rm opt}|\vec{\frak{q}},\vec{\frak{p}}])\psi_{\rm in}(q_{\rm in}).
\label{eq:class-approx}
\end{equation}
We imagine doing a final measurement of $\hat{q}$ on the outgoing wave function $\psi_{\rm out}$ and will be interested in evaluating Eq.~(\ref{eq:meas}). For the classical approximation we have
\begin{align}
\mathbb{P}(\overline{Z}=(-1)^b)=\int_{I_b} dq |\psi_{\rm out}(q)|^2= \notag \\
\int_{I_b} dq \int dq_{\rm in} \int q'_{\rm in} 
e^{- S[\vec{q}_{\rm opt}(q \leftarrow q_{\rm in})|\vec{\frak{q}},\vec{\frak{p}}]- S^*[\vec{q}_{\rm opt}(q \leftarrow q'_{\rm in})|\vec{\frak{q}},\vec{\frak{p}}]} \psi_{\rm in}(q_{\rm in}) \psi_{\rm in}^*(q'_{\rm in}),
\label{eq:interfer}
\end{align}
where we have explicitly indicated how the optimal path depends on the initial position $q_{\rm in}$ and the final position $q$. We note that the action in Eq.~(\ref{eq:action})
has the property $S^*[q_0, \ldots, q_M|\vec{\frak{q}}, \vec{\frak{p}}]=S[q_M, \ldots, q_0|\vec{\frak{q}}, \vec{\frak{p}}]$, i.e. it is the action of the time-reversed path. Thus when
$q'_{\rm in}=q_{\rm in}$, we have 
$e^{- S[\vec{q}_{\rm opt}(q \leftarrow q_{\rm in})|\vec{\frak{q}},\vec{\frak{p}}]- S^*[\vec{q}_{\rm opt}(q \leftarrow q_{\rm in})]}=e^{-2{\rm Re}(S[\vec{q}_{\rm opt}(q \leftarrow q_{\rm in})|\vec{\frak{q}},\vec{\frak{p}}])}$. In words: when we take a path along a closed loop from $q_{\rm in} \rightarrow q \rightarrow q_{\rm in}$, there is no phase accumulation. However, in Eq.~(\ref{eq:interfer}) there are certainly contributions when $q_{\rm in} \neq q'_{\rm in}$. 
For the purpose of developing an efficient classical decoder, we apply a stochastic approximation to Eq.~(\ref{eq:interfer}), keeping only the diagonal terms
i.e. 
\begin{equation}
\mathbb{P}_{\rm class}(b| \psi_{\rm in}, \vec{\frak{q}}, \vec{\frak{p}})=N \int_{I_b} dq \int dq_{\rm in} 
e^{- 2 {\rm Re}(S[\vec{q}_{\rm opt}(q \leftarrow q_{\rm in})|\vec{\frak{q}},\vec{\frak{p}}])}
|\psi_{\rm in}(q_{\rm in})|^2,
\label{eq:stoch}
\end{equation}
where $N$ is a normalization to make $\mathbb{P}_{\rm class}()$ a probability. Since this normalization does not play a role in the use of this expression in decoding, we do not need to determine it.  To evaluate Eq.~(\ref{eq:stoch}), one can generate a $q$ uniformly at random in the interval $I_b$ and $q_{\rm in}$ is generated according to $|\psi_{\rm in}|^2$. Given $q$ and $q_{\rm in}$, {\em if} we can evaluate the classical path $\vec{q}_{\rm opt}$ between these points and hence compute the weight $e^{- 2 {\rm Re}(S[\vec{q}_{\rm opt}(q \leftarrow q_{\rm in})|\vec{\frak{q}},\vec{\frak{p}}])}$ corresponding to this path, then we can stochastically estimate Eq.~(\ref{eq:stoch}). However, as was observed in Ref.~\cite{vuillot+:GKP} this classical-path approach is not computationally simple as the dynamics of the $q$-variable can be chaotic due to it taking place in a random potential induced by the measurement outcomes $\vec{\frak{q}}, \vec{\frak{p}}$. Hence, instead of sampling $q_{\rm in}$ and the endpoint $q$, we will sample $q_{\rm in}$ from $|\psi_{\rm in}(q_{\rm in})|^2$ and then apply a forward minimization technique, similar as in Ref.~\cite{vuillot+:GKP}, on the function ${\rm Re}(S)[.]$ to find  an approximately optimal path $\vec{q}_{\rm opt}$ given $q_{\rm in}$, leading to a final value for $q_{\rm out}$. We then determine whether $q_{\rm out}$ lies in $I_b$ and repeat to gather statistics. In fact, we observe numerically that this strategy shows the same performance as fixing an initial $q_{\rm in}=\underset{q}{\mathrm{argmax}} |\psi_{\rm in}(q)|^2$ to which we apply the forward minimization. The probability to land in the $I_b$ interval is denoted as $\mathbb{P}_{\rm forward}(b| \psi_{\rm in}, \vec{\frak{q}}, \vec{\frak{p}})$.

To adapt this forward decoder to the corrective displacement in each round as shown in Fig. \ref{fig:EC_GKP}, the action in Eq.~\eqref{eq:action} is modified by substituting $q_m \rightarrow q_m+\frak{q}_m$ for each round (and the addition of a term $i q_m \frak{p}_m$ which is irrelevant for this decoder). 

%Note that each iteration of the minimization is independent of the measurement result of the previous round, i.e. if $S[(q_m,q_{m-1})|\frak{q}_m, \frak{p}_m]$ is the action for a single round without final displacement, in each step we minimize for $q_m$ over $S[(q_m+\frak{q}_m,q_{m-1})|\frak{q}_m, \frak{p}_m]$.

The memoryless decoder presented in Fig.~\ref{fig:sim} in the main text is implemented by decomposing each measurement outcome $\frak{q}_m=\frak{l}^q_m\sqrt{\pi}+\frak{n}^q_m 2\sqrt{\pi}+\frak{e}^q_m$ and $\frak{p}_m=\frak{l}^p_m\sqrt{\pi}+\frak{n}^p_m 2\sqrt{\pi}+\frak{e}^p_m$ with $\frak{l}^{q/p}_m \in \mathbb{Z}_2$ (yes/no logical shift) $\frak{n}^{q/p}_m \in \mathbb{Z}$ (number of stabilizer shifts), $|\frak{e}^{q/p}_m|<\frac{\sqrt{\pi}}{2}$ (minimal shift error) and applying a corrective shift of $\alpha_m = \frac{-\delta_{q,m}+i\delta_{p,m}}{\sqrt{2}}$ with $\delta_{q/p,m}=\frak{n}^{q/p}_m 2\sqrt{\pi}+\frak{e}^{q/p}_m$ after the QEC round. Note that this is not the same correction as in Fig.~\ref{fig:EC_GKP}. This correction is motivated by a stochastic-shift error model and ensures that we keep the photon number low by applying an appropiate number of stabilizer displacements while correcting the perceived error.

%In case of additional photon loss between the QEC rounds, we modify this strategy using the update of the probability distribution over $q$ in Eq.~(\ref{eq:OU}) due to photon loss as follows. After each QEC round forward minimization gives us the $q_m$ for this round. We move to a $q'_m$ with transition probability in Eq.~(\ref{eq:OU}) for a given $\gamma=1-e^{-\kappa t}$ and apply the forward minimization of the next round.However this happens at the level of probability distributions, not wave functions..so the forward decoder would have to work at the level of probability distributions....

%As a side remark, on the fact that we could have used another approximation for the input ancilla's, e.g. Eq.~(\ref{eq:approxE}) which would have kept the path integral a Gaussian integral with an additional summation over integer variables, similar as Eq. (22) in \cite{vuillot+:GKP}: this approach however becomes computationally expensive with increasing rounds of measurements.
% CV, JC check whether you agree

Let us now further discuss how we test and evaluate our decoders.

\paragraph{Decoding of Repeated GKP Error Correction}

The experimentalist implementing the repeated rounds of error correction learns the value of $\vec{\frak{p}},\vec{\frak{q}}$ and the outcome of the final perfect homodyne measurement of $\hat{q}$, but in general she should make decoding decisions {\em without knowing the input state} \footnote{The reason is that this state may be part of a complicated quantum computation and we assume that we cannot simulate this quantum computation.}. However, she can play the game in which she assumes that the input wave state is either $\psi^0_{\rm in}(q)$ (or $\psi^1_{\rm in}(q)$) and then determine whether the dynamics, --given her measurement data $\vec{\frak{q}},\vec{\frak{p}}$--, would lead to a read-out of $\overline{Z}=1$ or $\overline{Z}=-1$. Naturally this may not a deterministic process, so she can calculate the probabilities
$\mathbb{P}(0|\psi_{\rm in}^0,\vec{\frak{q}},\vec{\frak{p}})$ ($\overline{Z}=1$) and $\mathbb{P}(1|\psi_{\rm in}^0,\vec{\frak{q}},\vec{\frak{p}})$ ($\overline{Z}=-1$). This defines a maximum-likelihood full-density matrix decoder. When using this decoder, the experimentalist evaluates the (normalized) Green's function in Eq.~(\ref{GtoM}), given her measurement data. She then flips the final experimentally measured logical outcome bit (modeled as a perfect homodyne measurement of $\hat{q}$) whenever $\mathbb{P}(1|\psi_{\rm in}^0,\vec{\frak{q}},\vec{\frak{p}})> \mathbb{P}(0|\psi_{\rm in}^0,\vec{\frak{q}},\vec{\frak{p}})$.
As logical error rate estimate for this decoder we take
\begin{equation}
\overline{P}_{\rm MLD}=\int d\vec{\frak{q}} \int d\vec{\frak{p}} \;\mathbb{P}(\vec{\frak{q}},\vec{\frak{p}})\min\left(\mathbb{P}(0|\psi_{\rm in}^0,\vec{\frak{q}},\vec{\frak{p}}),\mathbb{P}(1|\psi_{\rm in}^0,\vec{\frak{q}},\vec{\frak{p}})\right).
\label{eq:MLD}
\end{equation} 
This estimate assumes that the error rate is (roughly) the same when the input wavefunction is $\psi_{\rm in}^1(q)$ and that a superposition of such inputs behaves similarly, without logical interference. 

For systems of many modes, this decoding method will not typically be efficient (even for a single mode it comes down to a sizable computational effort), hence the goal of decoding is to infer errors without tracking the entire wavefunction. As we argued before, it turns out to be most computationally efficient to use a forward strategy and estimate $\mathbb{P}_{\rm forward}(0|\psi^0_{\rm in},\vec{\frak{q}},\vec{\frak{p}})$. With this probability in hand, let $f(\vec{\frak{q}},\vec{\frak{p}})=\underset{b}{\operatorname{argmin}} (\mathbb{P}_{\rm forward}(b|\psi^0_{\rm in},\vec{\frak{q}},\vec{\frak{p}}))$ be the indicator bit whether to flip.The logical error estimate that we consider is then
\begin{equation}
\overline{P}_{\rm forward}=\int d\vec{\frak{q}} \int d\vec{\frak{p}} \;\mathbb{P}(\vec{\frak{q}},\vec{\frak{p}}) 
\left[\delta_{f(\vec{\frak{q}},\vec{\frak{p}}),0}\mathbb{P}(1|\psi_{\rm in}^0,\vec{\frak{q}},\vec{\frak{p}})+\delta_{f(\vec{\frak{q}},\vec{\frak{p}}),1}\mathbb{P}(0|\psi_{\rm in}^0,\vec{\frak{q}},\vec{\frak{p}})\right].
\label{eq:class}
\end{equation}
Again, in using only this estimate, it is assumed (and not a priori given) that the evolution of the input state $\psi^1_{\rm in}(q)$ or an arbitrary superposition will have the same logical error rate, and that assuming a different input state would not lead the decoder to different conclusions. The logical error probability for the passive decoder which does not use any syndrome measurement data, but declares that output state is the same as input, is defined as
\begin{equation}
\overline{P}_{\rm passive}=\int d\vec{\frak{q}} \int d\vec{\frak{p}} \;\mathbb{P}(\vec{\frak{q}},\vec{\frak{p}}) \mathbb{P}(1|\psi_{\rm in}^0,\vec{\frak{q}},\vec{\frak{p}}).
\label{eq:zeno}
\end{equation}
Thus, similar as in Eq.~(\ref{eq:zeno}), the logical error is determined by the probability to get outcome 1 in the final $q$ measurement.
We have numerically estimated the logical errors rates of these three decoders, given in Eqs.~(\ref{eq:MLD}, \ref{eq:class}, \ref{eq:zeno}), as a function of $M$ for $\Delta=0.3$ and $\Delta =0.4$, see Fig.~\ref{fig:sim} in the main text. The logical error rates for adapted versions of these decoders including corrective displacements after each round are displayed in Fig.~\ref{fig:simdisp}.

\section{Hamiltonian Engineering Via Rotating Wave Approximations}
\label{sec:RWA}

The goal of this Appendix is to discuss the underpinnings and the `beyond' of the commonly-invoked rotating wave approximation of a Hamiltonian of the form $H=H_0+V$ with
\begin{equation}
H_0=\omega_a (a^{\dagger} a+\frac{1}{2})+ \omega_b (b^{\dagger} b+\frac{1}{2})+\omega_c (c^{\dagger} c+\frac{1}{2}),\;V=\sum_k \lambda_k \Phi^k, \Phi=\alpha \hat{q}_a+\beta \hat{q}_b+ \gamma \hat{q}_c.
\label{eq:examH}
\end{equation}
First, in the absence of applying time-dependent drives, the physical basis for an RWA approximation of a $\Phi^k$-term can be motivated in at least two ways. 
One is to move to a rotating frame in which the Hamiltonian remains time-independent, but is a form amenable to Schrieffer-Wolff degenerate perturbation theory so that terms which either (1) contain an unequal number of creation and annihilation operators, and/or (2) contain an equal number of creation and annihilation operators of modes which are sufficiently far detuned, are seen as perturbations to detuned Fock energy levels. For example, in Eq.~(\ref{eq:examH}), to argue about the perturbative effect of terms which contain an unequal number of creation and annihilation operators, we observe that they act as off-diagonal elements in the Fock basis of $H_0$, changing the number of total excitations. Hence if view $H_0$ as block-diagonal with blocks formed by a given total number of excitations in either mode a, b or c, separated by a gap $\min(\omega_a, \omega_b,\omega_c)$, then the effect of such excitation-number changing terms can by examined by Schrieffer-Wolff perturbation theory. In lowest-order perturbation theory, one projects $V$ onto these blocks labeled by the number of excitations, so that off-diagonal terms have no effect, Kerr and cross-Kerr terms remain, as well as excitation-number preserving terms (e.g. $a^{\dagger} b^{\dagger} c^2$). Given the GHz frequencies of the $\omega_i$s versus the relative strength of Josephson-induced coupling, keeping things to this lowest-order is a good approximation and we replace $V$ by $V_{\rm eff}$ which omit these terms which do not preserve the number of excitations. This strictly speaking is the rotating-wave-approximation.

As a next approximation, to handle terms which do not preserve excitation number in any of the {\em particular} modes, we go to the rotating frame at $\omega_c$ for all three modes so that $\tilde{H}_0=\Delta_{ac} (a^{\dagger} a+\frac{1}{2})+\Delta_{bc} (b^{\dagger} b+\frac{1}{2})$ with $\Delta_{ic}=\omega_i-\omega_c$ and $\tilde{V}_{\rm eff}=V_{\rm eff}$. We imagine that $\tilde{V}_{\rm eff}$ is expanded in terms which are products of creation and annihilation operators of the modes. 
The Hamiltonian $\tilde{H}_0$ has energy eigenspaces $\ket{x}_a \otimes \ket{y}_b \otimes \ket{\psi}_c$ with Fock states $x,y=0,1, \ldots$, each of which has the degeneracy of the oscillator (c) space, and each space is at least $\min(|\Delta_{ba}|, |\Delta_{ca}|, |\Delta_{bc}|)$ away from another space. The perturbation $\tilde{V}_{\rm eff}$ has both diagonal parts with respect to these eigenspaces (e.g. Kerr and cross-Kerr) as well as off-diagonal terms which map between the spaces. In lowest-order perturbation theory, one again projects $\tilde{V}_{\rm eff}$ onto these eigenspaces, so that off-diagonal terms have no effect and Kerr and cross-Kerr terms remain. This is then the full (RWA) approximation. To go beyond this, the effect of the off-diagonal terms can be estimated in second or higher-order perturbation theory to obtain an effective Hamiltonian $H_{\rm eff}$ which is diagonal in the Fock basis using Schrieffer-Wolff perturbation theory (see e.g. \cite{blais2004, SW:gadget, BDL:SW, consani:SW}). The spectrum of $H_{\rm eff}$ approximates that of $H$ when we are in the perturbative regime with $\min(\Delta_{ij}) \ll ||V_{\rm eff}||/2$. In second and higher-order perturbation theory, i.e. beyond RWA, $U H_{\rm eff}U^{\dagger}$ provides an approximation for $\tilde{H}$ where the unitary $U$ is the perturbatively expanded Schrieffer-Wolff transformation which provides a correction to the Fock eigenbasis. To get an approximation to the original Hamiltonian, one would finally rotate back to the original frame.

An alternative analysis could be based on the Magnus expansion: to apply this, we move to a rotating frame for each mode at its own frequency for the Hamiltonian in Eq.~(\ref{eq:examH}) such that the previously mentioned off-diagonal terms  become rapidly time-dependent: as a consequence their effect averages out over sufficiently long times. To observe the dynamics in this rotating frame, we consider the Schr\"odinger equation for the state $\ket{\tilde{\psi}(t)}=U_0^{\dagger}(t) \ket{\psi(t)}$ with $U_0=\exp(-i H_0 t)$, while $\ket{\psi(t)}$ obeys the Schr\"odinger equation with $H$. The state $\ket{\tilde{\psi}(t)}$ will evolve according to the rotating-frame (or interaction-frame) Hamiltonian $\tilde{H}=U_0^{\dagger} H U_0+i \frac{d U_0^{\dagger}}{dt} U_0$. For example, when $V= \lambda [a^2 b^{\dagger} c^{\dagger}+ h.c]$, one has 
\begin{equation}
\tilde{H}=\lambda [a^2 b^{\dagger} c^{\dagger} e^{i (2\omega_a-\omega_b-\omega_c)t}+ h.c].  
\end{equation} 
For time-dependent Hamiltonians, the Magnus expansion \cite{blanes+:magnus} or Magnus-Taylor expansion \cite{zeuch+:RWA} then forms a convenient representation of the effective dynamics. For the time-evolution (in the rotating frame) from an initial time $t=0$ to a final time $T$, the Magnus expansion reads
\begin{equation}
U(T, 0)={\cal T} \exp(-i \int_{0}^{T} dt \tilde{H}(t))=\exp(-i \overline{H}(T)), 
\end{equation}
with $\overline{H}(T)=\sum_{k=1}^{\infty}\overline{H}^{(k)}(T)$ and the first two terms equal 
\begin{align*}
\overline{H}^{(1)}(T)=\int_{0}^{T} dt \tilde{H}(t), \;\; \overline{H}^{(2)}(T)=-\frac{1}{2} \int_{0}^{T} dt' \int_{0}^{t'}  dt [\tilde{H}(t'),\tilde{H}(t)], 
\end{align*}
while for $k \geq 2$ one gets increasingly higher-order commutators \cite{blanes+:magnus}. For the Hamiltonian in Eq.~(\ref{eq:examH}), clearly terms diagonal in the Fock basis, are time-independent and will be present in $\overline{H}^{(1)}(T)$. As an example of a rotating term, consider a simple time-dependent Hamiltonian $\tilde{V}(t)=A \exp(i\Delta t)+h.c$ where $A$ is some product of creation and annihilation of some modes and $\Delta$ is a detuning. For this Hamiltonian, the strength of this lowest-order term decays inversely with $\Delta$, i.e. 
\begin{equation}
||\overline{V}^{(1)}(T) || \leq \frac{2}{\Delta} (|| A|| +||A^{\dagger}||) \sin(\Delta T/2)\leq \frac{4}{\Delta} || A|| , 
\end{equation}
scaling with the perturbative parameter $||A||/\Delta$. 
%The strength of the next term \begin{equation}||\overline{V}^{(2)}(T) || =|| \int_{0}^{T} dt' \int_{0}^{t'}  dt [A,A^{\dagger}]\sin(\omega(t'-t))||=||[A,A^{\dagger}]|| \frac{\omega T-\sin(\omega T)}{\omega^2}.\end{equation}
%grows with $T$ but is also suppressed by $1/\omega$.  
%When we use 3- or 4-wave ($\Phi^3$ or $\Phi^4$) mixing elements on 3 modes, we have some $\tilde{H}(t)=\sum_i  A_i \exp(i\omega_i t)+h.c$ where $A_i$ is some product of creation and annihilation of these 3 modes and each mode operator occurs at most 3 or 4 times. 
% BMT Does the Magnus expansion determinate {\em at a finite order}?  No

In general, a sufficient condition for the convergence of the Magnus expansion is that $\int_0^T dt ||\tilde{H}(t)|| < \pi$, but, similar as in degenerate perturbation theory, it can still serve as useful asymptotic series expansion.

We thus see that both methods give us perturbative expansions whose validity depends on the strength of the perturbative parameter.

Let us now discuss the case when we actively drive one of the modes, say mode c. As before one can apply a RWA making dropping terms which do not preserve total excitation number to obtain an effective $V_{\rm eff}$. If we assume that the only effect of the drive term is to create a coherent state with $\langle c(t) \rangle={\cal E} e^{-i \omega_p t}$ in oscillator c, we can remove $\omega_c c^{\dagger} c$ from $H$ and replace $c$ by $\langle c(t)\rangle$ everywhere. Given this time-dependence it then seems more convenient to use a Magnus expansion to analyze the effect of the higher-order effects of the non-resonant terms. For this we move to a rotating frame for the oscillators a and b at their own frequency, such that depending on the choice of the drive frequency $\omega_p$ some terms become time-independent. For example, a term like $a^{\dagger} b c$ is time-independent for the choice $\omega_p=\omega_a- \omega_b$.

%Assume that we have already neglected excitation-number changing terms in the Hamiltonian, that is, their relative weakness is somewhat affected by driving as it can increase the norm of the non-excitation preserving perturbations versus $\omega_i$ but we assume that this is a small effect for the moment. When we actively drive one of the modes, say drive mode c at frequency $\omega_d\equiv \omega_a-\omega_b$, we can model it by adding a term $H_{\rm drive}= c e^{i \omega_d t}+c^{\dagger} e^{-i \omega_d t}$ to $H$ in Eq.~(\ref{eq:examH}) and move to a rotating frame for modes a and b at this frequency $\omega_d$, so that $\tilde{H}_0=\omega_b (a^{\dagger} a+\frac{1}{2})+(2\omega_b-\omega_a)(b^{\dagger} b+\frac{1}{2})+\omega_c (c^{\dagger} c+\frac{1}{2})$. If we assume that the only effect of the drive term is to create a coherent state with $\langle c(t) \rangle=P e^{i (\omega_a-\omega_b)t}$ in oscillator c, then the perturbation $\tilde{V}_{\rm eff}=V_{\rm eff}(P)$ where each occurrence of $c$ (resp. $c^{\dagger}$) has been replaced by $P$ (resp. $P^*$) in $V_{\rm eff}(P)$. Hence $V_{\rm eff}(P)$ is time-independent and this can be understood from the fact that it only contains excitation-number preserving terms.
%Now the perturbation $V$ is not necessarily small in strength as the occupation of mode $c$ diagonal elements. 

A more thorough quantitative analysis of the error induced by the RWA approximation through perturbative or Magnus expansions would be desirable, as it plays into the accuracy of the CZ gate and is influenced by the number of photons in a GKP mode (as the latter influences the strength of the perturbation).

%Usually, there are two parts in a rotating-wave approximation. One is the omission of the non-secular terms, that is, terms which intrinsically do not conserve energy and contain a different number of annihilation and creation operators, such as $(a^{\dagger})^4$. The second approximation is based on neglecting terms which do contain an equal number of creation and annihilation operators, like $(a^{\dagger})^2 b c$ but modes $a$, $b$ and $c$ have different angular frequencies $\omega_a,\omega_b,\omega_c$ and none of them is actively pumped. 

\subsection{Time-Dependent Displacement Frame}
\label{sec:displace}
 
Imagine a Hamiltonian (in a rotating frame of the mode $a$) of the form $H=H_1(a, a^{\dagger})+i{\cal E}(t)a^{\dagger}-i {\cal E}^*(t)a$ where ${\cal E}(t)$ is some time-dependent envelope of the drive and $H_1(a, a^{\dagger})$ has some functional form on $a$ and $a^{\dagger}$. 
We consider the time-evolution of the vector $\ket{\tilde{\psi}(t)}=U_0^{\dagger}(0, t) \ket{\psi(t)}$ 
with $U_0(0,t)={\cal T} \exp(-i \int_{0}^t dt'[ i{\cal E}(t') a^{\dagger}-i {\cal E}^*(t')a])$ which evolves with the Hamiltonian $\tilde{H}(t)=U_0^{\dagger} H U_0+i\frac{\partial U_0^{\dagger}}{\partial t} U_0$. In words, we consider the evolution in the time-dependent displacement frame given by the time-dependent drive. 
When after some final time $T$, the total frame evolution is $U_0(0,T)=I$, the time-independent Hamiltonian $\tilde{H}(t)$ will describe the time-evolution of the actual Schr\"odinger state $\ket{\psi(t)}$ over the entire period of time $T$. 
% When this is convenient, $U_0(0,t)$ can also dependent on some ${\cal E}'(t)$ which is not identical to the time-dependent drive envelope in $H$.
We can write $U_0(0,t)=D(\beta(t))$ where $\beta(t)=\int_0^t dt' {\cal E}(t')$, so that we evaluate $\tilde{H}(t)=H_1(a+\beta, a^{\dagger}+\beta^*)+i {\cal E}(t)\beta^*-i {\cal E}^*(t)\beta$ and the last term can be omitted as it give rise to an irrevelant phase. 

% paper https://arxiv.org/pdf/1412.4633.pdf in which you drive one of the resonators..Yurke idea.
%Imagine we have the interaction $H=\omega_a a^{\dagger} a +\omega_b b^{\dagger} b +g ((a^{\dagger})^2 b^2+h.c)+ H_{\rm drive}$ with $H_{\rm drive}=\epsilon XX$where we drive $b$ both at its resonant frequency $\omega_b$ as well as at a pump frequency $\omega_p=2\omega_a-\omega_b$. If we replace $\langle b \rangle =\alpha_p \exp(-i \omega_p t)+\alpha_b \exp(-i \omega_b t)$, we see that in the rotating frame of two oscillators, we can make the term resonant.

%\cite{Vool2016}oi       = {10.1103/PhysRevA.73.012325},

\clearpage
\bibliographystyle{iopart-num}
\bibliography{bos-review-refs.bib}
\end{document}